\documentclass[aps, twocolumn, superscriptaddress]{revtex4-1}

\usepackage{graphicx}
\usepackage{amsmath}
\usepackage{hyperref}

\newcommand{\vdW}{{\mbox{\scriptsize vdW}}}
\newcommand{\LDA}{{\mbox{\scriptsize LDA}}}
\newcommand{\nl}{{\mbox{\scriptsize nl}}}
\newcommand{\eff}{{\mbox{\scriptsize eff}}}
\newcommand{\He}{{\mbox{\scriptsize He}}}
\newcommand{\homm}{{\mbox{\scriptsize hom}}}
\newcommand{\london}{{\mbox{\scriptsize London}}}
\newcommand{\heg}{{\mbox{\scriptsize HEG}}}
\newcommand{\self}{{\mbox{\scriptsize self}}}
\newcommand{\bfr}{{\mathbf{r}}}
\newcommand{\nr}{{$n(\mathbf{r})$}}
\newcommand{\Tr}{\mathrm{Tr}}
\newcommand{\bk}{\mathbf{k}}
\newcommand{\bq}{\mathbf{q}}
\newcommand{\diff}{\mathrm{d}}
\newcommand{\eps}{\epsilon}
\newcommand{\xc}{{\rm xc}}
\newcommand{\x}{{\rm x}}
\newcommand{\p}{{\rm p}}
\newcommand{\omp}{\omega_{\rm p}}
\newcommand{\Enl}{E_{\rm c}^{\nl}[n]}
\newcommand{\bqm}{\mathbf{q'}}
\newcommand{\vv}[1]{{\mathbf{#1}}}
\newcommand{\abs}[1]{{|{#1}|}}
\newcommand{\grad}{\nabla}
\newcommand{\rr}{{\bf r}}
\newcommand{\RR}{{\bf R}}
\newcommand{\rt}{{\bf\tilde{r}}}
\newcommand{\redgrad}{{\bf s}}
\newcommand{\kF}{k_{\text{F}}}
\newcommand{\dr}{\diff^3\mspace{-2mu}r}
\newcommand{\rk}{{\bf k}}
\newcommand{\dk}{\diff^3\mspace{-2mu}k}

\newcommand{\Oslo}{Centre for Materials Science and Nanotechnology, SMN, University of Oslo, NO-0318 Oslo,  Norway}

\newcommand{\MCtwo}{Microtechnology and Nanoscience, MC2, Chalmers
University of Technology, SE-412 96 G{\"o}teborg, Sweden.}
\newcommand{\LBNL}{Molecular Foundry, Lawrence Berkeley National
Laboratory, Berkeley, California 94720, USA.}
\newcommand{\UCB}{Department of Chemical and Biomolecular Engineering,
University of California, Berkeley, California 94720, USA.}
\newcommand{\Wake}{Department of Physics, Wake Forest University,
Winston-Salem, North Carolina 27109, USA.}
\newcommand{\APP}{Department of Applied Physics, Chalmers University of
Technology, SE-412 96 G{\"o}teborg, Sweden.}
\newcommand{\ORNL}{Materials Science and Technology Division, Oak Ridge
National Laboratory, Oak Ridge, Tennessee 37831-6114, USA.}

\graphicspath{{./figures/}}
\begin{document}
%\title{van der Waals interactions in density functional theory}
\title{van der Waals forces in density functional theory: The vdW-DF method}
\author{Kristian Berland}    \affiliation{\Oslo}
\author{Valentino R. Cooper} \affiliation{\ORNL}
\author{Kyuho Lee}           \affiliation{\LBNL}\affiliation{\UCB}
\author{Elsebeth Schr\"oder} \affiliation{\MCtwo}
\author{T. Thonhauser}       \affiliation{\Wake}
\author{Per Hyldgaard}       \affiliation{\MCtwo}
\author{Bengt I. Lundqvist}  \affiliation{\APP}
\date{\today}

\begin{abstract}
A density functional theory (DFT) that accounts for van der Waals (vdW)
interactions in condensed matter, materials physics, chemistry, and
biology is reviewed. The insights that led to the construction of the
Rutgers-Chalmers van der Waals Density Functional (vdW-DF) are presented
with the aim of giving a historical perspective, while also emphasising
more recent efforts which have sought to improve its accuracy. In
addition to technical details, we discuss a range of recent applications
that illustrate the necessity of including dispersion interactions in
DFT. This review highlights the value of the vdW-DF method as a
general-purpose method, not only for dispersion bound systems, but also
in densely packed systems where these types of interactions are
traditionally thought to be negligible.
\end{abstract}

\maketitle

\tableofcontents
%%%%%%%%%%%%%%%%%%%%%%%%%%%%%%%%%%%%%%%%%%%%%%%%%%%%%%%%%%%%%%%%%%%%%%%%

\section{Introduction}
%%%%%%%%%%%%%%%%%%%%%%%%%%%%%%%%%%%%%%%%%%%%%%%%%%%%%%%%%%%%%%%%%%%%%%%%

The field of van der Waals (vdW) interactions has grown enormously since its infancy \cite{luanetal95}. This is particularly true when considering how such forces can be included within electronic structure methods. As such it is not possible to cover the entire field in a single review paper. This review focuses on the history, development and application of one specific approach for including dispersion interactions within the framework of density functional theory; the van der Waals density functional (vdW-DF). For reviews on other methods and approaches we direct the reader to references \cite{rev1,rev2,rev3,rev4,rev5,rev6,Cooper10p1417,rev8,BeckePerspective}.

%Our late colleague David Langreth was invited by reports on progress in physiscs (ROPP) to  review our vdW-DF method. The whole field of van der Waals  (vdW) interactions in electron-structure calculations have  grown enormously ever since its infancy \cite{luanetal95}. 
%It was early clear that we would not be able to cover the whole field in a single review paper. In David's case he barely got time to start on it. Fortunately, ROPP invited us to inherit the task of reviewing the vdW-DF method. We owe him and the community a gathering of our somewhat scattered contributions.

Identified in 1873, there is a force that today attracts more interest
than ever. It was first introduced in a doctoral thesis by Johannes
Diderik van der Waals ``on the continuity of the gaseous and liquid
state" at Leiden University \cite{vdW1873}. The existence of the van der
Waals (vdW) force \cite{MaKe69} is today well established. It is present
everywhere, but its variation from one environment to another and its
complex manifestations still pose challenging questions nearly one
hundred years after van der Waals was awarded the Nobel Prize in
physics. These questions are relevant for such varied systems as soft
matter, surfaces, and DNA, and in phenomena as different as
supramolecular binding, surface reactions, and the dynamic properties of
water. Long-standing observations together with a flow of improved
experiments challenge existing theories, and a general theoretical
framework that can describe small molecules as well as extended systems
is needed.

The vdW interaction is a true quantum phenomenon \cite{EiLo30,Lo37}.
Asymptotic models and interpretations, such as those setting the
interactions of point particles with separation $R$ and molecules at a
distance $z$ outside a surface as $R^{-6}$ \cite{EiLo30,Lo30} and
$z^{-3}$ \cite{LJ32}, respectively, 
 and the fact that the multitude of such power laws
 for this microscopic force \cite{persson2008} depends on the microscopic shapes, were given early. 
Today's 
challenges might concern more subtle observations, such as detailed analyses of diffractive
scattering of H$_2$ and D$_2$ off single-crystal surfaces or the properties of DNA and liquid water.

In physical-chemistry terminology the term vdW includes the following
forces between molecules: (i) two permanent dipoles (Keesom force), (ii)
a permanent dipole and a corresponding induced dipole (Debye force), and
(iii) two instantaneously induced dipoles (London dispersion force)
\cite{EiLo30,Lo30}. In the condensed-matter community, typically just
the latter, which has a nonclassical origin, is referred to as the vdW
force.

A proper theory for atoms and molecules should account for all forces at
play, including covalent bonds, hydrogen bonds, and electrostatic
interactions, because they are all 
relevant in typical materials and systems. Density functional theory (DFT)
\cite{HoKo64,KoSh65,JoGu89,Burke} is such a general framework for
bonding and structure. vdW interactions
emanate from dynamic electron correlations, causing a net
attraction between fragments of electrons in many-electron systems. Like
all non-relativistic electronic effects, the vdW interactions are
present in the exact DFT functional \cite{luanetal95}.
Proper inclusion of vdW interactions in DFT calculations
requires that the total energy functional depends on the 
electron density \nr\, in a manner that reflects both the long-ranged and medium-ranged nature of vdW interactions. 
The usual implementation of DFT
involves the solution of the Kohn-Sham equations \cite{KoSh65}, which
are one-electron Schr\"odinger equations in the presence of an effective
one-electron self-consistent potential $V_\eff(\bfr)$. This potential is
the sum of the Coulomb potential from the density {\nr}, the external
potential---typically the Coulomb potential of the cores---and the
functional derivative of the exchange-correlation density (xc)
functional $E_{\xc} [n(\rr)]$, which describes many-particle effects. The
latter is universal and can in principle be derived from a diagrammatic
expansion in many-body perturbation theory. This functional is often
evaluated in the local-density approximation (LDA)
\cite{KoSh65,HeLuJPC1971,BaHe72,GuLu76,JoGu89} or extensions thereof,
such as the generalised-gradient approximation (GGA)
\cite{LaMePRL1981,PeEtAlPRB1982,PeBuEr96}. With the LDA and GGA, DFT is
successful in a broad range of applications \cite{JoGu89,Burke}.

From a technology perspective, the significance of LDA, e.g.\ for
semiconductor physics and thus the development of electronics,  cannot be
overestimated \cite{MaCohenReview}. The timing of the Nobel Prize for
DFT to Kohn closely followed that of its breakthrough in chemistry---a
breakthrough that is linked to its ability
in the GGA \cite{LaMePRL1981,PeEtAlPRB1982,PeBuEr96} to accurately describe covalent
bonds.

Until the start of this century, the route to extend the DFT approximations to long-range
forces has at best been a vision. By construction, LDA and GGA neglect
the long-range, nonlocal correlations that give rise to the vdW forces.
There are still many papers with claims like ``vdW accounted for by
LDA.'' Such statements are wrong from a formal perspective and it is
easy to give counterexamples in applications
\cite{Rydberg03p606,BeHy13}.

Being a correlation effect, vdW interactions are included in $E_\xc[n(\bfr)]$
\cite{luanetal95}. 
In practice, however, approximate forms of $E_\xc$ have to be
found. Studies of interacting inert atoms, molecules, and surfaces, with
analysis of the polarisabilities of the participating species
\cite{Bo36,Ha37}, give the well-known asymptotic $R^{-6}$ form of London
for atomic and molecular dimers \cite{Re12,Wa27,EiLo30,Lo30}, the
Lennard-Jones $z^{-3}$ law for a neutral molecule on a surface
\cite{LJ32}, and the $d^{-2}$ interaction law \cite{Li56,DzLiPi61} for
pairs of solids. There are also studies
\cite{BohrLindhard,Lindhard,PinesNozieres} showing that the
frequency-dependent polarisabilities, to a good approximation, can be
described by a one-pole formula, with one frequency characterising each
atom or molecule. Studies like these can be helpful in the search for
an approximate $E_\xc$ that includes vdW forces. In the literature, an
extensive knowledge of density fluctuations in general \cite{MahansBok}
and the role of plasmons in particular
\cite{BohrLindhard,Lindhard,PinesNozieres} has been developed. An
important tool in this context is the ``adiabatic connection formula,''
which connects $E_\xc$ and the density-density correlation function
\cite{LaPe75,GuLu76,LaPe77}.

The field of vdW interactions in DFT was practically absent before
around 1990, but picked up at the end of the previous century, grew
immensely during the first decade of the present one, and increased
exponentially thereafter. Overall, there are now several kinds of
approaches to the theme ``Use investments in traditional DFT and add
an account for vdW interactions.'' Several of these are based on
calculating atom-based pair potentials
\cite{RA,AnLaLu96,Scoles,Grimme1,Grimme2,TS09,Grimme3,TS2,ts-mbd}, some
of those also with the inclusion of advanced screening mechanisms
\cite{HuAnLuLa96,hurylula99,TS2,ts-mbd}. 
This review emphasises 
one orthodox, that is, a first-principles DFT treatment
\cite{RyLuLaDi00,
2001SurfScience, RydbergThesis,Paper6-01, Rydberg03p606,Rydberg03p126402,Dion, dionerratum, Langreth05p599,Thonhauser,Lee10p081101} of the
long-to-medium-ranged forces between fragments across regions with low
densities, the Rutgers-Chalmers vdW-DF method which includes van der Waals forces by using a nonlocal exchange-correlation functional. 
Inspired by this functional,
there are also other nonlocal
density functionals such as those proposed by Vydrov and Van Voorhis
\cite{VV09,VV10}. 

There are already several review papers on vdW interactions in electron
systems. In addition to our own 2005 \cite{Langreth05p599} and 2009
reviews \cite{langrethjpcm2009}, we can list the reviews
\cite{rev1,rev2,rev3,rev4,rev5,rev6,Cooper10p1417} and perspective
papers \cite{rev8,BeckePerspective}.
Considering the significant impact of the Rutgers-Chalmers vdW-DF on the field, we believe a review article that thoroughly covers this method---from its prehistory to the successive developments from the early 90s \cite{luanetal95} and the broader activity in the long-standing Rutgers-Chalmers collaboration and the current status of vdW-DF---is in order. 
Together with
David C.~Langreth \cite{david_passing}, the authors of this review have
worked within this collaboration. Other methods, particularly those closely related to vdW-DF, will be discussed as well. However the aim of this discussion is to put the development in context and to highlight the nature of vdW-DF.
For a more complete review of the other methods, we recommend that the reader consults other review articles on the topic \cite{rev1,rev2,rev3,rev4,rev5,rev6,Cooper10p1417,rev8,BeckePerspective}.

In the beginning of the program to include vdW forces in DFT, contact
was established with earlier developments. Initial attempts were made
with the nonlocal average density approximation (ADA) and weighted
density approximation (WDA) density functionals \cite{ADAWDA},
unfortunately with limited success. Next, asymptotic functionals were
derived for atomic and molecular dimers
\cite{AnLaLu96,AnHuRyApLuLa97,AnRy99,hurylula99} in part by modifying a
Rapcewicz-Ashcroft \cite{RA} concept. Similar functionals were also
developed for free molecules, molecules outside of a surface
\cite{HuAnLuLa96}, and for two parallel surfaces
\cite{HuAnLuLa96,AnHuApLaLu98,HuHyRoLu01}.  
The 90s involved
development of conceptual ideas, implementations, and adaptation of
existing codes, new codes, and exchange functionals.

The new century began with the development of two complete vdW
functionals, first in a two-dimensional configuration \cite{RyLuLaDi00,
Rydberg03p606} and then in a general geometry \cite{RydbergThesis,
DionThesis, Dion, Lee10p081101}. There are applications involving the
interactions of atoms \cite{AnLaLu96, AnRy99, Dion, Thonhauser},
molecules \cite{AnRy99}, solids \cite{Dion, Thonhauser,
Ziambaras07p155425, Londero10p054116, Rohrer11p165423}, molecular solids
\cite{Kleis08p205422, Berland10p134705, Berland11p1800,BeLoScHy13},
surfaces \cite{Chakarova-Kack06p146107, Johnston08p121404}, adsorption
\cite{Chakarova-Kack06p146107, Johnston08p121404, Moses09p104709,
Wellendorff10p378}, graphene \cite{2001SurfScience, Dion,
Langreth05p599, Chakarova-Kack06p146107, Ziambaras07p155425,
Lee10p081101, Chakarova-Kack10p013017, Rohrer11p165423, Berland11p1800,
berland11p135001, Bergvall11p155451, Akesson12p174702, Le12p424210},
metals \cite{lee11p193408, lee12p424213}, oxides
\cite{Chakarova-Kack06p155402, Londero10p054116, Londero11p1805},
polymers \cite{Kleis05p164902, Kleis05p192, Kleis07p100201}, nanosystems
\cite{Schroder03p721, Kleis05p192, Kleis08p205422, Bergvall11p155451,
Wyrick11p2944}, adsorbate interactions \cite{Schroder03p880,
Berland09p155431, Sun10p201410, Wyrick11p2944}, clusters
\cite{Kelkkanen11p113401}, DNA \cite{Cooper08p1304, LiCoThLuLa09,
LoHySc13}, nanotubes \cite{Kleis08p205422}, the carbon nanotube (CNT)
morphology \cite{Schroder03p880, Schroder03p721,Kleis05p192}, water
\cite{Mogelhoj11p14149}, and the list goes on. 

The objective of vdW-DF is to provide within DFT
an efficient method for calculations of vdW effects in all kinds of
electron systems based on many-body physics and general physical
laws. In this regard, the vdW-DF method differs from methods that use
empirical, semi-empirical, and ad hoc assumptions for such calculations.
By semi-empirical, one typically refers to methods that rely on optimization to reference systems for which data from accurate, computationally expensive methods are available. 
So far, we have published general nonlocal functionals in 2004 (vdW-DF; also referred to as vdW-DF1~\cite{Dion}) and in 2010 (vdW-DF2~\cite{Lee10p081101}). 
Based on physical principles, we have also developed  progressively more consistent exchange functionals (i.e. vdW-DF-C09 (2010)~\cite{cooper10p161104} and vdW-DF-cx (2014)~\cite{behy14,BeArCoLeLuScThHy14}) to complement the vdW-DF1 nonlocal correlation. Together these works demonstrate that the vdW-DF method~\cite{RydbergThesis,Paper6-01,DionThesis,Dion,Langreth05p599,Thonhauser,Lee10p081101,behy14,BeArCoLeLuScThHy14,hybesc14} provides a good framework for developing successively improved functionals.

%%%%%%%%%%%%%%%%%%%%%%%%%%%%%%%%%%%%%%%%%%%%%%%%%%%%%%%%%%%%%%%%%%%%%%%%
\section{The beginnings}
%%%%%%%%%%%%%%%%%%%%%%%%%%%%%%%%%%%%%%%%%%%%%%%%%%%%%%%%%%%%%%%%%%%%%%%%

Sparse matter has strong local bonds, as well as vdW bonds, and other
weak bonds. A proper description must include all. Numerous treatises
(\cite{Burke} is a recent one) have been devoted to  the
chemical or valence bond. Thus it is fair to focus on just the vdW bond
here, keeping in mind the whole set of bonds present. There are
many different configurations where vdW forces act between atoms or
fragments of electron densities  separated by empty space.
Extreme voids are provided by gas-solid interfaces, which lead us to a discussion of the  early surface-physics work important for functional development,
with contributions from the Ashcroft group, Langreth \& Vosko, and
others.

%%%%%%%%%%%%%%%%%%%%%%%%%%%%%%%%%%%%%%%%%%%%%%%%%%%%%%%%%%%%%%%%%%%%%%%%
\subsection{Surface-physics background and experimental aspects}
\label{Sec:SPexp}
%%%%%%%%%%%%%%%%%%%%%%%%%%%%%%%%%%%%%%%%%%%%%%%%%%%%%%%%%%%%%%%%%%%%%%%%

A typical introduction to vdW forces starts with molecule-molecule
interactions. To reach the vdW-DF functional we choose a
condensed-matter and surface physics perspective, as our background is
in these fields. 

%First, we review some early model descriptions of vdW interactions
%\cite{MaKe69}. Early on, calculations on the asymptotic interaction
%between two neutral molecules gave the well-known $R^{-6}$ potential of
%London \cite{EiLo30}. The power of the microscopic vdW force depends 
%on the macroscopic shape \cite{LJ32, Li56,DzLiPi61} and there are multiple such power laws \cite{MaKe69}.

Surface potentials can be obtained by bombarding atoms or molecules
against surfaces and studying the scattering. 
In the early days,
studies on metals were lagging behind those on, for instance, ionic
crystals. On metal single crystals, diffraction spots are much weaker.
This reflects the much weaker corrugation of close-packed metal
surfaces \cite{Boato} than of, {\it e.g.}, an ionic crystal, like LiF(100)
\cite{Garcia}. In the 70s, experimental techniques improved, and metal
surfaces started to drive the development \cite{Boato}. 
On the theory side, jellium and smooth surfaces were studied.

In the early 90s, the stage for describing the physisorption on metal
surfaces using the jellium model was set by the Zaremba-Kohn (ZK) theory
\cite{ZarembaKohn1976,ZarembaKohn1977}. The ZK theory provides key
concepts, such as repulsive walls, vdW attraction, induced
surface charges, and dynamic image or vdW planes. It also provides a
semi-quantitative framework for analysing accurate experimental results.
This is the ``traditional picture'' of physisorption, where the
interaction potential $V(z)$ between a metal surface and an inert
adparticle at a separation $z$ apart is approximated by a superposition
\cite{LJ32,ZarembaKohn1977,HarrisNordlander1984,persson2008},
\begin{equation}
V(z) = V_R (z) + V_\vdW(z)\;.
\label{eq:old1}
\end{equation}
The short-range Pauli repulsion, $V_R(z)$, is due to the overlap between
orbital tails of the metal conduction electrons and the closed-shell
electrons of the adparticle. In the %{early} 
Lennard-Jones potential
\cite{LJ32} it was expressed either as $R^{-12}$ or an exponential, as
in
\begin{equation}
V_R (z) = V_{RO} \exp(- \alpha z)\;,
\label{eq:old2}
\end{equation}
where the constants $V_{RO}$ and $\alpha$ determine the strength and the range
of the repulsive potential. There are schemes to calculate $V_R (z)$
from the shifts of the one-electron energies of the metal induced by the
adparticle, for instance calculated with perturbation theory, where the
adparticle can be described with pseudopotentials and the metal surface
in a jellium model. As the local density of metal-electron states decays
exponentially away from the surface, the exponential form of the
repulsive wall (\ref{eq:old2}) follows.

The long-range vdW attraction, $V_\vdW (z)$, arises from adsorbate-substrate
electron correlations. A common approximate form is
\begin{equation}
V_\vdW(z) = - \frac{C_\vdW }{(z - z_\vdW)^3} f\big(2k_{\rm c}(z - z_\vdW)\big)\;,
\label{eq:old3}
\end{equation}
where $z_\vdW$ is the dynamic image-plane location
\cite{ZarembaKohn1976}. The magnitude of the asymptote $C_\vdW$ and the
position of the vdW plane $z_\vdW$ depend on the dielectric properties
of the metal substrate and the polarisability of the adsorbate
\cite{ZarembaKohn1976,Li86}. The function $f(x)$ saturates the vdW term
at atomic-scale separations. In some accounts it has the form $f(x) =
1-(1+x+x^2/ 2) \exp(-x)$. This saturation lacks a rigorous prescription,
which leaves a level of arbitrariness for $V_\vdW(z)$. These empirical
saturation functions, like those in
\cite{andersson1993,perandersson1993,HarrisNordlander1984,persson2008},
resemble the damping functions used in semi-empirical DFT descriptions
of dispersive interactions \cite{Grimme3}.

The development of experimental techniques often makes once canonised
theoretical results appear insufficient. An example is gas-surface
scattering, in particular diffractive scattering in the elastic
backscattering mode which can exhibit resonance structures. For instance,
beams of light molecules scattered off single-crystal Cu surfaces can be
used to deduce physisorption potentials. An example  is the physisorption-well depth of 30.9 meV, determined
from scattering of H$_2$ and D$_2$ molecules off the Cu(111) surface
\cite{andersson1993,perandersson1993}. This is substantially larger than
the ZK value of around 22~meV \cite{andersson1993,perandersson1993}.
%\sout{Motivating experiments might be of a more general nature as well, like
%the structure of liquid water with its extraordinary density dependence.
%As such, a unified treatment of the whole spectrum of vdW interactions
%is desirable.}\KBnote{This is a good sentence, but feels like intro or conclusion material.}

Inert atoms and molecules undergo no significant change in their
electronic configurations when they physisorb. Even in metallic
physisorption, the coupling to electronic excitations is weak, which
makes the adsorption essentially electronically adiabatic
\cite{SchAndGunnars1980}. The energy transfer occurs through the phonons
of the solid lattice \cite{Brenig1987}. The internal molecular degrees
of freedom add further details to the particle-surface interaction.
Contact between theory and experiment can be established when the
potential energy surface governing the gas-surface collision
process is known. For well-defined impact conditions,
molecular-beam-scattering experiments can provide such information
and also data on energy transfer and sticking probabilities
\cite{persson2008}.

\begin{figure}
\begin{center}
\includegraphics[width=\columnwidth]{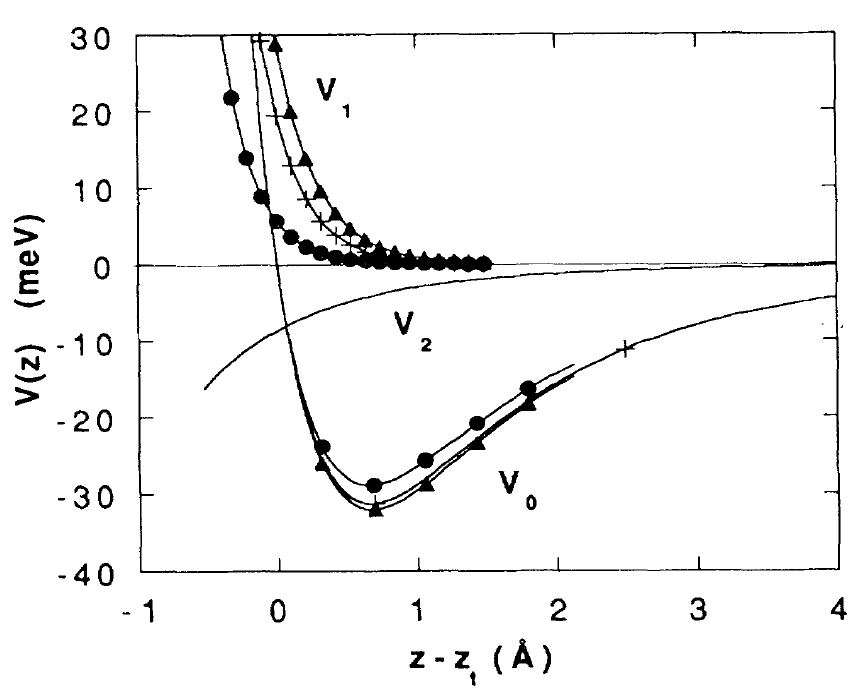}
\caption{\label{fig:pot} Physisorption interaction potentials for H$_2$
(D$_2$) on Cu(111) (circles), Cu(100) (squares) and Cu(110) (triangles)
\cite{andersson1993}. $V_0(z)$, $V_1(z)$ and $V_2(z)$ are the laterally
averaged physisorption potential, the corrugation potential, and the
laterally averaged min-to-max variation of the rotational anisotropy
$V_2$ potential functions, respectively. The position $z$ of the
molecular centre of mass is given with respect to the classical turning
point $z_t$ at $\eps_i=0$. Reprinted with permission from
\cite{andersson1993},~\copyright\ 2008 American Physical Society.}
\end{center}
\end{figure}

The benchmark provided by resonant elastic backscattering of light
molecules, like H$_2$ on Cu single-crystal surfaces, is extraordinary.
The data provide (i) quantum-mechanical energy eigenvalues, $\eps_i$,
for H$_2$ in the potential energy well, which are directly tied to
measured intensities, (ii) the laterally averaged physisorption
potential $V_0$, which is derived from measured data, (iii) the
corrugation $V_1$, also derived from measured data, and (iv) the
laterally averaged min-to-max variation of the rotational anisotropy
$V_2$ \cite{persson2008}.

Figure \ref{fig:pot} shows the potential energy curves for H$_2$
on Cu(111), (100), and (110), deduced from resonant elastic
backscattering experiments. These potential wells can trap H$_2$
\cite{andersson1993,perandersson1993,persson2008}. The traditional
theoretical picture of the interaction between an inert adsorbate and a
metal surface \cite{ZarembaKohn1977, HarrisNordlander1984, 
HarrisAndLiebsch1982b} is used to deduce the potential energy curves
based on the experimental energy-level values.
The resulting physisorption potentials based on (\ref{eq:old1}) and
(\ref{eq:old3}) provide a good fit. The curves in figure \ref{fig:pot}
have potential-well depths of 29.5, 31.4, and 32.3 meV and a potential
minimum located at 3.50~{\AA} outside the topmost layer of copper ion
cores on Cu(111), (100), and (110), respectively \cite{andersson1993,
perandersson1993,persson2008}.

\begin{figure*}
\begin{center}
\includegraphics[width=0.87\textwidth]{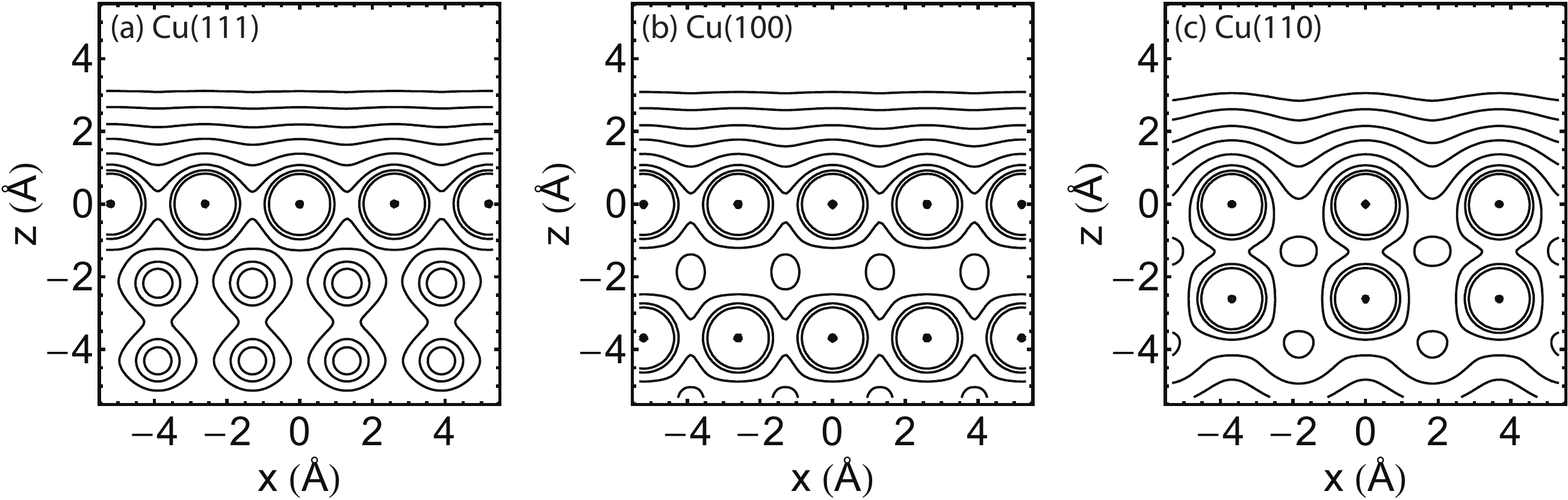}
\caption{\label{fig:density}Electron density profiles of the clean
Cu(111), (100), and (110) surfaces calculated with the vdW-DF2
functional \cite{Lee10p081101}. The density contours take values in a
nonlinear fashion. Reprinted with permission from 
\cite{lee12p424213}, \copyright\ 2012 Institute of Physics.}
\end{center}
\end{figure*}

The physisorption potential $V(z)$ (\ref{eq:old1}) depends
on the details of the surface electron structure, via both the electron
spill out ($V_R$) and the centroid of fluctuations of exponentially
decaying surface charges ($V_\vdW$). For a given adparticle, this
results in a crystal-face dependence of $V(z)$ \cite{AnderssonPeHa96}.
From the ZK theory one gets no strong hints about the dependence on $n(\bfr)$, needed in DFT. However, the weak dependence of the diffraction of, {\it e.g.}, the He atom on metal surfaces  was
observed early \cite{Boato}. It is explained in terms of a simple link
between the scattering potential and the electron-density profile. This 
gives a hint for approximate DFT: The He-surface interaction energy $E_\He(\bfr)$ can be reasonably well
expressed as \cite{Esbjerg}
\begin{equation}
E_\He(\bfr) \simeq E_\He^\homm\big(n_o(\bfr)\big)\;,
\label{eq:old4}
\end{equation}
where $E_\He^\homm(n)$ is the energy change on embedding a free He atom
in a homogeneous electron gas of density $n$. In this case, $n_o(\bfr)$
is the host electron density. On close-packed metal surfaces the
electron distribution $n_o(\bfr)$ is smeared out along the surface
\cite{Smoluchowski1941}, resulting in weak corrugation. The crude
proposal (\ref{eq:old4}) might be viewed as the precursor to the
effective-medium theory \cite{EMT}.

Figure \ref{fig:density} shows the calculated density profiles for the
Cu(111), (100), and (110) surfaces, using density functional theory. It
illustrates the point that the corrugations on these facets are small,
but increasing when going from (111) to (100) to (110).
For the scattering experiment the density contours far away from the
surface, in the tails of the wavefunctions, are particularly important.

The kinematical condition for a diffraction resonance involving a
surface reciprocal lattice vector $\mathbf{G}$ is
\begin{equation}
\eps_i=\eps_n + \frac{\hbar^2}{2m_p} \left(\mathbf{K}_i + \mathbf{G}\right)^2\;,
\label{eq:old5}
\end{equation}
where $\eps_n$ is the bound-state energy, $m_p$ the particle mass, and
$\eps_i$ and $\mathbf{K}_i$ the energy and wavevector component
parallel to the surface of the incident beam. When the resonance
condition (\ref{eq:old5}) is fulfilled, weak periodic lateral
corrugations greatly affect diffracted beam intensities. The
resonances are usually observed as narrow features in the spectra of the
diffracted beam intensities. This sharpness allows a number of levels to
be uniquely determined. A single accurate gas-surface potential curve
can then be constructed according to the Rydberg-Klein-Rees method of
molecular physics \cite{Roy}. Detailed mapping of the bound level
spectrum and the gas-surface interaction potential by resonance
scattering measurements has only been performed for hydrogen
\cite{PerrauAndLapujoulade1982, YuEtAl1985, ChiesaEtAl1985,
HartenEtAl1986, AnderssonEtAl1988, andersson1993, AnderssonPeHa96}.
Having two isotopes, H$_2$ and D$_2$, with significantly different
masses available as well as the different rotational populations of para-H$_2$,
ortho-D$_2$, and the normal species, is extremely useful in data analysis.

%%%%%%%%%%%%%%%%%%%%%%%%%%%%%%%%%%%%%%%%%%%%%%%%%%%%%%%%%%%%%%%%%%%%%%%%
\subsection{Surface physics and theoretical aspects}
%%%%%%%%%%%%%%%%%%%%%%%%%%%%%%%%%%%%%%%%%%%%%%%%%%%%%%%%%%%%%%%%%%%%%%%%

To describe the excitation spectrum of electrons in the framework of
noninteracting particles, Bohr and collaborators already realised
\cite{BohrLindhard} that a simple assumption one can make is to
associate each point in the atom with a single frequency, which is a
function only of the local density---a model used in particular by
Lindhard and Schaff~\cite{LiSc53}. In the simplest version, one chooses
the frequency equal to the classical plasma frequency
\begin{equation}
\omp = \sqrt{4\pi n e^2 /m}\;.
\label{eq:old6}
\end{equation}
The choice of this resonance frequency is equivalent to assuming that
the local response to the field is the same as that of a uniform
electron gas of a density equal to the local density $n(\rr)$.

For atoms, this type of approximation has been used to calculate the
response function for external electrical fields with long wavelengths
\cite{BrLu63}. For metals there are famous publications that rely on
this model \cite{BohrLindhard,Lindhard}. The extensive experience in studying 
density fluctuations by this Copenhagen school
\cite{BohrLindhard,LiSc53,Lindhard} has influenced our thinking, in
particular bringing attention to the role of plasmons
\cite{BohrLindhard,Lindhard}. We call this ``the plasmon description.''
It is common not only for atoms, metals, and stopping power, but also
for models such as the ``Swedish Electron Gas" \cite{LarsStigSSP}, and
for the modelling of plasmonics, plasmaronics \cite{Lu67,CaBlNi12}, and
transition metals. There is thus a rich variety of physical systems
that have nurtured the conceptual development behind the vdW-DF method.
Many of these appeal to the density of electrons \nr, but, like for the
GGA, results that depend on the gradient of the density are of key
interest. Results for inhomogeneous electron systems in general, and
surfaces in particular, come into focus.

Surfaces are thus important for the development of electron-structure
theory. The first descriptions of interacting electrons in condensed
matter were made for the homogeneous electron gas. The many-body aspects
of this model system of charged fermions were relatively successfully
treated with diagrams, which emerge in perturbation theory to infinite
order or with a corresponding quantum-field theory. Going beyond the
homogeneous electron gas, a functional that performs well on surfaces is
likely to be useful even for other classes of systems.

To account
for vdW forces by DFT one considers how charge density fluctuations that
are basically dynamic in nature can be accounted for with a static
quantity like the density \nr. This problem together with the nature
and form of the vdW interaction in DFT are treated in three many-body
articles from the late 80s \cite{MA,LaVo87,RA}.

vdW bonds emanate from nonlocal electron correlations. This is
illustrated in early dipole-model descriptions of the interactions in
noble-gas crystals \cite{StigAlf64,Jerry65}. The electrodynamical
coupling between the atomic dipoles gives a shift in the dipolar
oscillator frequencies, and the sum of all these shifts gives the vdW
binding energy \cite{StigAlf64,Jerry65}.

%%%%%%%%%%%%%%%%%%%%%%%%%%%%%%%%%%%%%%%%%%%%%%%%%%%%%%%%%%%%%%%%%%%%%%%%
\subsection{Work of the Ashcroft group and Langreth \& Vosko}
\label{sec:ALV}
%%%%%%%%%%%%%%%%%%%%%%%%%%%%%%%%%%%%%%%%%%%%%%%%%%%%%%%%%%%%%%%%%%%%%%%%

A particular type of dipolar oscillations are present in noble metals.
The question of effective interatomic potentials in noble metals arising
from quantum fluctuations in the atom-centred $d$ electrons can be
addressed with diagrammatic perturbation theory \cite{MA}. With a
philosophy of incorporating the many-body interaction between the core
states from the outset, many of the cohesive properties of noble metals
are found to be directly linked to fluctuation effects. Maggs and Ashcroft
(MA) \cite{MA} identified large contributions to the potentials that
originate in certain diagrams for the homogeneous electron gas, which
had been overlooked in the linear response of homogeneous systems.
These are the ones with the screened Coulomb interaction between ions
and they lead to the possibility of recovering vdW forces in nonlocal
functional theories of the electron gas. The class of diagrams corresponding to vdW interactions
between the core electrons are screened by the intervening electron gas.
This screening is given by the frequency-dependent dielectric function
for the homogeneous electron gas. The core density-density response
function (``vertices'') depends on the frequency-dependent core
polarisability. To a good approximation the internal
lines can be replaced with a simplified expression involving the
plasma frequency~(\ref{eq:old6}) and wavevector $q$, as follows
\begin{equation}
4\pi\omega^2/q^2(\omega^2+\omega^2_{\rm p})\;.
\label{eq:old7}
\end{equation}
The simplest model assumes the core fluctuations to be dominated by a
single excited-state frequency, $\Delta$, which makes the integrals over
frequency straightforward, giving an approximate formula for the
screened vdW interaction energy for a pair of atoms separated by $r$ in
a polarisable metal \cite{MA},
\begin{equation}
E_{\rm b}(r) = \frac{3\Delta}{4}\frac{\alpha^2(0)}{r^6}
\left( \frac{\Delta}{\Delta+\omp} \right)^3\;.
\label{eq:old8}
\end{equation}
This result relates to DFT via the Hohenberg-Kohn-Sham \cite{HoKo64,
KoSh65} energy-response kernel $K_\xc(q)$, a key property in the
description of responses in electron systems. In a dense homogeneous
electron gas it is defined by
\begin{equation}
\delta E_\xc = \sum_q K_\xc(q)|\delta n_q|^2\;,
\label{eq:old9}
\end{equation}
where $n_q$\ is the electron-gas density in planewave representation.
The kernel defines the static dielectric function $\eps(q)$ of the
electron gas,
\begin{eqnarray}
\eps(q) &=& 1-(4\pi e^2/q^2) \chi (q)\;, \label{eq:old10}\\
\chi(q) &=& \chi_0(q)/\big[1-2 K_\xc(q) \chi_0(q)\big]\;.
\label{eq:old11}
\end{eqnarray}

\begin{figure}[t]
\includegraphics[width=\columnwidth]{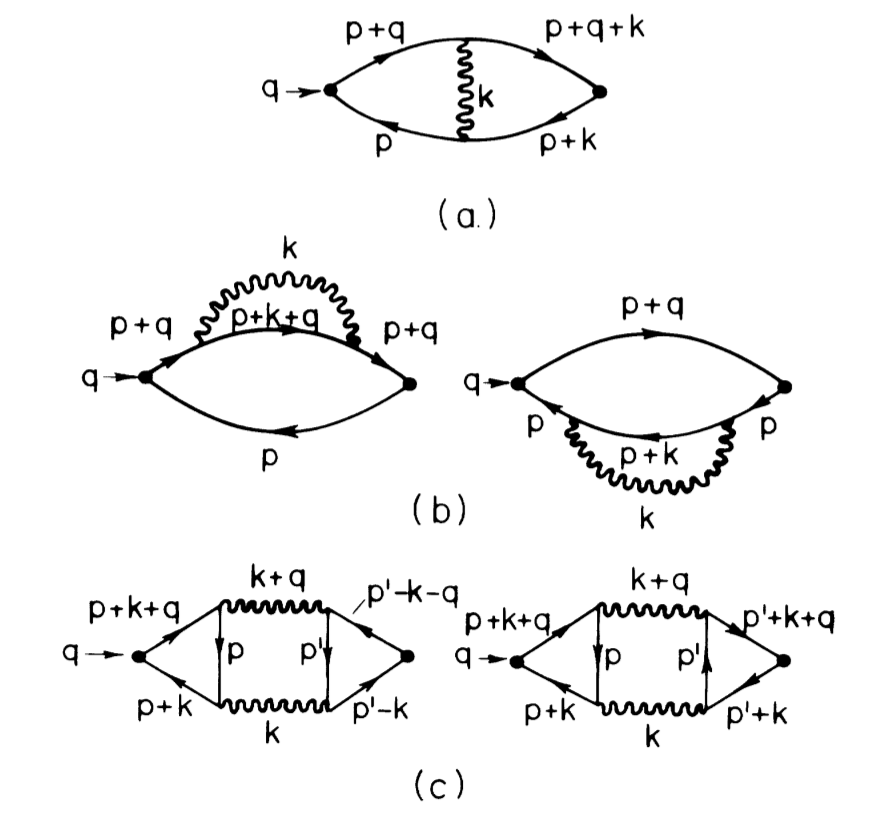}
\caption{\label{fig:LV}Diagrams for the leading order corrections to the
response function $\chi$. The wiggly lines are the screened Coulomb
interaction in the random-phase approximation. Diagrams in (a) and (b)
contribute to $Z_x$, whereas the diagrams in (c) contribute to $Z_c$.
Reprinted with permission from \cite{LaVo87}, \copyright\ 1987 American
Physical Society.}
\end{figure}

Figure~\ref{fig:LV} shows diagrams corresponding to corrections to the
free-electron response function $\chi_0(q)$. The approximation $K_\xc =
0$ is the random-phase approximation (RPA) and corresponds to
disregarding these higher-order diagrams. The $q =0$ component
corresponds to the LDA. The expansion
\begin{equation}
K_\xc(q) = K_\xc(0) + \frac{\pi e^2 }{8k_F^4} Z(q) q^2
\label{eq:old12}
\end{equation}
defines the dimensionless quantity $Z(q)$. Here
\begin{equation}
K_\xc(0) = -\frac{\pi e^2}{2k_F^2}[1+\lambda (1-\ln 2)]\;,
\label{eq:old13}
\end{equation}
where, to sufficient accuracy \cite{LaVo87},
\begin{equation}
\lambda= (\pi a_0 k_F)^{-1} = 0.521 r_s/\pi = (k_{\rm TF}/2k_F)^2\;.
\label{eq:old14}
\end{equation}

For $q \to 0$, diagrammatic perturbation theory has given important
results to leading order in $r_s(n)$: $Z(0)$ is a number independent of
$e^2$ \cite{MaBr}; the value is $Z(0) = 1.98 - 7/9$. There are two types of contributions:
\begin{equation}
Z(q)= Z_c(q) + Z_x(q)\;.
\label{eq:old15}
\end{equation}
For finite $q$, it has been shown that the exchange part $Z_x(q)$
remains close to its zero-$q$ value of $-7/9$ \cite{LaVo87}. For small
$q$, the correlation part is given by
\begin{equation}
Z_c(q) = 1.98 + 0.77 \,Q\,\ln Q -1.25 \,Q + \ldots\,,
\label{eq:old16}
\end{equation}
to within 1\% for $Q$ up to about 1, where $Q=q\sqrt{3}/k_{\rm TF}\ $.

These results are relevant for density functional theory \cite{LaVo87},
providing input to the Langreth-Mehl functional \cite{LaMePRL1981} as
well as indicating ways to improve it. They also provide the
leading-order correction to the static Lindhard screening function
\cite{Lindhard}.

Rapcewicz and Ashcroft (RA) \cite{RA} considered the coupling between
fluctuations giving rise to an attractive interaction. This interaction
originates in the lowest-order fluctuation term of the interacting
electron gas. The corresponding diagram is similar to that of
figure~\ref{fig:LV}, but the two three-point functions are connected by
two internal screened Coulomb lines \cite{LaVo87}.

Seeking a DFT account of vdW forces, one has to face the fact that vdW
forces are linked to density fluctuations, while DFT is linked to
densities. RA exploit the analogy between the
correlation hole and the electron charge localised around nuclei in
condensed atomic systems \cite{RA} to connect vdW forces arising from
the fluctuations in the electron liquid to the electron density \nr.
Clearly, such a reformulation is helpful for designing nonlocal
correlation functionals in DFT.

RA raise ``the question of vdW interactions from fluctuations in an
otherwise static response charge" in the same way as it had been done
for the atomic case. From the standard argument for the atomic case
\cite{Barash} together with dimensional analysis \cite{MA}, Maggs and Ashcroft  \cite{MA}
expected an attractive pair interaction of the form \cite{RA}
\begin{equation}
\sim -\hbar \omp (R_s/r)^6\;.
\label{eq:old18}
\end{equation}

The work of RA illustrates that a plasmon picture can be used to
formulate a theory of vdW forces in an electronic liquid. Thus we can benefit from decades of experience with LDA and GGA, which can also be formulated in terms of plasmons.

%%%%%%%%%%%%%%%%%%%%%%%%%%%%%%%%%%%%%%%%%%%%%%%%%%%%%%%%%%%%%%%%%%%%%%%%
\section{Asymptotic functionals}
\label{sec:asymptotic}
%%%%%%%%%%%%%%%%%%%%%%%%%%%%%%%%%%%%%%%%%%%%%%%%%%%%%%%%%%%%%%%%%%%%%%%%

Traditionally, the asymptotic form at large separations has attracted the most interest. The first attempts by us and others to 
capture vdW behaviour in approximate forms of $E_\xc[n(\bfr)]$ 
also concerned asymptotic functionals describing the interaction between widely separated fragments of electrons. 
Both local and semilocal
approximations (GGAs)  have the wrong
asymptotic dependences on separation. To retain the vdW interactions in
approximate DFT methods, there are, however, several previous ideas and 
results for approximate nonlocal functionals, and even for local or semilocal approximations to benefit from.

Several not so successful attempts were first made. One started from the
so-called weighted-density approximation \cite{ADAWDA}. With the
Gunnarsson-Jones expression for the xc hole \cite{GuJo80}, the leading term for two widely separated neutral objects
becomes $-C_5 R^{-5}$, whereas for a neutral point-like object outside a
metal surface it goes like $- C_2 z^{-2}$ \cite{luanetal95}. In
addition, the $C$-coefficients take unphysically high values.

A dipole-dipole type of weighted-density approximation for $E_\xc$ has
also been attempted. It appeared to be able to retain the image
potential ($-1/4z$) but failed to the connect long- (i.e.\ $-R^{-6}$) and
short-range parts \cite{luanetal95}.

%%%%%%%%%%%%%%%%%%%%%%%%%%%%%%%%%%%%%%%%%%%%%%%%%%%%%%%%%%%%%%%%%%%%%%%%
\subsection{The functional of Rapciewicz and Ashcroft}
%%%%%%%%%%%%%%%%%%%%%%%%%%%%%%%%%%%%%%%%%%%%%%%%%%%%%%%%%%%%%%%%%%%%%%%%

The attention of vdW-DF developers then turned to the RA work
\cite{RA}. In their study of the fluctuation attraction in condensed
matter, the lowest-order fluctuation term corresponds to a diagram with
two three-point functions, connected by two internal screened Coulomb lines---see figure \ref{fig:LV}(c). The lowest order fluctuation term
shown in figure \ref{fig:LV} leads to a static vdW attraction between
electrons \cite{Re12,EiLo30}. Its physical significance is greater than
its formal order in perturbation theory might imply, which is related to the dynamical screening. The effective interaction between electrons at
$\bfr_1$ and $\bfr_2$ is \cite{RA,luanetal95}
\begin{align}
&- \frac{3}{4} \hbar \left(\frac{e^2}{m}\right)^2 \times
\frac{1}{[\omp(\bfr_1,\bfr_2)]^3|\bfr_1 - \bfr_2| ^6}\;.
\label{eq:old19}
\end{align}
Here, the dielectric function is given by the plasmon-pole approximation
for a homogeneous electron gas with an effective density given by the
geometrical mean of the densities
\begin{equation}
n_\eff = \left[n^{(1)}(\bfr_1)\:n^{(1)}(\bfr_2)\right]^{1/2}
\label{eq:old20}
\end{equation}
of the two fragments, so that the effective plasma frequency becomes
\begin{equation}
\omp(\bfr_1,\bfr_2) = \left[\omp(\bfr_1)\:\omp(\bfr_2) \right]^{1/2}\;.
\label{eq:old21}
 \end{equation}
An approximate formula for the screened vdW interaction energy for a
pair of atoms separated by $r$ in a polarisable metal is given by
(\ref{eq:old8}). However, to get a density functional that is valid in both the uniform gas and separated atom limit \cite{AnLaLu96} we must have a form that is viable and physically motivated in both limits, as seen in section B.

%%%%%%%%%%%%%%%%%%%%%%%%%%%%%%%%%%%%%%%%%%%%%%%%%%%%%%%%%%%%%%%%%%%%%%%%
\subsection{Improvement by Andersson, Langreth, and Lundqvist}
%%%%%%%%%%%%%%%%%%%%%%%%%%%%%%%%%%%%%%%%%%%%%%%%%%%%%%%%%%%%%%%%%%%%%%%%

It is desirable that the approximate density functional is valid, viable,
and physically motivated in both the uniform-gas and separated-atoms
limits. Such a situation arises when we consider an effective density
in the kernel $K_\xc$ defined by the expression for the
exchange-correlation energy of a slightly nonuniform system
\cite{AnLaLu96}:
\begin{equation}
\delta E_\xc=\int \, \diff^3r_1 \int \diff^3r_2 \;
K_\xc(\bfr_1,\bfr_2)\;\delta n(\bfr_1)\;\delta n(\bfr_2)\;,
\label{eq:old24}
\end{equation}
as given by the real-space representation of (\ref{eq:old9}).
The interaction between two small but distant charge perturbations in a
uniform electron gas is described through the limiting behaviour of this
linear response kernel $K_\xc$ \cite{luanetal95}. This has implications
for the effective plasmon frequency $\omp$ in equation (\ref{eq:old21}).
In the formulation by Andersson, Langreth, and Lundqvist (ALL)
\cite{AnLaLu96}, the effective density is
\begin{equation}
n_\eff = \left[ \sqrt{n(\bfr_1)\;n(\bfr_2)}
\left(\sqrt{n(\bfr_1)} + \sqrt{n(\bfr_2)} \right)\right]^{2/3}\,,
\label{eq:old25}
\end{equation}
and the total fragment density is used instead of $\delta n$ in the
isolated fragment limit, following \cite{LaMePRL1981, PeWa86,RA}. This
gives an effective long-range interaction of the form
\begin{equation}
\phi(\bfr_1,\bfr_2) \rightarrow \frac{3 e^4}{2m^2} \frac{1}{\omp(\bfr_1)\omp(\bfr_2)[\omp(\bfr_1)+\omp(\bfr_2)]|\bfr_1-\bfr_2|^6}\;,
\label{eq:old26}
\end{equation}
which differs from the RA expression (\ref{eq:old19}). This has the same form as the London expression for the vdW
interaction between two atoms A and B at separation $R$, for the case
where only one excitation frequency $\omega_{A/B}$ is considered for each atom \cite{EiLo30,MaKe69},
\begin{equation}
E_\vdW^\london = - \frac{3e^4}{2m^2}
\frac{Z_A Z_B}{\omega_A\omega_B(\omega_A+\omega_B)}
\frac{1}{R^6}\;.
\label{eq:old26b}
\end{equation}

The long-range interaction 
is related to the electric susceptibility $\chi_i(\omega)$ or polarisation
response of a uniform electron gas at density \nr\ to an external electrical field
\begin{equation}
E_\xc^{l-r} = - \frac{3}{\pi}
\int_0^\infty \!\diff u \int_{V_1}\! \diff^3 r_1 \int_{V_2} \! \diff^3 r_2
\frac{\chi_1(iu)\chi_2(iu)}{|\bfr_1 - \bfr_2 |^6}\;.
\label{eq:old27}
\end{equation}
For two atoms, widely separated by a
distance $R$,   (\ref{eq:old27}) gives $E_\xc^{l-r} = -C_6R^{-6}$, with the
standard expression for $C_6$ in terms of the atomic polarisabilities $\alpha_i(\omega)$ \cite{MaSt62,MaKe69}.

\begin{figure}
\begin{center}
\includegraphics[width=\columnwidth]{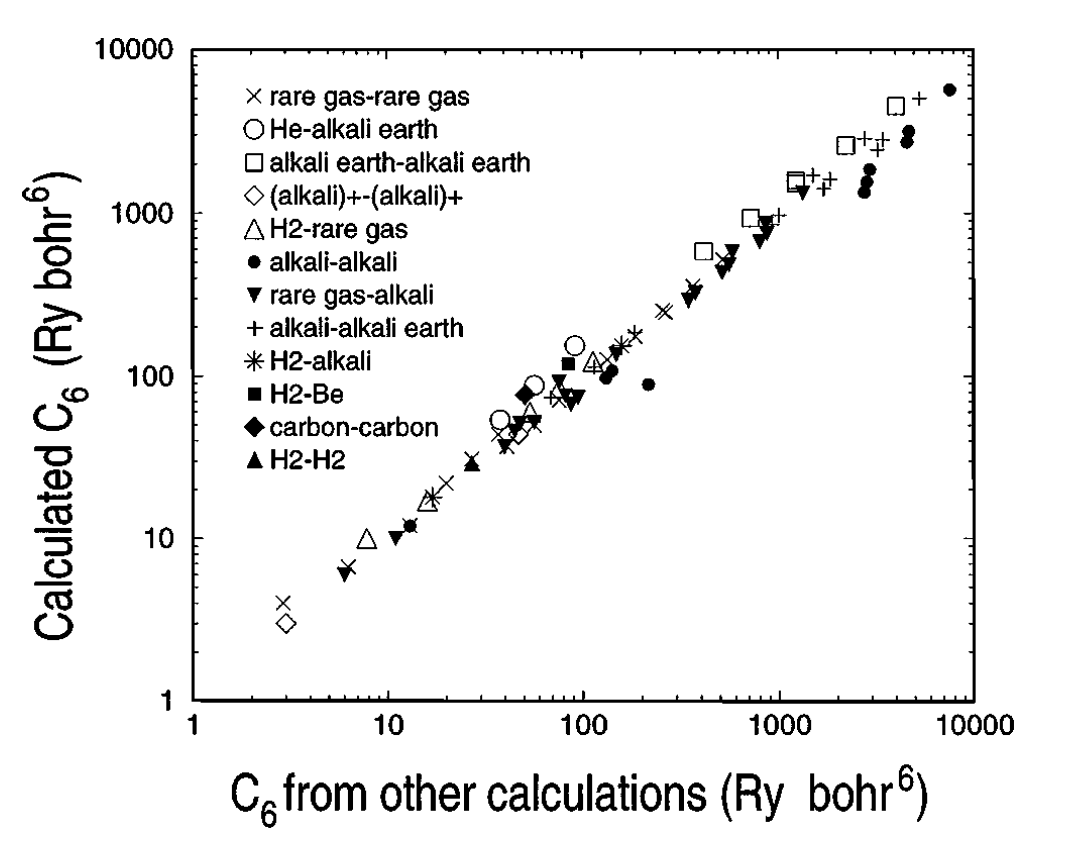}
\caption{\label{fig:milkyway}van der Waals coefficients $C_6$ (in Ry
atomic units) calculated from $E_\xc^{l-r} = -C_6R^{-6}$ and (26)
and (28), plotted against corresponding values from first-principles
calculations. Reprinted with permission from \cite{AnLaLu96},
\copyright\ 1996 American Physical Society.}
\end{center}
\end{figure}

The ALL theory is crude. Like RA, it contains a cutoff, specifying the
spatial regions where the response to an electric field is defined to
be zero. Results can be overly sensitive to the specific cutoff.
Nevertheless, results based on ALL compare well with those of
first-principles calculations, over wide classes of atoms and molecules~\cite{AnThesis,AnRy99}
(figure \ref{fig:milkyway}). 

ALL also provides one of the foundations for the more general vdW-DF, which was developed nearly a decade later. It shows
that a functional that is quadratic in the density works
\cite{AnLaLu96}. Once the interaction at large distance was understood, it took a decade before we could grasp the small-separation limit
(see chapter~\ref{sec:general-geometry}).

The ALL functional has had
multiple applications, sometimes with the addition of
empirical damping functions, which will be discussed in the next
chapter.

%%%%%%%%%%%%%%%%%%%%%%%%%%%%%%%%%%%%%%%%%%%%%%%%%%%%%%%%%%%%%%%%%%%%%%%%
\subsection{Improvement by Dobson et al.}
%%%%%%%%%%%%%%%%%%%%%%%%%%%%%%%%%%%%%%%%%%%%%%%%%%%%%%%%%%%%%%%%%%%%%%%%

In a parallel and independent study, Dobson and Dinte~\cite{DobDint96}
focused on constraint satisfaction in local and gradient susceptibility
approximations in the development of a vdW density functional. They show how
charge conservation and reciprocity, that is $ \chi(\vv{r},\vv{r'},iu;
\lambda) = \chi(\vv{r'},\vv{r},-iu; \lambda)$, can be built into local
density or gradient approximations for density-density response
functions (susceptibilities). Applying these ideas, they were able to
derive a variant of the RA formula for the vdW
interaction.

To describe vdW forces,  an approach is introduced \cite{PeBe95} that (i) simplifies the problem
of achieving hole normalisation, that is ensuring that the xc hole
 contains one charge unit \cite{GuLu76}, and (ii) facilitates the
derivation of vdW functionals. This expression for the vdW interaction
between nonoverlapping electronic systems is similar but not identical
to the RA one \cite{RA}. In the
denominator the geometric mean $(\omega_1\omega_2)^{1/2}$ has been
replaced by an arithmetic mean giving a kernel form similar to equation \ref{eq:old26}. This makes the result less sensitive to a low-density
cutoff. This choice was motivated partly by the desire to reduce the
noise sensitivity compared to RA. A cutoff, for example, like that in
RA, would certainly still be appropriate, because the uniform-gas-based
approximation seriously overestimates the response in the outer tails of
the electronic density.

The essential point, though, is that a formula for vdW asymptotic
interactions has been derived by a simple local-density approach, which
embodies suitable constraints. The satisfaction of charge conservation
is essential, because without it equation (\ref{eq:old27}) would
represent the second-order Coulomb interaction between spurious nonzero
charges, and would not give the correct $r_{12}^{-6}$ interaction.

%%%%%%%%%%%%%%%%%%%%%%%%%%%%%%%%%%%%%%%%%%%%%%%%%%%%%%%%%%%%%%%%%%%%%%%%
\subsection{Self-consistency and the inclusion of corrections}
\label{sec:unified}
%%%%%%%%%%%%%%%%%%%%%%%%%%%%%%%%%%%%%%%%%%%%%%%%%%%%%%%%%%%%%%%%%%%%%%%%

The ALL functional \cite{AnLaLu96} calculates the
frequency-dependent molecular polarisability as a perturbation in the
screened electric field,
\begin{equation}
  {\bf E}(\bfr,\omega) = {\bf E}_{\rm ext}(\bfr,\omega)/\epsilon(\omega;n(\bfr))\;.
\label{eq:screen}
\end{equation}
It thus uses a local-density screening account, that is a local
approximation for the appropriate response functions~\cite{HuAnLuLa96}.
This would be wrong for macroscopic objects, but gives surprisingly good
results for atoms and molecules \cite{AnLaLu96,AnRy99}.

In \cite{HuAnLuLa96}, a procedure for describing the interactions of a molecule with a surface which relies on a better account of the electrodynamics is developed. It starts from
Maxwell's equations and the standard electrodynamic treatment of the
electric field and the displacement vector.  
The factorisation of the electron
density  proposed in ALL (\ref{eq:screen}) is equivalent to assuming a
local-density response to the external electric field. 
In the improved procedure, to get
 the atomic
polarisability $\alpha(iu)$ and similar quantities, one uses a
local relationship between the polarisation ${\bf P}(\bfr,\omega)$ and the
total electric field ${\bf E}(\bfr,\omega)$,
\begin{equation}
  {\bf P}(\bfr,\omega) = \frac{1}{4\pi}[\epsilon(\bfr,\omega) - 1] {\bf E}(\bfr,\omega)\;,
\label{eq:polP}
\end{equation}
and solves the Poisson equation $\grad\cdot{\bf D}(\bfr,\omega) =
\grad\cdot[\epsilon(\bfr,\omega) {\bf E}(\bfr,\omega)] = 0$ in the presence of an
external electric field ${\bf E}_{\rm ext}(\bfr,\omega)$.
In this evaluation a diagonal
dielectric tensor is used:
\begin{equation}
\epsilon_{\alpha\beta}(\vv{r},\vv{r'};\omega) =
\delta_{\alpha\beta}\;\delta(\vv{r}-\vv{r'})\;\epsilon(\vv{r};\omega)\;,
\label{eq:epstens}
\end{equation}
with the standard dielectric form
\begin{equation}
\epsilon(\vv{r};\omega) = 1 - \frac{\omega^2_{\rm p}(\vv{r})}{\omega^2}\;.
\label{eq:epsdrude}
\end{equation}

Applications to polarisabilities and charge centroids show that these
successfully describe the asymptotic physisorption of He, Be, and H$_2$
on jellium and of H$_2$ on the low-indexed faces of Al
\cite{HuAnLuLa96}. Comparison is also made with results from
time-dependent LDA calculations \cite{Li86} and experiment
\cite{AnEtAl96}. Calculated trends in the vdW coefficient and the vdW
reference-plane position $z_{\rm ref}$ \cite{HuKi97} signal that this
reference plane depends strongly on the crystal facet. There are also
applications to interactions between macroscopic bodies, in particular
between two parallel surfaces \cite{AnHuApLaLu98}.

With the overall goal of a general functional, the treatments of the
electrodynamics need unification.  A  method with a proper electrodynamical account for all kinds of geometries has also been developed
\cite{hurylula99}. This development unifies several
earlier treatments \cite{AnLaLu96,HuAnLuLa96,AnHuApLaLu98} used for the
cases of molecular pairs, a molecule outside a surface, and parallel
surfaces, respectively. 

It can be noted that in the long-range limit the fully electrodynamical
account performs quite well, even for the buckyball C$_{60}$, where
standard ALL has some problems \cite{AnRy99}. Furthermore, recent studies have
been concerned with properly describing the C$_{60}$ polarisation
\cite{fullerenewisdom,AsymptoticSeries}. However, this was done quite well
even with the fully electrodynamical account of ALL \cite{AnRy99}.  The calculated
polarisabilities and vdW coefficients are in good agreement with results
in the literature. This makes it possible to easily calculate these
quantities for complex systems with useful accuracy \cite{hurylula99}. Finally, we note that the approximation in ALL turns out to be in good agreement with the self-consistent account for typical molecular geometries. This was an important clue in the development of the van der Waals density functional for general geometries described in chapter VI.

%%%%%%%%%%%%%%%%%%%%%%%%%%%%%%%%%%%%%%%%%%%%%%%%%%%%%%%%%%%%%%%%%%%%%%%%
\section{Other methods for including vdW interactions}
%%%%%%%%%%%%%%%%%%%%%%%%%%%%%%%%%%%%%%%%%%%%%%%%%%%%%%%%%%%%%%%%%%%%%%%%

%%%%%%%%%%%%%%%%%%%%%%%%%%%%%%%%%%%%%%%%%%%%%%%%%%%%%%%%%%%%%%%%%%%%%%%%
\subsection{Brief overview of methods}
%%%%%%%%%%%%%%%%%%%%%%%%%%%%%%%%%%%%%%%%%%%%%%%%%%%%%%%%%%%%%%%%%%%%%%%%

Other methods have also been proposed for studies of vdW systems. Traditional DFT is a natural starting
point. There are three main types of approaches: (i) explicit density
functionals, (ii) DFT extended with atom-pair potentials, and (iii) perturbation theory, typically in the random-phase approximation. This 
review focuses on explicit density functionals with an emphasis on the
path that led to the vdW-DF. Still, 
we briefly review other methods
because the accuracy of different methods for the inclusion of vdW
forces is compared in so many studies and because the contrast between
these methods helps to highlight the nature of vdW-DF. 

The approach of extending approximate DFT with pair-potentials has been
widely used, both in jellium-type surface studies \cite{Li86,
HartenEtAl1986} and in explicit electron-structure calculations
\cite{Scoles,ElHoFrSuKa01,Wu02p515,Grimme1,Grimme2,TS09,Grimme3,Gianturco00p3011,Meijer96p8684}; for a
review see \cite{GrAnScMu07}. The force fields used have often been
heavily parametrised either through fitting to experimental data sets or
to calculated results using more advanced methods, although less so in
the most recent forms~\cite{Grimme3}.

The simplest pair potential is the London $\sim-R^{-6}$ form. In modern
variants, this potential is ``dressed'' with a damping function $F(R)$,
which preserves the long-range behaviour of the dispersion interaction,
while preventing the singularity in the dispersion term from
overwhelming the repulsive term at short ranges. An early form for
$F(R)$ was proposed by Brooks \cite{Br52}, who warned against the
crudeness of the approximation. More recently, Nordlander and Harris
proposed a prefactor similar to the $f(x)$ in the ZK expression
(\ref{eq:old3})~\cite{HarrisNordlander1984}. Starting from the ZK
expression for the vdW attraction potential $V(z)$ of an atom outside a
surface, they argue that introducing such a damping function amounts to
introducing a wavevector cutoff $k_c$ in a wavevector analysis
\cite{HarrisNordlander1984}.

For molecular complexes, from small organic molecules to large and composite systems, like sparse materials and protein-DNA complexes,
traditional DFT calculations with pairwise vdW potentials are commonly
used. 
Two well-known methods are DFT-D \cite{Grimme1,Grimme2,Grimme3}
and TS-vdW~\cite{TS09}.
The idea is that strong covalent bonds are well
described by traditional approximations, like the GGAs
\cite{LaMePRL1981,PeEtAlPRB1982,PeBuEr96} and a pair-wise atomistic
correction can be used to account for the vdW forces. 
In these methods,
there are two kinds of parameters that must be specified. 
The first
kind are dispersion coefficients $C_6$, which characterise the asymptote,
and the second kind are cutoff radii $r_{\rm cut}$, which characterise
the damping functions.

The DFT-D method has been successively refined for higher accuracy and less empiricism. 
In the recent DFT-D3 version \cite{Grimme3}
dispersion coefficients and cutoff radii are 
computed from first principles. 
To distinguish between dispersion coefficients of atoms in different chemical environments, the method relies on the concept of fractional coordination numbers. 
The DFT-D3 pair-wise account of vdW forces can be tailored to different density functionals lacking such forces by adjusting two global parameters.  
Advantages of the method include the facts that it is simple, asymptotically exact
for a gas of weakly interacting neutral atoms, and atomic forces are easy to calculate.
The DFT-D3 \cite{Grimme3} framework does allow for three-body
nonadditivity terms to be included, though these are not in the standard version. 
Such effects  have been shown to be non-negligible for large molecular systems and
organic solids \cite{Lilienfeld10p125010}.

The TS-vdW methods is also an almost parameter-free method for
accounting for long-range vdW interactions. It relies on the summation
of interatomic $C_6R^{-6}$ terms that are updated for each configuration
based on the electron density of a molecule or solid by scaling
reference coefficients for the free atoms \cite{TS09}. The mean
absolute error in the $C_6$ coefficients is 5.5\%, when compared to
accurate experimental values for 1225 intermolecular pairs. The
effective atomic $C_6$ coefficients have been shown to depend strongly
on the bonding environment of an atom in a molecule \cite{TS09}.

Comparing huge data sets is becoming more and more common in
first-principles and semiempirical calculations. 
The same datasets are often used in many comparative studies for
semiempirical parametrisation, for instance, no less than 29 different
systems in \cite{Grimme1} as early as 2004. However, these datasets
are often biased towards small dispersion bound molecules. The goal is
to develop a functional that crosses the boundary for what is thought to
be traditional dispersion-bound complexes to covalent-bonded systems. As
such, the vdW-DF philosophy has been to try to benefit from what can be deduced
from basic theory. Irrespective of these differences, we acknowledge
that skilful users of, for instance, the DFT-D method, regularly
produce valuable results.

A different kind of approximate approach, which can be seen as going beyond a
density-based xc in DFT, is to use the RPA to calculate the correlation
energy. This method has become much more efficient \cite{kaltak14p54115}, but still carries
higher computational cost than DFT. Since the RPA describes the
response in terms of single-particle orbitals, it is presumably a
suitable complement to the exact exchange energy~\cite{JarlKresse}, particularly when  comparing systems in which 
the number of particle-hole  excitations remains unchanged~\cite{Eshuis12}. The
RPA  correlation energy \cite{PinesNozieres} incorporates a
screened nonlocal exchange term and long-range dynamic correlation
effects that underpin vdW bonding \cite{JarlKresse}. Several
suggestions for RPA corrections exist \cite{MahansBok}. A recent study
\cite{RPA_single} suggests a single-excitation extension for RPA
calculations in inhomogeneous systems, which lowers the mean average
error for noncovalent systems. RPA calculations have been used for
solids \cite{JarlKresse,RPA:sol}, molecular systems
\cite{RPA:mol,RPA:ads}, and layered materials \cite{BjorkmannLayered1}.

Relatively expensive methods, such as RPA, provide a rough benchmark for methods that include vdW forces, particularly for bulk materials and larger systems where
traditional quantum chemistry methods fall short. However, these
methods are approximate and will themselves require extensive
benchmarking to determine how accurate they are in general.

%%%%%%%%%%%%%%%%%%%%%%%%%%%%%%%%%%%%%%%%%%%%%%%%%%%%%%%%%%%%%%%%%%%%%%%%
\subsection{Case of H$_2$ on Cu(111)}
\label{sec:H2Cu111case}
%%%%%%%%%%%%%%%%%%%%%%%%%%%%%%%%%%%%%%%%%%%%%%%%%%%%%%%%%%%%%%%%%%%%%%%%

\begin{figure}
\begin{center}
\includegraphics[width=\columnwidth]{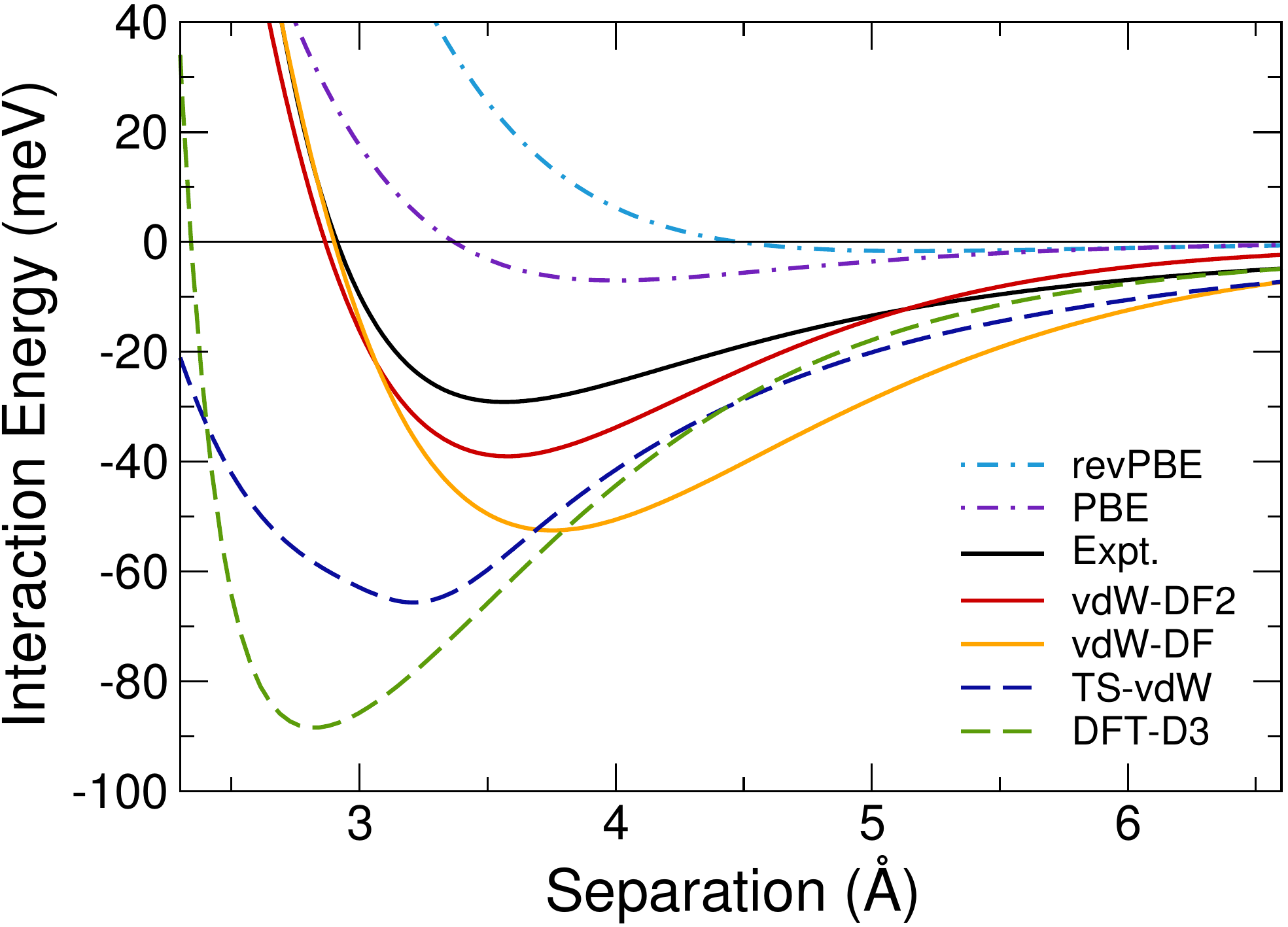}
\caption{\label{fig:PhysPotH2Cu} Comparison between experimentally
determined \cite{andersson1993} and calculated interaction energy curves
for H$_2$ on Cu(111) using different methods. Reprinted with permission
from \cite{lee12p424213}, \copyright\ 2012 Institute of Physics.}
\end{center}
\end{figure}

The accuracy of methods that include vdW forces are often assessed based
on how good the numbers are for a particular set of systems.
Quantitative comparisons are a common ingredient in many recent
publications in the field and there are often only small differences
between the methods, with modern vdW-DF variants typically faring well.
These comparisons are important, because accuracy is important, but that
does not mean that these studies always provide insight into how well a
method captures the physics of a problem. One can often learn more from
specific case studies. This is illustrated by a comparison
between modern calculated results on the interaction energy curve of
H$_2$ on Cu(111) \cite{lee11p193408,lee12p424213} with the results of
backscattering experiments detailed in section \ref{Sec:SPexp} and in
figure~\ref{fig:pot}. The comparison in figure \ref{fig:PhysPotH2Cu}
indicates a major shortcoming of typical pair-wise atomistic corrections
to DFT, namely their limited or complete inability to distinguish
between bulk and surface electrons. It is important to note here that
it does not suggest that such methods are inherently bad;
however, it does suggest that they are not good starting points for developing
general-purpose methods to handle all sorts of sparse and dense matter.
As such, it serves to motivate our emphasis on nonlocal correlation
functionals.

Figure \ref{fig:PhysPotH2Cu} shows the comparison between calculated
(DFT-D3 \cite{Grimme3}, TS-vdW \cite{TS09}, vdW-DF1, and vdW-DF2)
and
experimental results. vdW-DF2 gives interaction
curves in good agreement with the experimental physisorption curve.
vdW-DF1, exhibiting its expected overestimation of separation, is less
accurate but is still in qualitative agreement with experiment. DFT-D3 
on the other hand fails to predict a binding in reasonable agreement with experiment with a well depth four times deeper than
experiment and significantly shorter separations.
It should be noted that C$_6$ coefficients for every coordination number are not available for copper. 
However, using TS-vdW, which has built-in mechanisms to adjust the dispersion coefficients based on the density around the atoms does not resolve the issue. 

H$_2$ on Cu(111) is a particularly difficult system, because the
response of the tiny hydrogen molecule is so different from copper---a
coinage metal where $d$-shell electrons also come into play. The good
results of vdW-DF therefore present one of its major successes and are
strongly tied to its ability to distinguish between different density
regions---an ability that is very important. 
For many other cases, pair-potential based methods can produce numbers that are competitive with, and sometimes even better than, vdW-DF-based methods. This typically occurs for systems with a nature fairly similar to the datasets they were developed with. 
The failure of pair-potential based methods for the case of H$_2$ on Cu(111) however raises questions concerning the general transferability of such potentials, in particular to metallic systems. One can indeed question if the partitioning of vdW forces into contributions arising from specific atoms is justified for metallic systems.

%%%%%%%%%%%%%%%%%%%%%%%%%%%%%%%%%%%%%%%%%%%%%%%%%%%%%%%%%%%%%%%%%%%%%%%%
\subsection{Asymptotic functionals with damping functions }
%%%%%%%%%%%%%%%%%%%%%%%%%%%%%%%%%%%%%%%%%%%%%%%%%%%%%%%%%%%%%%%%%%%%%%%%

\label{sec:asympDamp}

The ALL functional, presented in the previous chapter, has served as an
important milestone on the path to vdW-DF. Others have further developed
ALL along a different path, using the ALL
functional to determine the dispersion coefficients for traditional DFT
with force-field potentials.
Some of these extensions start with an analysis of
the range-separated contributions to the full xc energy
\cite{Kurth99p10461,Fuchs02p235109}. 
Sato and coworkers \cite{Sato05p104307,Sato07p234114} developed an approach that 
 adds ALL to a long-range-corrected
DFT \cite{Iikura01p3540} based on a formulation that merges GGA exchange with
the Hartree-Fock description. 
At large separations, it includes
both nonlocal exchange and correlation terms. 
This method has been applied to $\pi$-bonded aromatic complexes, as well as
dipole-dipole and hydrogen-bonded systems \cite{Sato05p104307,Sato07p234114}. 
Gr\"afenstein and Cremer \cite{rev5} combined GGA with an efficient evaluation of the ALL energy and force terms in a partitioning scheme. They found good agreement with coupled-cluster calculations for the benzene dimer \cite{rev5}.

The Silvestrelli approach  is similar in nature, but evaluates the ALL term
\cite{Silvestrelli08p53002} for localised orbitals
\cite{JoGu89}, specifically maximally localised Wannier functions
\cite{Marzari97p12847,Silvestrelli98p7}. 
The idea is to describe
ALL in terms of the partial electron densities of the occupied Wannier orbitals.
Expanding the density in such orbitals $n = \sum_l |w_l|^2$ in each
fragment, the ALL functional is used to determine asymptotic vdW
interaction coefficients C$_{6,l,m}$ for every pair $(l,m)$ that can be
formed from two orbitals of different fragments. 
The total vdW interaction
is evaluated by summing over all such pairs. An advantage of this
approach is that a damping function exclusively affects contributions of
inter-fragment $(l,m)$ pairs which have overlapping densities. The
approach is successful for a range of systems, including vdW solids,
molecular dimers, and aromatic complexes
\cite{Silvestrelli08p53002,Silvestrelli09p5224}.

More recently, two extensions of this  framework were designed
\cite{Andrinopoulos11p154105,ambrosetti12p73101}. The first includes
additional states in the Wannier representation
\cite{Andrinopoulos11p154105} to improve orbital
localisation and symmetry which, for example, is relevant 
for describing the benzene ring. 
The second \cite{ambrosetti12p73101}
corrects for the  density overlaps that exist within fragments and
replaces the ALL specification of the asymptotic vdW coefficients with
the simpler London account \cite{Lo37}. The first step eliminates
arbitrariness in the vdW description that can arise from symmetry
breaking, for example in weakly bonded noble-gas dimers
\cite{Andrinopoulos11p154105}. These extensions and adjustments of the
original formulation by Silvestrelli \cite{Silvestrelli08p53002} produce good
agreements with the binding energies of coupled-cluster calculations for
a range of organic complexes.

Finally, other approaches that rely on density dependent $C_6$ coefficients 
 have also been developed and used to include vdW forces in DFT.
 The Tkatchenko-Scheffler method
\cite{TS09} mentioned earlier is a hybrid between this approach and
traditional atomistic pair potentials. There are also the Vydrov-Van
Voorhis \cite{Vydrov10p62708} and the Becke-Johnson
\cite{Becke05p154101,Becke07p154108} formulations.
The Vydrov-Van Voorhis approach is a limiting case of the VV09
functional \cite{VV09} (section~\ref{sec:VV}) that is itself a fully nonlocal density
functional related to the general geometry vdW-DF.
The Becke-Johnson
formulation, which also provides multipole corrections in terms of $C_8$ and
$C_{10}$ coefficients, describes the vdW interactions as an electrodynamical multipole
coupling of nonisotropic exchange holes that are formed in the electron
gas around the nucleus. Emphasising the energy shifts arising with the hole
coupling, the approach is linked with the RA picture \cite{RA} of vdW forces and thus
related to the physical picture underpinning ALL.

%%%%%%%%%%%%%%%%%%%%%%%%%%%%%%%%%%%%%%%%%%%%%%%%%%%%%%%%%%%%%%%%%%%%%%%%
\section{Functionals for all distances -- functional for layered systems}
\label{sec:layered}
%%%%%%%%%%%%%%%%%%%%%%%%%%%%%%%%%%%%%%%%%%%%%%%%%%%%%%%%%%%%%%%%%%%%%%%%

This chapter is devoted to our first all-space vdW functional. 
The first four chapters gave an introduction to (i) the nature of vdW
interactions, (ii) their importance, (iii) their relevance for DFT, and
(iv) early work that stimulated further development. Here, we present the
early history of the vdW-DF method, with the first explicit general
functionals vdW-DF (more recently termed vdW-DF0) in 2003
\cite{RyLuLaDi00, 2001SurfScience, RydbergThesis, Rydberg03p606,
Rydberg03p126402}, vdW-DF1 (originally termed vdW-DF) in 2004 \cite{Dion,
Thonhauser}, and vdW-DF2 in 2010 \cite{Lee10p081101}. The vdW-DF method
has the potential for developing successively improved functionals
\cite{behy14, BeArCoLeLuScThHy14, hybesc14}. Their common roots allow
us to present a joint introduction to these functionals.
Following this introduction, we outline the derivation of vdW-DF0 and present examples of applications to illustrate its nature and its potential for future improvements. 

vdW forces are often associated with
asymptotic formulas and such formulas are used in many theoretical schemes. The singular behaviour that occurs at small separations has been dealt with by introducing saturation functions. 
However, vdW forces are important for all bonds and reactions as they originate from nonlocal correlations among electrons and are relevant in an extensive region of intermediate-sized separations. We seek to construct approximate vdW functionals with an account of sparse matter that
seamlessly extend the local (LDA) and semilocal approximations (GGA) for
exchange and correlation \cite{KoSh65,GuLu76,LaPe77,PeWa92,PeBuEr96}.

The adiabatic connection formula (ACF)~\cite{LaPe75,GuLu76,LaPe77} is
the starting point for developing vdW-DF. 
The ACF provides an expression for the exchange-correlation
energy $E_\xc$ in terms of a coupling-constant integration $\lambda$, as
follows
\begin{equation}
 E_\xc = - \int_0^1 \diff \lambda \int_0^\infty \frac{\diff u}{2 \pi} \,
\Tr \left\{ \chi(\lambda,i u) V \right\}
-E_{\self}\;.
\label{eq:ACF}
\end{equation}
Here, $\chi(\lambda,i u )$ is the density response function or the
reducible density-density correlation function in many-body theory, at a
coupling strength $\lambda$ with a density $n(\rr)$ set to that of the
fully interacting system. $V(\vv{r}-\vv{r'}) =
{1}/{\abs{\vv{r}-\vv{r'}}}$, and $E_{\self}$ is the Coulomb energy of all
electrons. The coupling-constant integration is computationally complex
and calls for simplification and approximations.

In the vdW-DF method \cite{RyLuLaDi00, RydbergThesis, Rydberg03p126402, Paper6-01, Dion, DionThesis, Langreth05p599, Thonhauser, Lee10p081101, behy14, BeArCoLeLuScThHy14}, we work with the local field response
function, or irreducible correlation function
$\tilde{\chi}(\lambda,iu)$, defined via a Dyson equation
\begin{equation}
\chi(\lambda,iu)=\tilde{\chi}(\lambda,iu)+\lambda
\tilde{\chi}(\lambda,iu) V \chi(\lambda,iu)\;.
\label{eq:chirelate}
\end{equation}
This function provides a full description of screening in the electron
gas at any given coupling constant $\lambda$. The  leading order
approximation for $\chi(\lambda,iu)$, i.e.\ the RPA,
sets $\tilde{\chi}(\lambda,iu) = \tilde{\chi}(0,iu)$. Since 
$\lambda$ then acts just as a prefactor of $V$ in the coupling-constant
integration, it can be performed analytically. Higher-order correlation
diagrams with internal Coulomb interactions, each proportional to
$\lambda$, may have intricate $\lambda$-dependences in the irreducible
correlation function $\tilde{\chi}(\lambda,iu)$ that prohibit analytical
solutions and complicate numerical ones.

Progress in the development of vdW-DF was made~\cite{RyLuLaDi00} by assuming that the
coupling dependence of $\tilde{\chi}$ can be neglected when performing
the $\lambda$ integration (\ref{eq:ACF}), even beyond the RPA. Since the
leading term in $\lambda$ is made explicit, we approximate
$\tilde{\chi}(iu)$ by a value at a characteristic coupling constant $0<
\lambda_c < 1$ and complete the integral to obtain
\begin{equation}
E_{\xc} = \int_0^{\infty} \, \frac{\diff u}{2\pi} \,
\Tr\{\ln[1-\tilde{\chi}(iu)V]\} - E_{\self}\;.
\label{eq:formalEval}
\end{equation}
The PhD thesis of H. Rydberg~\cite{RydbergThesis,Paper6-01} mentions
``RPA-like approximations,'' noting that approximations beyond the RPA
can also allow for an approximate evaluation of the coupling-constant
integration (\ref{eq:ACF}). This can for instance be achieved by using
a dielectric function, where the $\lambda$-dependence is absorbed into
the plasmon dispersion~\cite{Paper6-01}.

The layered vdW-DF in its early formulation~\cite{RyLuLaDi00} and
subsequent realisations \cite{2001SurfScience, RydbergThesis,
Rydberg03p606, Rydberg03p126402, Langreth05p599} are based on the
so-called full-potential approximation (FPA), as detailed in a later
publication \cite{Langreth05p599}. In this section, we further discuss
the FPA together with the layered-geometry vdW-DF version. This
approximation involves replacing $\tilde{\chi}_{\lambda}$ by
$\tilde{\chi}_{\lambda=1}$, leading to equation (\ref{eq:formalEval}).
In the vdW-DF design, a model dielectric function $\epsilon$ specifying
$\tilde{\chi}$ should not include nonlocal correlations
\cite{RydbergThesis, Paper6-01, DionThesis, Dion, Thonhauser, hybesc14}.
Reference \cite{Langreth05p599} discusses spectator
excitations~\cite{RA} in $\tilde{\chi}(\lambda,iu)$ and notes that if
 spectator contributions in $\tilde{\chi}_{\lambda=1}$ are discarded
in the FPA, one obtains the exact vdW asymptote.

The vdW-DF method differs from RPA-based methods, which specify the
response functions $\tilde{\chi}$ in terms of the single-particle
orbitals without taking many-body effects into account. The vdW-DF
method instead relies on a scalar model dielectric function. The
continuity equation relates this model dielectric function to
$\tilde{\chi}$~\cite{RyLuLaDi00, RydbergThesis, Dion, Langreth05p599,
Thonhauser}, as follows
\begin{equation}
\tilde{\chi} = \nabla \alpha \cdot \nabla \;,
\end{equation}
where the scalar susceptibility $\alpha$ satisfies $\epsilon=1+4\pi
\alpha$. In terms of the scalar dielectric function $\epsilon$, the
formal result (\ref{eq:formalEval}) can be written as
\begin{equation}
 E_{\rm xc} =
\int_0^\infty \frac{\diff u}{2 \pi} \,
\Tr \left\{ \ln\left(\nabla \eps\cdot \nabla G \right) \right\}
-E_{\self}\;,
\label{eq:Found2}
\end{equation}
where $G=-V/4\pi$ denotes the Coulomb Green function. The model
dielectric function $\epsilon$ is chosen to make $\tilde{\chi}$ reflect
essential features of an LDA and GGA description.

For the homogeneous electron gas (HEG), equation (\ref{eq:formalEval})
reduces to the following expression
\begin{equation}
E_{\xc}^{\heg} = \int_0^{\infty} \, \frac{\diff u}{2\pi} \,
\Tr\{\ln[\epsilon^{\heg}(iu)]\} - E_{\self} \;,
\label{eq:formalEvalHEG}
\end{equation}
where $\epsilon^{\heg}(iu)$ denotes the dielectric function of the
homogeneous system. This result will be helpful when we remove
short-ranged correlation effects from the full functional.

The expression for $E_{\xc}$ in the vdW-DF framework (\ref{eq:Found2})
can in principle be formulated as an exact
relation~\cite{Paper6-01,hybesc14}. The local field response functional
$\tilde{\chi}$ is then viewed as arising at a characteristic
coupling-constant value $0 < \lambda_{\rm char} < 1$. The argument is
that for every $E_{\xc}$ one can find a scalar dielectric function
that satisfies equation (\ref{eq:Found2})~\cite{RydbergThesis}. Such a
scalar dielectric function can formally be expressed~\cite{hybesc14} in
terms of
\begin{equation}
\chi(iu) V = 1 - \exp\left[\int_0^1 \, \diff \lambda \, \chi(\lambda,iu)
V\right]\;,\label{eq:acfeval}
\end{equation}
which provides a mean-value specification of $\chi(iu)$. In principle,
this expression can be used to determine the characteristic
$\tilde{\chi}(iu)$ of equation (\ref{eq:formalEval}). The different vdW-DF versions thus rest on a common formally exact framework \cite{Paper6-01,hybesc14}. The various versions differ in which approximations are made.

%%%%%%%%%%%%%%%%%%%%%%%%%%%%%%%%%%%%%%%%%%%%%%%%%%%%%%%%%%%%%%%%%%%%%%%%
\subsection{Development of the layered-geometry van der Waals density
functional}
%%%%%%%%%%%%%%%%%%%%%%%%%%%%%%%%%%%%%%%%%%%%%%%%%%%%%%%%%%%%%%%%%%%%%%%%

Planar-geometry problems can be viewed as quasi-one-dimensional. Treating them as such
was helpful in the formulation of the van der Waals density functional
for layered geometries. Though its general applicability is limited, this
functional is useful for key model systems such as jellium slabs and
some real materials such as vdW bonded layered systems.

The first formulation of the layered vdW-DF functional was proposed by
Rydberg, Dion, Langreth, and Lundqvist for strictly planar systems
\cite{RyLuLaDi00}. Results for the interaction between two parallel
jellium slabs  agree well with those of an earlier more
elaborate calculation \cite{DoWa99}. 
The correlation-interaction energy was found to saturate significantly with shorter separations.
The study concludes: ``generalisations
to three-dimensional systems are possible and (\ldots) there
should be a basis for applications to numerous physical, chemical, and
biological systems, (\ldots) such as crystals, liquids, adhesion, soft
condensed matter, and scanning-force microscopy.''

The layered-geometry vdW-DF~\cite{2001SurfScience,Rydberg03p126402,Rydberg03p606} of Rydberg, Dion, Jacobson, Schr{\"o}der, Hyldgaard,
Simak, Langreth, and Lundqvist can be applied to real layered systems,
such as graphite, boron nitride, and molybdenum disulphide, with the
breakthrough of a first predicted vdW-bond energy value of an extended
system, a graphene bilayer, in
2001 \cite{2001SurfScience}. In this functional, the
exchange-correlation energy functional
\begin{equation}
E_\xc^\text{vdW-DF}[n] = E_\xc^{0}[n] + E_{\rm c}^\nl[n]\;,
\label{eq:vdWDFsplit}
\end{equation}
is split into a semilocal functional part $E_{\xc}^0[n]$ and a nonlocal
part $E_{c}^{\nl}[n]$. This splitting is retained in later versions of
the functional. The full electron-density variation, including the
in-sheet variations, is retained in $E_{\xc}^0$, while the evaluation of
the smaller nonlocal correlation $E_c^{\nl}$ is based on a planar
average of the density. The assumption is that the vdW forces are less
sensitive to the microscopic details than other terms of the DFT
evaluation. As a convenient shorthand, we will denote this layered-geometry vdW-DF as vdW-DF0 in the following.

vdW-DF0 is valid at all separations. The semilocal functional $E_{\xc}^0$ is specified as the sum of
local correlation \cite{PeWa92} and gradient-corrected exchange given
within GGA using the Zhang and Yang reparametrisation \cite{ZhYa98} of
PBE \cite{PeBuEr96} (revPBE). This functional is chosen because it was
considered to give interaction energies in close agreement with exact
Hartree-Fock exchange at separations relevant for vdW bond lengths and
did not exhibit appreciable binding in vdW
systems~\cite{Langreth05p599}.

A central ingredient in vdW-DF0 is a simple plasmon-pole description of
the sheet response. The evaluation of the nonlocal correlation term is
adapted from \cite{RyLuLaDi00}, and the laterally-averaged density
$n(z)$ is used to define a characteristic $\tilde{\chi}$. The response
description is in turn given by a model dielectric function based
on a plasmon-pole form,
\begin{equation}
\eps_k(z,iu) = 1 + \frac{\omp^2(z)}{u^2 + (mv_{\rm F}(z)q_k )^2/3 +
q_k^4/4 }\;.
\label{eps:layered}
\end{equation}
Here $\omp^2(z)=4\pi n(z)/m$ and $mv_{\rm F}(z)=[3\pi^2 n(z)]^{1/3}$.
The  plasmon dispersion is  $q_k^2 = k^2 + q_\perp^2$,
where $\bk$ is the in-plane wavevector. 
In this model, the $q_\perp^2$ value has a special purpose, as it is designed to reflect the local-field susceptibility $\alpha$
corresponding to the dielectric function. It is 
set by matching the static polarisability  to that
of  an individual sheet as obtained in a regular  DFT calculation with an external field. 
Since the $q_\perp^2$  reflects the full density variation, it mitigates the error of computing
vdW forces from a laterally averaged density.

The dielectric function is local in the $z$-direction but includes
in-plane dispersion. The definition (\ref{eps:layered}) thus extends the
form $\epsilon=\epsilon^{\heg}$ used to define a plasmon propagator
$S\equiv 1-\epsilon^{-1}$, as used in early determinations of LDA
correlation \cite{Lu67,HeLuJPC1971,GuLu76}. The plasmon-pole form of
$\epsilon$ (\ref{eps:layered}) is inspired by the models of Lindhard and
Bohr \cite{Lindhard,BohrLindhard}. The construction for
layered-geometry vdW-DF0 calculations can be seen
as an approximation to treating $\tilde{\chi}$ in the FPA limit as $q_\perp^2$ is evaluated at the
full coupling strength.

\begin{figure*}
\begin{center}
\includegraphics[height=1.75in]{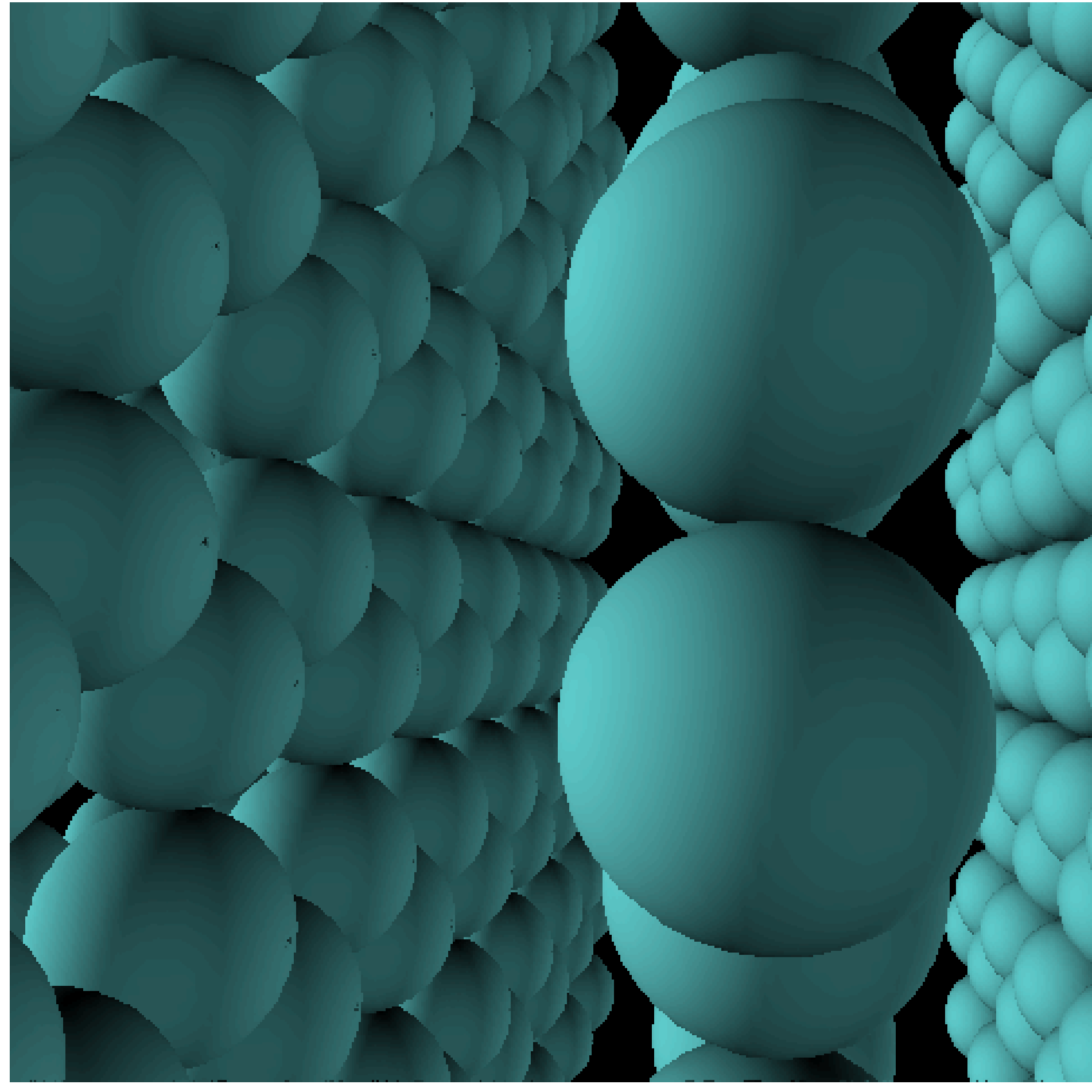}\hfill
\includegraphics[height=1.795in]{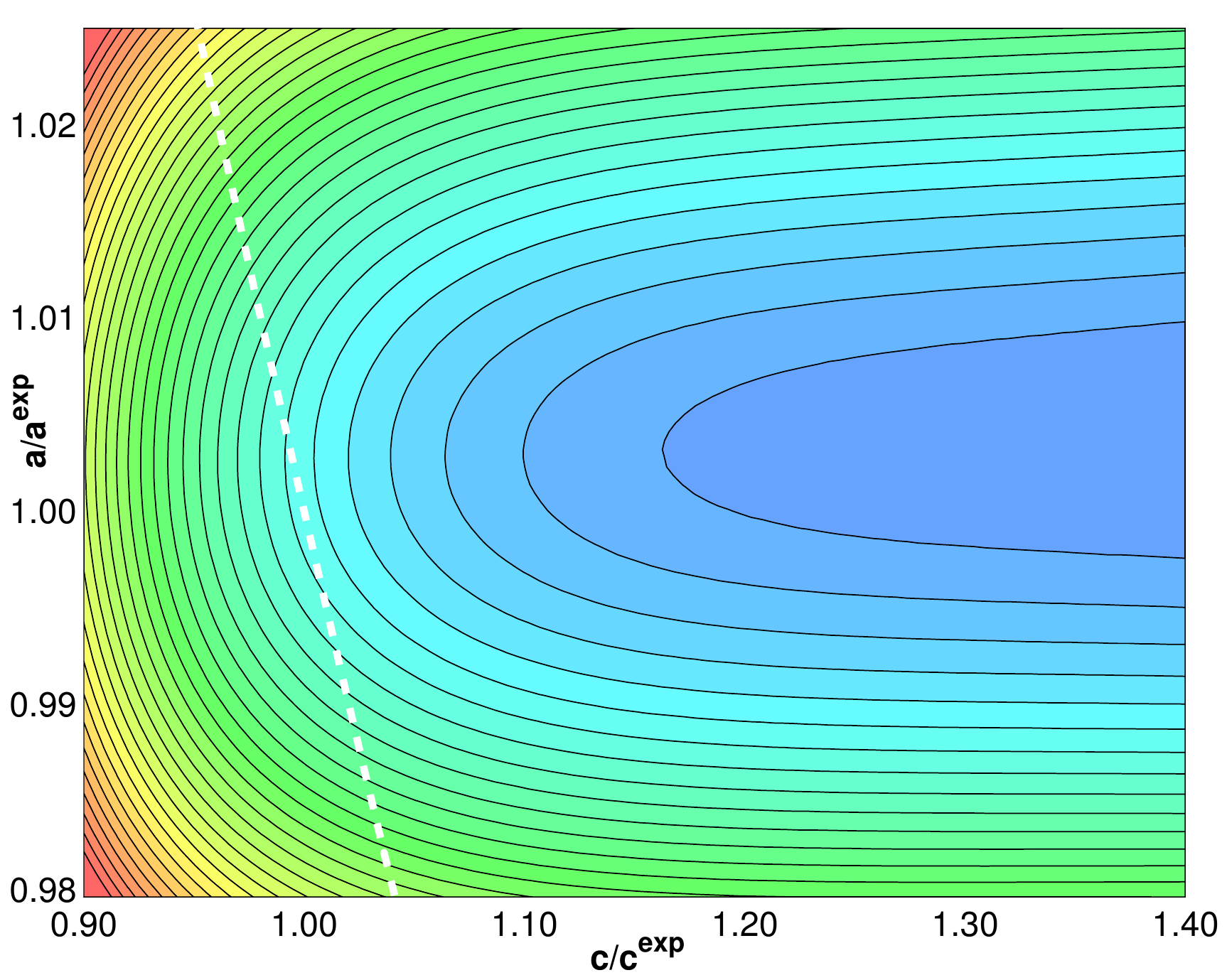}\hfill
\includegraphics[height=1.795in]{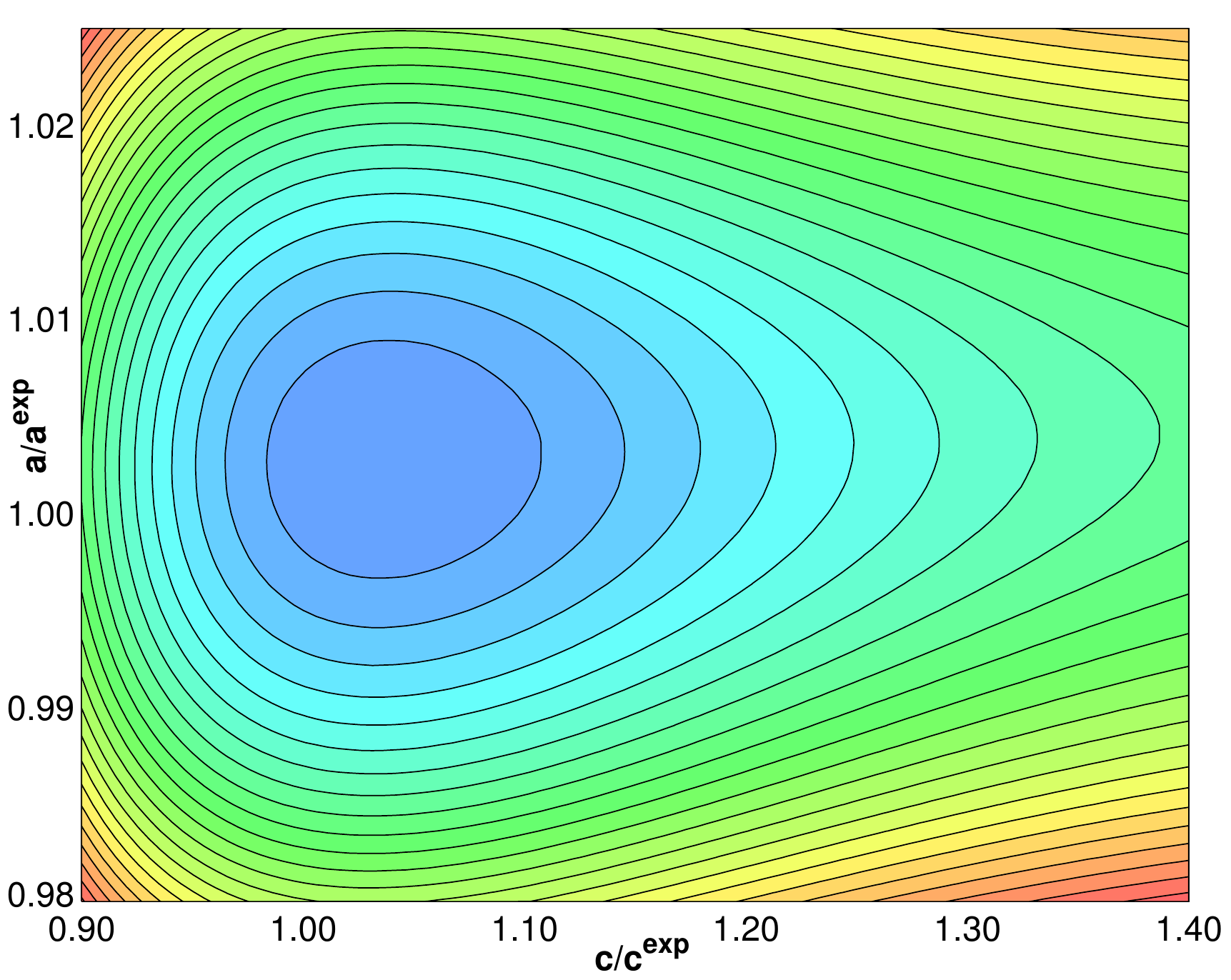}
\caption{\label{fig:layeredgraphite} The role of vdW forces in graphite.
The left panel shows the atomic configuration which comprises densely
packed graphene sheets with large inter-sheet separations, $c_{\rm exp}\approx
3.35$ {\AA}. A low electron density characterises the inter-sheet
regions. The nonlocal correlation acts across these regions, providing
material cohesion. The middle and right panels contrast the variation of
the interlayer interaction energy for our GGA (PW91 \cite{PeWa91}) and
layered-geometry vdW-DF0 \cite{Rydberg03p126402} studies, here plotted
as functions of both inplane lattice constant $a$ and interlayer
separation $c$, relative to the experimental values $(a^{\rm exp},c^{\rm
exp})$. %The dashed curve in the middle panel is discussed in the text.
The dashed curve in the middle panel identifies ($a$,$c$) values that are consistent with the experimentally observed unit-cell volume of graphene. 
The colour scale for the potential energy plots ranges from 0 (light green) 
to $-5$ eV (dark blue). Reprinted with permission from \cite{Rydberg03p606},  
\copyright\ 2003 Elsevier.}
\end{center}
\end{figure*}

To obtain an explicit expression for the nonlocal term $E_c^{\nl}$ \cite{RyLuLaDi00,Rydberg03p126402}, we
need to remove the semilocal 
part of equation~(\ref{eq:Found2}).  The HEG result (\ref{eq:formalEvalHEG}) suggests the
following expression
\begin{align}
E_\xc^{0} =
\mathrm{Re} \int_0^\infty \, \frac{\diff u}{2 \pi} \,
\Tr \left\{ \ln( \eps) \right\}-E_{\self}\;.
\label{eq:FPAsr}
\end{align}
In effect, we obtain an expression for the nonlocal correlation
\begin{equation}
E^{\nl}_{c} \equiv \int_0^\infty \frac{\diff u}{2 \pi}
\Tr \left[ \ln \right(\nabla \eps \cdot \nabla G\left)
-\ln \eps \right] \;. \label{eq:nlstart}
\end{equation}

The evaluation of $E_c^{\nl}$ proceeds by viewing equation
(\ref{eq:nlstart}) as a one-dimensional electrodynamical
problem~\cite{RyLuLaDi00,RydbergThesis,Rydberg03p606,Langreth05p599}.
One solves a Poisson equation
\begin{equation}
\frac{\diff }{\diff z} \left(\epsilon_k \frac{\diff }{\diff z} \phi_k\right) - k^2 \epsilon_k \phi_k= 0
\end{equation}
to obtain the solution $\phi_k(z,iu)$, given that a charged sheet is
introduced at $z=0$. One also determines the solution
$\phi_{k,0}(z,iu)$ for empty space, i.e., with $\epsilon=1$. 
Finally, one extracts the nonlocal correlation energy per area $A$
\cite{RyLuLaDi00,Langreth05p599}
\begin{equation}
E_c^{\nl}/A = - \lim_{L\to \infty} \int_0^{\infty} \, \frac{\diff u}{2\pi} \,
\int \, \frac{\diff^2k}{(2\pi)^2} \, \ln\left(\frac{\phi_k'(L)}{\phi'_{k,0}(L)}\right)\;.
\end{equation}

A valuable feature of vdW-DF0 is that it treats the screening exactly by
solving the Poisson equation. The idea of mapping the $E_{\rm c}^{\nl}$
evaluation onto an electrodynamics problem
\cite{RyLuLaDi00,RydbergThesis} could be more widely applied. One might
conceive calculating the nonlocal correlation in a similar manner for
other systems with high symmetry, for example, filled and hollow
spheres and cylinders.

%%%%%%%%%%%%%%%%%%%%%%%%%%%%%%%%%%%%%%%%%%%%%%%%%%%%%%%%%%%%%%%%%%%%%%%%
\subsection{Applications of the layered functional}
%%%%%%%%%%%%%%%%%%%%%%%%%%%%%%%%%%%%%%%%%%%%%%%%%%%%%%%%%%%%%%%%%%%%%%%%

The major features of the vdW bond in a prototype DFT application can be
illustrated with graphite. Figure~\ref{fig:layeredgraphite} shows the
atomic configuration with dense and sparse regions (left),
binding-energy contours \cite{Rydberg03p606} obtained with traditional DFT in the
GGA (middle), and with the layered-geometry vdW-DF0 (right) as functions
of both the inplane (intra-sheet) lattice constant $a$ and inter-sheet
separation $c$ divided by their experimental reference values $(a_{\rm
exp},c_{\rm exp})$. The dashed curve in the middle panel identifies 
$(a,c)$ values that are consistent with the experimentally observed
unit-cell volume $V_{\rm Gr}^{\rm exp}$.

In the GGA description, interlayer binding is
absent except at an unphysically large separation and with minuscule
binding energy, a few meV per graphene atom. An early GGA study of graphite
structure found a relevant value of $c$ only because it constrained the 
search to the curve specified by $V_{\rm Gr}^{\rm exp}$ \cite{Rydberg03p606}.  
The vdW-DF0 results reflect a solution to a 
long-standing challenge in traditional DFT: the state of the art DFT, 
which was GGA at the time, should not rule out the
existence of the most stable carbon allotrope, i.e.\ graphite.

With the development of the layered-geometry vdW functional, vdW-DF0, 
in 2001 \cite{2001SurfScience,Rydberg03p606}, we could announce ``a binding-energy value of 10.3~meV/{\AA}$^2$''
\cite{2001SurfScience}, which corresponds to 43 meV per atom
\cite{Rydberg03p606} in good agreement with the experimental reference at the time  of 34
meV/atom \cite{Benedict98p490}. In contrast, LDA calculations yielded 20
meV \cite{YiCo84,ScMa92,Furthmuller94p15606,ChGoMi94,Charlier95p43} and
GGA calculations barely gave binding. 
The simple fact
that vdW-DF had a number, a correct sign, and a reasonable magnitude spurred much enthusiasm among the developers in the field and this optimism was communicated by the title of the report  ``Hard Numbers for Soft
Matter'' \cite{Rydberg03p606}. The early vdW-DF0 applications indeed signalled
the start of an era of tremendous activity by us and others. 

Early applications of vdW-DF0 include the binding between two graphene
sheets~\cite{2001SurfScience,Rydberg03p606,Rydberg03p126402} and the
interlayer binding in graphite, boron nitride, and molybdium disulphide,
MoS$_2$ \cite{Rydberg03p606,Langreth05p599}. The results were judged
promising for the vdW-DF method \cite{Rydberg03p606}. Overall, the
performance was encouraging for binding distances, binding separations,
and elastic response, considering its nonempirical basis and the
approximations made.

The first applications of vdW-DF0 for real materials relied on the PW91
functional to describe semilocal exchange effects, which later was replaced with revPBE to avoid spurious binding. This
substitution results in smaller binding energies and larger interlayer
separation distances and therefore generally reduces the agreement with
experiments. The choice was nevertheless important to firmly establish
the role of vdW forces in layered materials.

\begin{figure}
\centering
\includegraphics[width=\columnwidth]{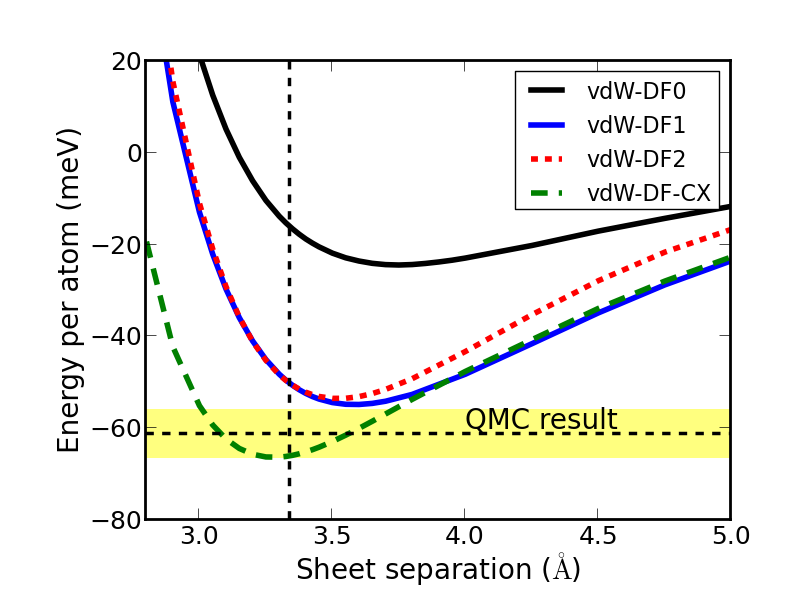}
\caption{\label{fig:graphite} Comparison between the results for the interlayer binding energy of graphite obtained with the layered-geometry vdW-DF~\cite{Rydberg03p606},  later general-geometry vdW-DF functionals, and a quantum Monte Carlo (QMC)~\cite{qmc_graphite, Spanu09p196401} calculation. Here, the covalent bond length in the layer is fixed to the experimental value. The dashed vertical line indicates the experimental reference for the sheet separation. The dashed horizontal line indicates the results of QMC, with error bars indicated by the yellow region.}
\end{figure}

Figure \ref{fig:graphite} compares results of vdW-DF0 and later versions of the
vdW-DF for general geometries for graphite. As a modern reference for
the cohesion energy, we use Quantum Monte Carlo (QMC) results
\cite{Spanu09p196401}. Thus, we avoid experimental uncertainties and
contributions from zero-point and thermal motion that are not a part of
standard DFT in the Born-Oppenheimer approximation. The figure shows a
successive improvement in cohesion energy and separation with the
biggest step being taken when going from vdW-DF0 to vdW-DF1. The shallow
binding of vdW-DF0 and the overestimation in separation can partly be
attributed to the choice of exchange, as revPBE is quite repulsive in
this region.

The comparison of results from different functionals in figure~\ref{fig:graphite} indicates that the plasmon model of
vdW-DF1 and vdW-DF2, described in chapter~\ref{sec:general-geometry}, is more suited for describing nonlocal correlations
than that of vdW-DF0. An advantage of vdW-DF0 is its
exact handling of screening, although subject to the limitations imposed
by its more limited description of local-field effects.

In addition to the early progress for the layered vdW system, the vdW-DF0 has
also been adapted to provide an early DFT account of
 polyaromatic hydrocarbons (PAHs) dimers in selected geometries
\cite{ChSc05a,ChSc05b,Chakarova-Kack10p013017}. The adaptation rests on
defining an effective area per PAH molecule \cite{ChSc05b}. This step
can be viewed as an analogy to deducing estimates of the graphite interlayer binding energy
 from the measured binding energy of increasingly larger
PAH dimers~\cite{Zacharia}. This effective-area approach gives at least
a qualitative account of vdW binding between PAH molecules
\cite{ChSc05a,ChSc05b}.

vdW-DF0 fairly well describes the cohesion
in layered systems like graphite and other layered materials such as
boron nitride and molybdenum disulphide. This is in spite of the fact
that the nonlocal correlation term in vdW-DF0 is based on a laterally
averaged density with an emphasis on a FPA representation of the
local-field response $\tilde{\chi}$. Perhaps the most important aspect of the vdW-DF0 development was the strong encouragement it provided for further developments.

%%%%%%%%%%%%%%%%%%%%%%%%%%%%%%%%%%%%%%%%%%%%%%%%%%%%%%%%%%%%%%%%%%%%%%%%
\section{General-geometry vdW-DF functionals}
\label{sec:general-geometry}
%%%%%%%%%%%%%%%%%%%%%%%%%%%%%%%%%%%%%%%%%%%%%%%%%%%%%%%%%%%%%%%%%%%%%%%%

%%%%%%%%%%%%%%%%%%%%%%%%%%%%%%%%%%%%%%%%%%%%%%%%%%%%%%%%%%%%%%%%%%%%%%%%
%%%%%%%%%%%%%%%%%%%%%%%%%%%%%%%%%%%%%%%%%%%%%%%%%%%%%%%%%%%%%%%%%%%%%%%%
A general density functional can not be limited to layered geometries. The focus thus moved quickly to designing a general-geometry version \cite{RydbergThesis,Dion}. This chapter describes the overall design of such a functional, outlines its derivation, and discusses its underlying properties. A  few applications and results that were important for the development work are described, as is the development of self-consistent implementations. Some valuable refinements of the method are also discussed. Finally, we discuss two functionals closely related to vdW-DF.

\subsection{Design of general-geometry vdW-DF versions}

A general-geometry vdW-DF functional should obey general
physical laws, be physically transparent, transferable, and simple enough to allow efficient computations. 
Previous studies in the late 20th century 
taught us that (i) nonlocal correlations among the electrons are
essential for describing sparse matter; (ii) vdW forces emanate from
dynamic electron correlations; (iii) vdW forces relate to the static
electron density $n(\bf{r})$ via, for instance, the classical plasma
frequency $\omp = \sqrt{4\pi n e^2 /m}$; and (iv) the asymptotic vdW
potentials can at large separations be derived from the small wavevector ($q\to 0$) limit
of the plasmon dispersion $\omega_{\bq}$ and give a reasonable magnitude
and correct form for the interaction between atoms and molecules
\cite{AnLaLu96,AnRy99}, between neutral molecules and insulating
surfaces \cite{HuAnLuLa96}, as well as between surfaces
\cite{AnHuApLaLu98}. This experience was incorporated into the
layered-geometry functional, vdW-DF0 \cite{RyLuLaDi00, 2001SurfScience, RydbergThesis,
 Rydberg03p126402, Rydberg03p606}, as was the value of a
plasmon-pole description for the electron-gas dielectric function
\cite{Lu67,La70}.

The vision for a new nonlocal functional for general geometries was, and still 
is, that it should be a general-purpose one; able to describe all kinds of materials and molecular systems from the dense to the sparse~\cite{RydbergThesis, DionThesis,behy14,BeArCoLeLuScThHy14,hybesc14}. This philosophy emphasises binding separations in favour
of the asymptote. While the asymptotic behaviour is determined by 
long wavelength, i.e. $q \rightarrow 0$, excitations in the polarisable
medium, nonlocal correlations at shorter separations arise from many
different electron-hole pair excitations and plasmon modes.
To describe vdW forces at typical binding separations, 
a plasmon model that covers the overall effect of
many different $q$ values is needed.
For this aim, the GGA can be used as a
guide, as this approximation provides an excellent account of
interactions within dense matter. 
The plasmon model should adhere to known constraints for the electron
gas \cite{Paper6-01, Dion, Thonhauser, Lee10p081101, BeHy13,
behy14, BeArCoLeLuScThHy14, hybesc14}. 
The functional should also
connect seamlessly between the description at van der Waals binding separations and the regime of covalent bonds.

The preferred procedure used to derive the explicit vdW-DF functionals
has varied with time. In this chapter we outline a procedure, 
similar to the one appearing in H.~Rydberg's thesis
\cite{Paper6-01}, which has been successively refined and
complemented \cite{Dion,Langreth05p599,Thonhauser,hybesc14}. This procedure is based
on a mean-value evaluation of the integral of the dimensionless coupling
constant $\lambda$, as detailed in chapter V. 
This starting point is exact and is thus more natural for deriving the
explicit functionals~\cite{Paper6-01,hybesc14} than relying on the FPA
outlined in the seminal paper by Dion and
coworkers~\cite{Dion}.

To design a general geometry functional, an essential step is to define
an expansion parameter for the coupling-constant averaged expression in (\ref{eq:Found2}) and for the nonlocal correlation (\ref{eq:nlstart}).
In this step, we use the propagator concept of many-body theory,
building on simple forms like $1/{(\omega - \omega_q)}$. For the
functional development from (40), a suitable choice is $S_\xc(i u)
\equiv \ln(\epsilon(iu))$, which can be seen as an auxiliary, spatially
nonlocal, polarisation function \cite{Paper6-01}. This dielectric
function is formally defined in terms of the coupling-constant
integration of the density response function as discussed in the
chapter~\ref{sec:layered}. To better understand the connection to the MA,
RA, and ALL picture of vdW interactions and nonlocal correlations
\cite{MA,RA,AnLaLu96} (discussed in chapter~\ref{sec:asymptotic}),
we may instead consider $S(iu)\equiv 1- \epsilon(iu)^{-1}$,
that is $S_\xc$ to lowest order in $S$. Its connection
to the propagator form is easily seen in the case where
$\epsilon(\omega) = 1 - \omp^2/\omega^2$.

In practice, an ansatz is made for $S_\xc$ in terms of a single-pole
plasmon description that is constrained by physical laws, including
charge conservation, cancellation of self-correlation, and time-reversal
symmetry. The plasmons that give rise to the nonlocal correlation in
vdW-DF, described in terms of the poles of the polarisation function
$S_\xc(iu)$, are parametrised by connecting the dielectric function to
the so-called internal functional \cite{Paper6-01,Dion,Lee10p081101},
\begin{equation}
E_\xc^{\rm int} = \int_0^{\infty} \, \frac{\diff u}{2\pi}
\hbox{Tr} \{S_\xc\} - E_{\self}\;,
\label{eq:internalxc}
\end{equation}
which describes short-range exchange and correlation effects and can be
described through a semilocal exchange-correlation functional. The
polarisation function $S_{\rm xc}(i u)$ bears strong resemblance to the
coupling-constant averaged density response function defined in
(\ref{eq:ACF}) and is equivalent in the homogeneous electron gas limit
(\ref{eq:formalEvalHEG}). It is closely related to the plasmon
propagator that describes the collective response of the electron gas.
By neglecting the longitudinal projection in (\ref{eq:Found2}) the
polarisation function becomes short ranged and no effects associated
with long-range van der Waals forces are included within this quantity.
It can thus be viewed as a short-range external field
response function.

The advantages of the nonempirical general-geometry functionals,
vdW-DF1, vdW-DF2, and vdW-DF-cx include an increasingly improved
consistency between the semilocal functional $E_\xc^0$ that describes
the energetic contributions to the total energy arising from short-range
exchange and correlation effects, and the internal functional
parametrising the plasmons~\cite{Dion,Thonhauser,Lee10p081101}. Such
consistency is related to charge conservation of the
exchange-correlation hole, and
improving this aspect of vdW-DF has been one driving force for successive
refinements of the functional~\cite{behy14,BeArCoLeLuScThHy14}.

%%%%%%%%%%%%%%%%%%%%%%%%%%%%%%%%%%%%%%%%%%%%%%%%%%%%%%%%%%%%%%%%%%%%%%%%
\subsection{Outline of derivation and approximations made }
\label{sec:outline}
%%%%%%%%%%%%%%%%%%%%%%%%%%%%%%%%%%%%%%%%%%%%%%%%%%%%%%%%%%%%%%%%%%%%%%%%

The first presentation of the general-geometry vdW-DF concept
\cite{Paper6-01} was a manuscript in Rydberg's PhD thesis
\cite{RydbergThesis}. It contained several of the key steps required to
obtain the general-geometry vdW-DF, a framework for formal improvements,
and an implementation that enabled efficient computations of the
nonlocal correlation energy in terms of a six-dimensional integral,
\begin{equation}
E_c^\nl = \frac{1}{2}
\int\, \diff^3r\, \diff^3r' \; n(\vv{r})\; \Phi_0(\vv{r},\vv{r'})\;
n(\vv{r'}) \;. \label{eq:enlstd}
\end{equation}
The formal results meant great progress, and together with references
\cite{Langreth05p599} and \cite{DionThesis}, which includes time-invariance, they form the basis for the final vdW-DF1
paper \cite{Dion,dionerratum}. Just like with the layered
functional, the full xc functional in vdW-DF1 is given by the sum of a semilocal part $E^0_\xc$ and the nonlocal correlation
(\ref{eq:vdWDFsplit}).

\begin{figure}
\includegraphics[width=\columnwidth]{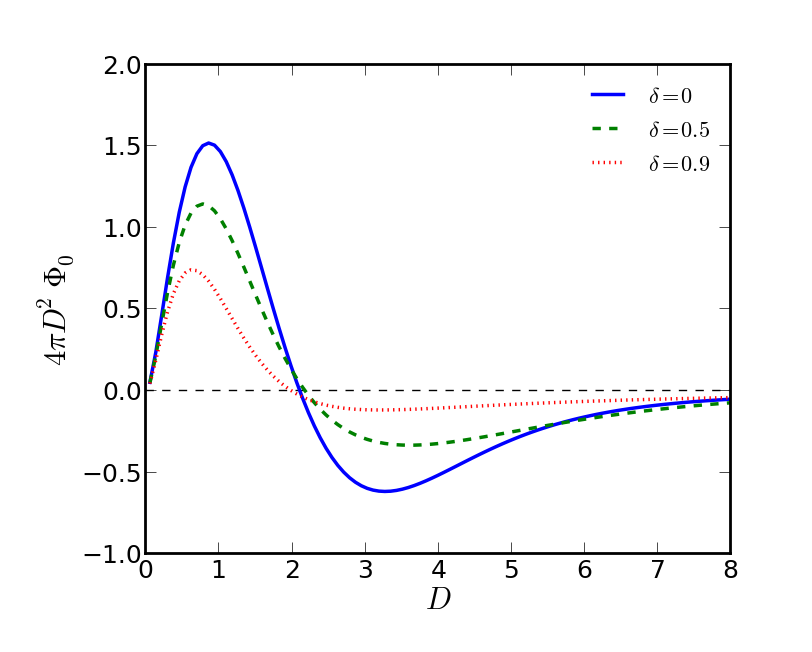}
\caption{\label{fig:kernel}The kernel (in a.u.) as a function of the
dimensionless separation $D$ for three values of the asymmetry parameter
$\delta$. Figure similar to that appearing in the erratum
\cite{dionerratum}.}
\end{figure}

The kernel (\ref{eq:enlstd}) describing the nonlocal correlation is
shown in figure~\ref{fig:kernel}, expressed in terms of the
dimensionless parameters $D$ and $\delta$. The parameter $D$ 
can be viewed 
as an effective scaled separation between two density regions, while
$\delta$ describes how different the response of these two density
regions are. It is interesting that the kernel in its universal form
keeps its shape in successively improved approximations. Thus, how $D$
and $\delta$ are scaled determines the nonlocal correlation energy, for
instance, when going from vdW-DF1 to vdW-DF2.

The nonlocal correlation energy functional $E_c^{\nl}$ (\ref{eq:enlstd})
is only one part of the xc-energy functional. There are also the local
correlation and the semilocal exchange. A
glance at vdW-DF-type functionals proposed by us~\cite{Dion, Lee10p081101,
cooper10p161104,behy14} and others~\cite{optX, vdWsolids, Hamada14}
shows that improvements beyond vdW-DF1 have been primarily concerned
with the exchange part $E_x$, which so far has been treated in GGA.

The original vdW-DF1 paper \cite{Dion} targeted the vdW bond, which motivated using the revPBE exchange
functional.  When vdW-DF1 is used for other bonds, such as mixes of vdW with covalent, ionic,
$d$-electrons, and hydrogen bonds, this can give results close to the
mark, such as for covalent bond lengths in carbon
systems~\cite{Ziambaras07p155425};  in other cases, the results are
further away, such as for bond lengths within coinage metals
\cite{vdWsolids,BeHy13}. Some have expressed disappointment, perhaps
unfairly, considering that  the vdW bond was the target.
However, the overall goal of the vdW-DF is to describe general
matter, and with successive refinements, as discussed in
section~\ref{sec:refinements}, it is approaching that goal. 

Assessing vdW methods by benchmarking against accurate quantum chemistry
results for a particular set of systems is common but not enough. For
instance, results for a set of molecular dimers exhibiting vdW and
hydrogen bonds might not be helpful for studies of transition metals
with $d$-shell electrons. This example emphasises the value of both a
broader benchmarking, including comparisons with extensive experimental
data sets \cite{lee11p193408}, and of designing functionals based on
general criteria, such as those given by fundamental physical laws. An
important example is the law of charge conservation, which is
essential in many-body theory \cite{BaymKadanoff} and DFT. A central concept is the xc hole: the depletion of the mean
density surrounding a given electron. A special case which is easy to
envision is the charge depletion left behind by an electron leaving a
metal surface. The failure of the gradient expansions in the density
functional proposed in the original Kohn-Sham formulation of
DFT~\cite{HoKo64,KoSh65}, which is inferior to the LDA,
can be blamed on the violation of the charge sum rule~\cite{GuJo80};
that is, its xc hole does not integrate to unity~\cite{GuLu76}. Likewise, the success
of LDA for the hydrogen atom, with its far from homogeneous electron
distribution~\cite{GuLu76}, is credited to compliance with the charge
sum rule. In the design of general-geometry vdW-DFs charge
conservation is important. In the early work, results for the almost
homogeneous electron gas were carefully selected from many-body theories
honouring conserving approximations.

The derivation outlined here is close to that of
the original one~\cite{Paper6-01}, but updated
by the insights that led to the final vdW-DF1 version~\cite{Dion}, as well as recent
analysis~\cite{hybesc14}. Details for the more familiar derivation
relying on the FPA-based approach~\cite{Dion}, which uses
$S=1-1/\epsilon$ as an expansion parameter, can be found
elsewhere~\cite{DionThesis,BerlandThesis}.  Since the exact expression for
$S_{\rm xc}$ is not known (\ref{eq:internalxc}), an ansatz is made in
the planewave representation $S_\xc(\bq,\bq')$ based on four
constraints: time-reversal symmetry; the $f$-sum rule, which is linked
to current conservation \cite{VV09comment}; cancellation of the
self-energy; and charge conservation.

Time-reversal symmetry is built into the
construction by defining $S_\xc$ as a symmetric function of a related
quantity $\tilde{S}$, as follows
\begin{align}
S_\xc(\bq,\bqm) & = \frac{1}{2}\left[\tilde{S}(\bq,\bqm)+\tilde{S}(-\bq,-\bqm)\right] \;.
\label{eq:revSform}
\end{align}

The second constraint, the {\it f}-sum rule, implies that in the
high-frequency $\omega$ limit, $S_{\rm xc} \rightarrow -4\pi/(m
\omega^2)n(\bq-\bq')$ \cite{Dion}. With the following form of $\tilde{S}$, this
constraint is ensured,
\begin{align}
\tilde{S}(\bq,\bqm) & = \int \diff^3r e^{-i(\bq-\bqm)\cdot \rr} \frac{\omega_\p^2(n(\rr))}
{[\omega+\omega_\bq(\rr)][-\omega+\omega_{\bqm}(\rr)]}\;.
\label{eq:plasmonSform}
\end{align}
Here, the classical plasmon frequency in the numerator can be viewed as
an amplitude in a spectral function with poles at $-\omega_\bq(\rr)$ and
$\omega_{\bqm}(\rr)$. The high-frequency limit of $S_{\rm xc}$
corresponds to the dielectric function taking on its familiar high
frequency limit, $\epsilon(\omega) = 1 - \omp^2/\omega^2$.

The third constraint is that the divergence in the self-energy term
(\ref{eq:internalxc}) must cancel out. This constraint requires that
the dispersion $\omega_q$ crosses over to a free-electron behaviour
$\omega_q\rightarrow q^2/2$ in the large $q$ limit---a natural limit, as
it reflects that fast moving electrons have no time to interact with
each other. To see how this cancellation occurs, we use the specific form of $S_{\rm xc}$ to write equation
(\ref{eq:internalxc}) in the form
\begin{equation}
E^{\rm int}_\xc = \int \diff^3r\; n(\rr)\;\varepsilon_{\xc}^{\rm int}(\rr) \;,
\end{equation}
with
\begin{equation}
\varepsilon_{\xc}^{\rm int}(\rr) = \int \diff^3 q\, \varepsilon_{\xc}^{\rm int}(\rr,\bq) = \pi\int \diff^3 q\, \left(\frac{1}{\omega_q(\rr)}-\frac{2}{q^2}\right) \;,\label{eq:epsInt}
\end{equation}
which makes the cancellation evident. The term $\varepsilon_{\xc}^{\rm
int}(\rr)$ is identified as the xc energy density of the internal
functional. This expression also provides a wavevector
decomposition \cite{LaPe77} of the internal functional for a given
plasmon dispersion $\omega_\bq$.

The fourth constraint is charge conservation. The polarisation function
$S_\xc$ remains bounded in the upper right quadrant of the
complex-frequency plane. As a result of the longitudinal projection of
$\epsilon$, the coupling-constant averaged response function is
given by $\chi = \nabla \epsilon \nabla G$. This property ensures that the
expression for the full xc energy (\ref{eq:Found2}) is charge
conserving~\cite{Dion,behy14,hybesc14}. The corresponding xc hole
integrates to unity~\cite{BeArCoLeLuScThHy14,hybesc14}. 
The intricacies of the charge conservation in vdW-DF is discussed further after the full functional has been laid out.

The specific form chosen for the plasmon frequency $\omega_q$ should be
simple yet capture the overall plasmon response. The
following form was chosen
\begin{equation}
\omega_q(\rr) = \frac{q^2}{2} \frac{1}{h(q/q_0)} \label{eq:omq}\;,
\end{equation}
with $h(y) = 1-\exp(-\gamma y^2)$ and $\gamma=4\pi/9$. The $q_0(\rr)$
parameter sets an inverse length scale and determines the plasmon
dispersion in a spatial region. With the added benefit of being simple,
this form cancels out the divergence in the self-energy term and ensures
a finite plasmon frequency in the $q\rightarrow 0$ limit.

The plasmon dispersion should reflect the overall response in a spatial
region. This property is established by identifying the internal
functional with a specific semilocal xc functional, which determines a
specific value of $q_0(\rr)$ at each $\rr$. For this step, it is
helpful that the xc-energy density of the internal functional
(\ref{eq:epsInt}) can be evaluated analytically. The result,
\begin{equation}
\varepsilon_{\xc}^{\rm int}(\rr) = \frac{3}{4 \pi} q_0(\rr)\;,
\end{equation}
resembles the exchange energy density of LDA: $\varepsilon^\LDA_x(n) = -
(3/4 \pi)\;k_{\rm F}(n)$. In fact, the homogeneous electron gas
expression (\ref{eq:formalEvalHEG}) given in the previous chapter
indicates that in the homogeneous limit, it is natural to set the
internal functional equal to the xc energy of the LDA. Thus, we can
identify $q_0^{\rm HEG}(n)= k_{\rm F}(n)\;
\varepsilon^\LDA_\xc(n)/\varepsilon^\LDA_x(n)$.

In extending the procedure for setting the inverse length scale
$q_0(\rr)$ to inhomogeneous systems, two specific choices are made in
vdW-DF1. First, the internal functional is identified as a
semilocal functional that combines local correlation effects in an LDA
description with GGA level exchange effects. 
This results in \cite{Dion}
\begin{align}
 q_0(\rr) & = k_{\rm F}(\rr) \frac{ \varepsilon^\LDA_x(\rr) F^{\rm int}_x(s)+ \varepsilon^\LDA_c(\rr) }{ \varepsilon^\LDA_x(\rr)}\;.
\end{align}
Here, $F^{\rm int}_x(s)$ is the exchange enhancement factor and depends
on the reduced gradient $s(\rr)= |\nabla n |/(2k_F(\rr)n(\rr))$.
Second, the Langreth-Vosko \cite{LaVo90} form is used for the
exchange enhancement factor $F^{\rm int}_x$. This enhancement form is a
simple quadratic function of $s$~\cite{LaVo87,Dion}, i.e.\
\begin{equation}
F_x^{\rm int}(s) = 1-\left(\frac{Z_{ab}}{9}\right) s^2\;.
\label{eq:Fex}
\end{equation}
Here, $Z_{ab}=-0.8491$. We postpone a discussion of why these choices
were made until after we have laid out some more details of the derivation.

The close resemblance between $q_0$ and $k_{\rm F}$ is a consequence of
a carefully chosen value for $\gamma$. This does not make $\gamma$ an
adjustable parameter, rather it is an arbitrary parameter in the true
sense of the word. A different value of $\gamma$ would result in a
different scaling of $q_0$, but the plasmon dispersion (\ref{eq:omq})
and thus the nonlocal correlation would end up the same.

Now that an ansatz for the expansion parameter $S_{\rm xc}$ has been
specified, we turn to expanding the expression for the xc energy
(\ref{eq:Found2}). Since $S_\xc$ can be formally represented by a
matrix, we can use the relation $\eps = \exp[S_\xc]$ in the expansion of
(\ref{eq:Found2}). The first order term is the internal functional
(\ref{eq:internalxc}). Since (\ref{eq:internalxc}) is intended to include all xc effects
except the nonlocal correlations, the second order expression is
identified as the vdW-DF approximation for the nonlocal correlation,
\begin{equation}
E_c^{\nl} =
\int_{0}^{\infty}\frac{\diff u}{4\pi}
\Tr\{S_\xc^2-(\nabla S_\xc \cdot \nabla G)^2\}\;.
\label{eq:secondWwithS}
\end{equation}
To obtain this result, one can use partial integration and $\nabla^2 G=
1$. In a planewave representation the nonlocal correlation can be
written as
\begin{align}
E_c^{\nl} =&\int_{0}^{\infty}\frac{\diff u}{4\pi}\int \frac{\diff^3 q\,
   \diff^3 q' }{(2\pi)^3 (2\pi)^3 } \nonumber \\ &
\times\left[ 1 - (\hat{\bq}\cdot \hat{\bqm})^2 \right] S_\xc(\bq,\bqm)
   S_\xc(\bqm,\bq)\;.
\label{eq:secondWwithSb}
\end{align}
By design, this expression vanishes in the homogeneous limit. 
Once $S_{\rm xc}$ becomes diagonal, only terms with $\bqm=\bq$ contribute, for which the term within the bracket vanishes. The expansion (\ref{eq:secondWwithSb}) implies that the popular vdW-DF versions, unlike RPA and
unlike the vdW-DF framework \cite{hybesc14}, cannot account for
many-body dispersion effects \cite{Barash88,Sernelius98,qmc_graphite,ts-mbd,fullerenewisdom} across all length scales. The many-body dispersion effects often involve the screening impact on the lower-energy plasmons and are primarily relevant in the
asymptotic-interaction regime \cite{qmc_graphite,DobsonMB14,hybesc14}; however,
at binding separations there are many plasmons that contribute to the
nonlocal-correlation attraction \cite{BeHy13}. As such, building upon the
underlying account of the GGA response, the popular (expanded)
vdW-DF versions do indeed reflect many-body dispersion effects at binding
separations \cite{hybesc14}.

The nonlocal correlation in the familiar form (\ref{eq:enlstd}), with
two spatial integrals over a kernel $\Phi_0(\rr,\rr')$ that connects two
density regions $n(\rr)$ and $n(\rr')$, is obtained after integrating
over the imaginary frequency and both planewave coordinates $\bq$ and
$\bq'$. Of these integrals the one over $u$ and the angular parts of
$\bq$ and $\bq'$ can be performed analytically. This leaves us with two
one-dimensional integrals over $q$ and $q'$, which are performed
numerically. The resulting kernel can be tabulated once and for all in
terms of $d=q_0(\rr) |\rr-\rr'|$ and $d'=q_0(\rr') |\rr-\rr'|$ or
related quantities such as $D=(d+d')/2$ and $\delta = (d-d')/2D$. The
two spatial integrals that remain come from $S_\xc$
(\ref{eq:plasmonSform}). That the kernel depends on merely two
dimensionless parameters follows from the judiciously chosen
ansatz for $S_\xc$ and the fact that only a single function sets
the effective inverse length scale $q_0(\rr)$ that describes the plasmon
dispersion.

To discuss choices in the design of vdW-DF1 and its properties, 
it is convenient to express 
the xc functional of vdW-DF as
\begin{equation} 
E_\xc^{\text{vdW-DF}}[n] =E^0_{\xc}[n] + E^\nl_c[n]\;,
 \label{eq:full}
\end{equation}
that is, with explicit nonlocal correlation and a term with local correlation and exchange, $E^0_{\xc}$. 

In vdW-DF1, the $E^0_\xc$ is given by the sum of LDA correlation and
the revPBE \cite{ZhYa98} variant of GGA exchange. Conceivably, one could consider including semilocal correlation terms within $E^0_\xc$;
however, the nonlocal correlation term also has significant semilocal
contributions \cite{Paper6-01} and double counting is undesirable. 
A strict derivation in terms of the $S_\xc$
expansion starting from equation~(\ref{eq:Found2}) implies that
$E^0_\xc[n]$ equals the internal functional $E^{\rm int}$.
However, relying on the internal functional to describe the semilocal xc
energy implies that one trusts the simple plasmon model in $S_{\rm xc}$
to accurately describe not only the overall plasmon response, but also
the exchange energy, the biggest part of the xc energy. Given
its simple form, it is natural that correction terms could arise. For
one, the Langreth-Vosko expression (\ref{eq:Fex}) that was chosen for the
internal functional to parametrise the plasmon dispersion is a poor
description in the regime of large density variations (high-$s$).
However with correction terms, the conserving expression (\ref{eq:Found2}), involving a
longitudinal projection of $\epsilon$, would no longer describe the full xc
energy. None of the vdW-DFs have an exact agreement between 
$E^{\rm int}_\xc$ and $E^0_\xc$ which would make them free of implicit correction terms,
but there is a successively better agreement
when going from vdW-DF1 to vdW-DF2 and to the newest development vdW-DF-cx
\cite{behy14}. The latest version is guided by the aim of using a
consistent exchange choice; thus achieving this goal (\ref{eq:formalEval}) for the most relevant density regions.

Charge conservation nevertheless remains an essential design principle for
vdW-DF. The difference between $E^{\rm int}_\xc$ and $E^0_\xc$ only
affects the short range part of the functional. Further, charge
conservation is also imposed by relying on conserving approximations for
$E^0_\xc$ and $E^{\rm int}_\xc$ separately. For $E^0_\xc$ the
construction of the explicit form of the exchange functional can be
traced to the construction of a numerical GGA that is designed by
imposing conservation on the xc hole that arises in a gradient expansion
\cite{PeWa86,PeBuWa96,PeBuEr96,MuLeLa09}. This is particularly true for
the exchange functional used in vdW-DF2, the PW86r functional \cite{MuLeLa09}, which is
fitted directly to such a numerical GGA construction.

For the internal functional, charge conservation \cite{hybesc14} follows
from the wavevector form~(\ref{eq:epsInt}). It provides an expression
for the angle-averaged xc hole $n_{\xc}^{\rm int}(\rr,\bq) = q^2
\eps^{\rm int}(\rr,\bq)$ around a given $\rr$, as expressed in momentum
space. The relation follows since the spherically averaged xc hole
$\bar{n}_{\xc}^{\rm int}(\rr,q)$ \cite{GuLu76,ADAWDA} also defines a
natural wavevector decomposition \cite{LaPe77,LaPe80,LaVo87} for the
energy per particle. Since $\omega_\bq$ is finite in the $\bq
\rightarrow 0$ limit, $n_{\xc}^{\rm int}(\rr,0) = -1$. Thus, the
integral over the xc hole in spatial coordinates gives $-1$, i.e.\ the
depletion of a single electron.

Next, we discuss the choices made for the internal functional
$E_{\xc}^{\rm int}$. Langreth, one of the vdW-DF architects, also had a
central role in the developments that led to GGA with a diagram-based
foundation \cite{LaPe77,LaPe80,LaVo87,LaVo90,LaMePRL1981}. Some of this
development is reviewed in section \ref{sec:ALV}. 
The diagram-based
foundation means that the starting point is the terms that arise in a
perturbative many-body expansion of the almost homogeneous electron gas.
Such terms can be neatly visualised by Feynman diagrams, as in
figure~\ref{fig:LV}. 
vdW-DF1 uses a Langreth-Vosko (LV) \cite{LaVo90,Dion,Thonhauser} form for the exchange
enhancement factor (\ref{eq:Fex}).  The LV value of $Z_{\rm
ab}$ that is used in vdW-DF1 represents an updated result compared to
the value $Z_x$ that appeared in the earlier discussion of the
gradient-expanded exchange in section~\ref{sec:ALV}.

The LV form attempts to capture screened \cite{Rasolt, LaPe80, LaVo87,
LaVo90} rather than pure exchange effects \cite{KlLe88,LaVo90}. It
involves an evaluation of a $Z_x(q=0)$ response contribution.
Figure~\ref{fig:LV} shows the corresponding
diagrams. The \lq a\rq\ and \lq b\rq\ of $Z_{\rm ab}$ indicate that
certain classes of diagrams, indicated in the figure labels, contribute
\cite{Rasolt,LaVo90}. The third class of diagrams \lq c\rq\ can be
identified as arising from nonlocal correlations and is not part of the
screened exchange. The choice of not including these contributions in
the internal functional motivates why only local correlations are
included in the semilocal part of the xc energy of vdW-DF.

Standard GGA functionals have an enhancement factor $F_\x$ (\ref{eq:Fex}) that softens at larger $s$ values compared to the aggressive quadratic enhancement factor within vdW-DF, see figure~\ref{fig:Fs_x} for some examples. 
Such a quadratic form is chosen to make the vdW-DF construction as simple as feasible (the quadratic form is described by a single parameter), but also to ensure that very low density regions barely contribute to the nonlocal correlation energy.  
The aggressive enhancement factor acts as a cutoff because 
at low densities the
scaled gradient $s$ diverges, causing $q_0$ to diverge even more strongly. In turn, the effective
dimensionless separation $D$ diverges, thereby overwhelming the nonlocal correlation energy.
This particular kind of cutoff mechanism  observes the $f$-sum rule \cite{DionThesis,behy14}, one of the constraints of vdW-DF.

Finally, we discuss the handling of screening in vdW-DF1. A common
misconception is that because vdW-DF1 connects only two different density
regions it lacks screening. However, $S_{\rm xc}$
is a semilocal approximation to the external field response. It has a screening account based on the density in its vicinity, much like in the ALL functional \cite{AnLaLu96}, that
provides a good account of asymptotic vdW interactions.  vdW-DF even includes gradient
corrections and therefore reflects broader density variations.
Related to this discussion is the fact that the evaluation of equation
(\ref{eq:secondWwithS}) and hence (\ref{eq:enlstd}) can be viewed as the
electrodynamical interaction between two semilocal xc holes
$n_{\xc}^{\rm int}$. This is in line with the Rapcewicz and Ashcroft
interpretation \cite{RA,AnLaLu96,hybesc14} of the fluctuation diagram
that the second-order expansion of $E_c^{\nl}$ represents. We also point
out that several applications of vdW-DF show that vdW-DF to a large extent
captures screening effects such as image-plane and collective effects at typical binding separations
\cite{Kleis08p205422, Berland10p134705, Berland11p1800, lee11p193408,
lee12p424213, BeHy13}.

%%%%%%%%%%%%%%%%%%%%%%%%%%%%%%%%%%%%%%%%%%%%%%%%%%%%%%%%%%%%%%%%%%%%%%%%

%%%%%%%%%%%%%%%%%%%%%%%%%%%%%%%%%%%%%%%%%%%%%%%%%%%%%%%%%%%%%%%%%%%%%%%%
\subsection{Self-consistency of vdW interactions}
%%%%%%%%%%%%%%%%%%%%%%%%%%%%%%%%%%%%%%%%%%%%%%%%%%%%%%%%%%%%%%%%%%%%%%%%

In the Kohn-Sham
scheme of DFT~\cite{KoSh65}, the interactions among electrons are accounted
for by the sum of the mean-field electrostatic contributions and the 
 xc potential
$V_\xc[n](\vv{r})=\delta E_\xc[n]/\delta n(\bf{r})$.
This potential acts locally on each
electron, and thus propagates these interactions only through its
dependence on the density.

So far, the development of vdW-DF has focused on  the xc energy functional $E_\xc[n]$ itself. The 
first implementations evaluated this functional in a post-processing
manner: In the first step, the system in question is brought to full
self-consistency with a standard functional---often PBE~\cite{PeBuEr96}
or revPBE~\cite{ZhYa98}---and the corresponding density is stored. In
the second step, this static density is then used to evaluate $E_\xc[n]$
in a ``one-shot'' calculation, performed by an
independent code or inside the code that performed the first
step. Clearly, this scheme is approximate as the vdW-DF is evaluated with
a density that corresponds to a different functional and it was not until recently that
 an expression for the error made with such a non self-consistent evaluation was derived \cite{BeLoScHy13}.

Self-consistency is important as it lays the foundation for the
calculation of forces and the stress tensor, both essential for an
efficient structural optimisation~\cite{Thonhauser,Sabatini12p424209}.
It is thus crucial for efficient calculations of structure, energies,
phase transitions, and elastic responses for bulk, layered, and molecular
materials~\cite{BeArCoLeLuScThHy14}.

To make vdW-DF self-consistent, an  expression
for the nonlocal correlation potential $V^{\text{nl}}_c(\rr)$ is required~\cite{Thonhauser}. The vdW-DF xc energy consists of semilocal parts and a
 nonlocal part~(\ref{eq:vdWDFsplit}). The
functional derivative of the semilocal part is well-established and we
only focus on the derivative of the nonlocal energy functional:
\begin{align}
V^{\text{nl}}_{\rm c}(\rt) &= \frac{\delta E^{\text{nl}}_c[n]}
{\delta n(\rt)}\;.\label{twopointv}
\end{align}
This functional derivative  is straightforward but tedious to evaluate \cite{Thonhauser} 
and results in
\begin{equation}\label{equ:vfinal}
V^{\text{nl}}_c(\rr) = \int \diff^3 r' \: n(\rr')\sum_{i=0}^3
\alpha_i(\rr,\rr') \: \Phi_i(\rr,\rr')\;,
\end{equation}
where the functions $\alpha_i(\rr,\rr')$ and $\Phi_i(\rr,\rr')$ are
given by:
\begin{subequations}\label{equ:alpha}
\begin{eqnarray}
\alpha_0 &=& 1\\
\alpha_1 &=&\frac{1}{q_0(\rr)}\Big[
   \frac{Z_{ab}}{9}\nabla\cdot\redgrad(\rr)+
   \frac{7}{3}\frac{Z_{ab}}{9}s^2(\rr)\kF(\rr)
   \nonumber\label{equ:alpha1}\\
&& \qquad\quad -\frac{4\pi}{3}n(\rr)
   \varepsilon_{\text{xc}}^{\text{LDA}}(\rr)\Big]\\
\alpha_2 &=& \frac{Z_{ab}}{9} \frac{\redgrad(\rr)\cdot\nabla q_0(\rr)}
   {q_0(\rr)^2}\label{equ:alpha2}\\
\alpha_3 &=& \frac{Z_{ab}}{9}\hat{\RR}_{\rr\rr'}\cdot\redgrad(\rr)
   \label{equ:alpha3}\label{equ:alpha4}
\end{eqnarray}
\end{subequations}
and
\begin{subequations}\label{equ:phi}
\begin{eqnarray}
\Phi_1 &=& d\phi_d(d,d')\label{equ:phi1}\\
\Phi_2 &=& d^2\phi_{dd}(d,d')\label{equ:phi2}\\
\Phi_3 &=& \phi_d(d,d') + d\phi_{dd}(d,d')+d'\phi_{dd'}(d,d')\label{equ:phi3}\;.
\end{eqnarray}
\end{subequations}
Here, $\hat{\RR}_{\rr\rr'}$ is a unit vector in the direction from $\rr$ to
$\rr'$ and subscripts of $d$ and $d'$ denote the corresponding partial derivatives.
The additional three kernel functions $\Phi_1(d,d')$ through
$\Phi_3(d,d')$ are the analogues to the single kernel $\Phi_0(d,d')$
used for $E_c^\text{nl}$. As pointed out earlier, $d=q_0(\rr) |\rr-\rr'|$
and $d'=q_0(\rr') |\rr-\rr'|$.

\begin{figure}
\includegraphics[width=0.9\columnwidth]{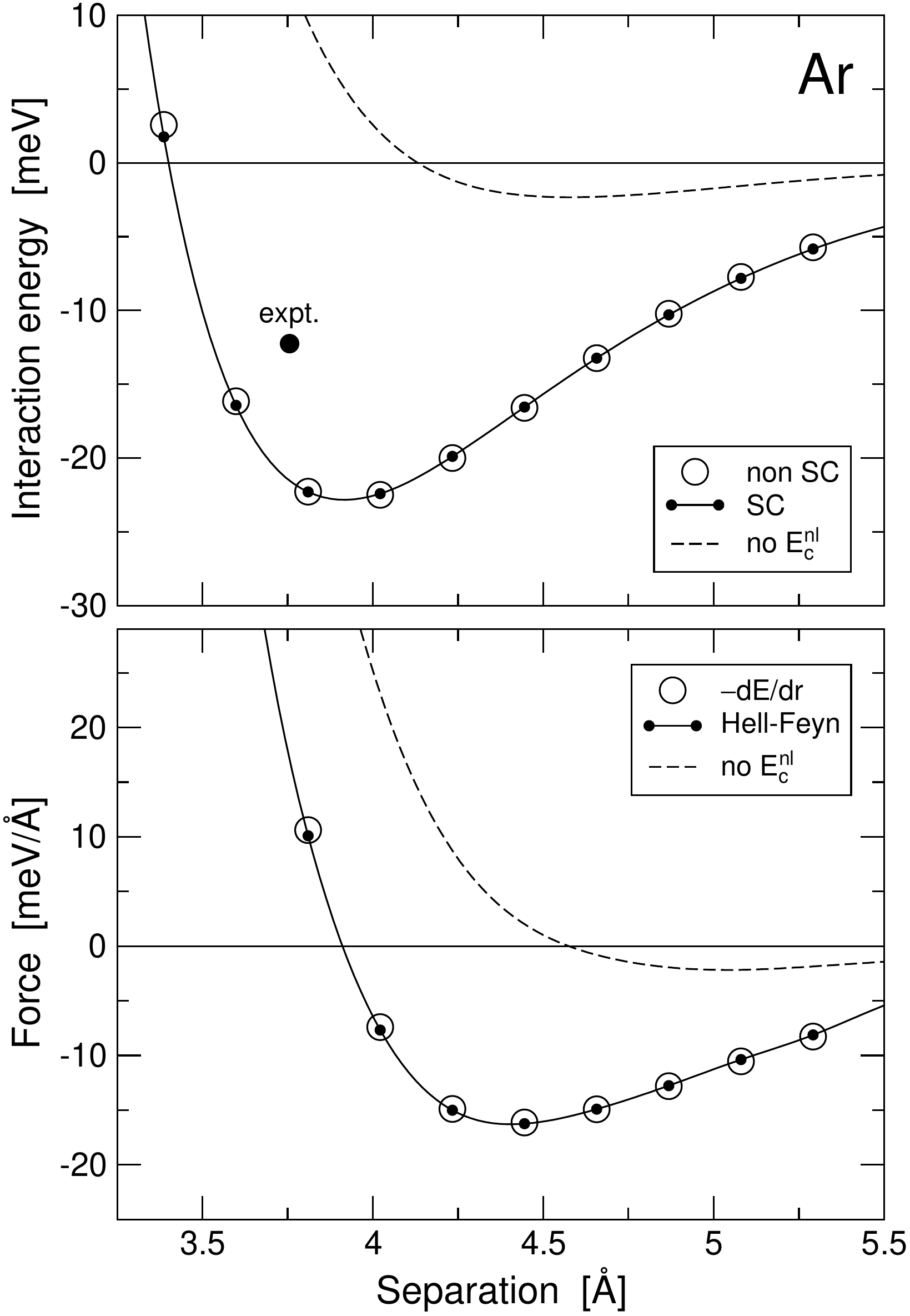}
\caption{\label{fig:Ar}(top) Interaction energy of the Ar dimer as a
function of separation for the self-consistent and non self-consistent
approach. In addition, we show results where $\Enl$ has been neglected.
(bottom) Forces calculated as the derivative of the energy ($-dE/dr$)
and the Hellmann-Feynman forces. Reprinted with permission from
\cite{Thonhauser}, \copyright\ 2007 American Physical Society.}
\end{figure}

Once it was developed, the self-consistent formulation was applied to a
number of simple test cases including the Ar dimer.
Figure~\ref{fig:Ar} shows its interaction energy as a function of separation. The differences between
the non self-consistent  and the fully self-consistent
results are minimal, at least at larger separations.  The bottom
panel shows that the  forces calculated self-consistently for vdW-DF through the Hellmann-Feynman theorem agree well with the numerical
derivative of the energy.

\begin{figure}
\includegraphics[width=0.8\columnwidth]{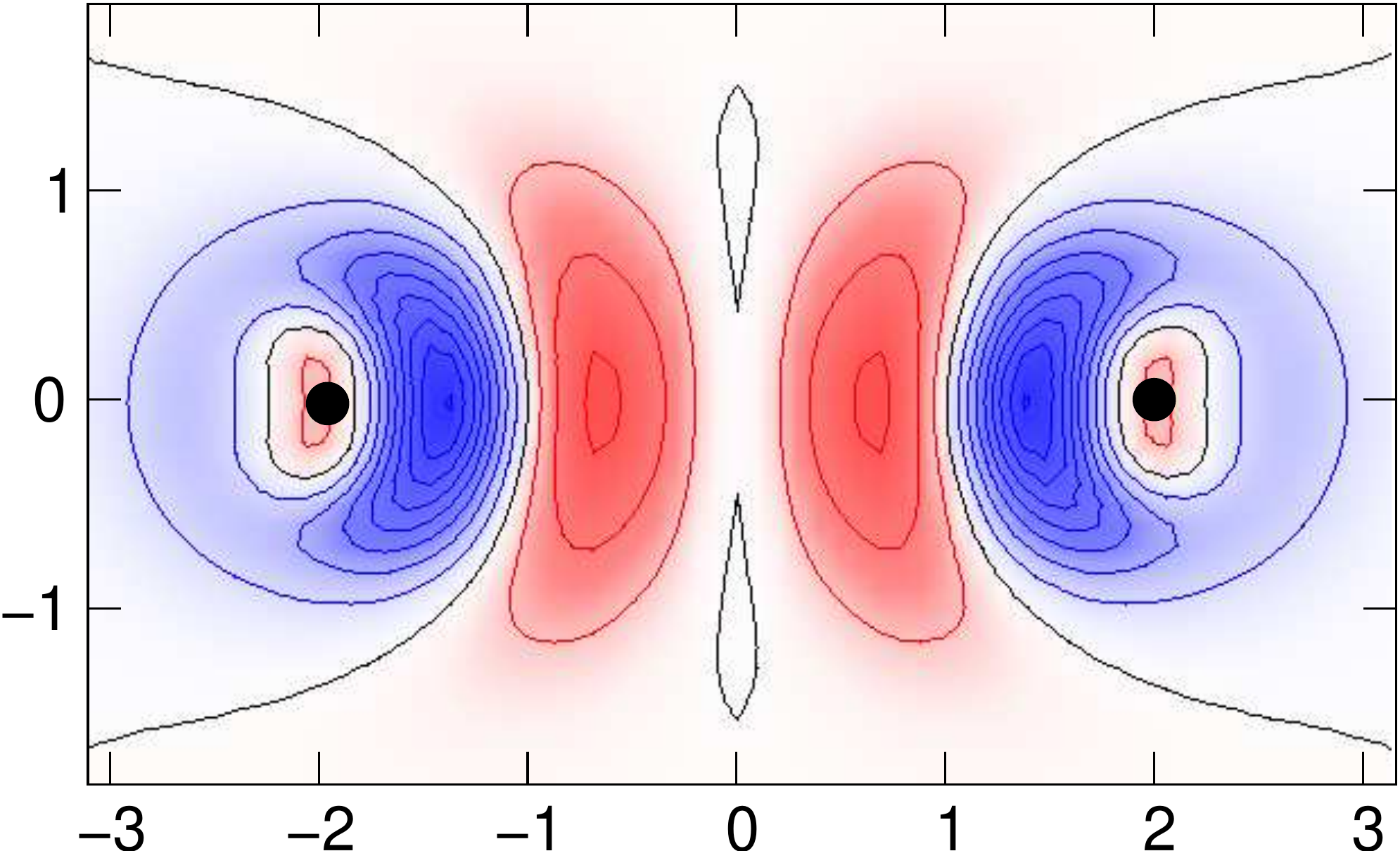}
\caption{\label{fig:dens}Bonding charge of the Ar dimer. Shown is the
difference in induced electron density. The scale is in \AA\ and the
black dots mark the position of the nuclei. The zero level is marked by
the black contour. Red areas represent areas of electron density gain
when the nonlocal part is included; conversely, blue areas indicate loss
of electron density. Increments between contour lines are
5$\times$10$^{-5}$ electrons/\AA$^3$. Reprinted with permission from
\cite{Thonhauser}, \copyright\ 2007 American Physical Society.}
\end{figure}

The self-consistent method can also be used to show how the density
evolves under the influence of vdW interactions. 
The density change in the case of the Ar dimer is tiny (figure~\ref{fig:dens}), but
nonetheless responsible for the binding of the dimer by pulling charge
in-between the nuclei. This captivating and illuminating picture quantitatively
displays the nature of the vdW bond.

%%%%%%%%%%%%%%%%%%%%%%%%%%%%%%%%%%%%%%%%%%%%%%%%%%%%%%%%%%%%%%%%%%%%%%%%
\subsection{Efficient FFT implementation, algorithm of
Rom\'{a}n-P\'{e}rez and Soler}
%%%%%%%%%%%%%%%%%%%%%%%%%%%%%%%%%%%%%%%%%%%%%%%%%%%%%%%%%%%%%%%%%%%%%%%%

The evaluation of the nonlocal contribution to the exchange-correlation
energy $E_c^\text{nl}[n]$ requires solving a six-dimensional spatial
integral; the evaluation of the corresponding potential
$V^{\text{nl}}_c(\rr)$ requires a three-dimensional integral for every
point $\rr$---see (\ref{eq:enlstd}) and (\ref{equ:vfinal}). A straightforward numerical evaluation of those integrals
required for vdW-DF is much more time-consuming than for simple LDA or GGA
functionals. As the vdW-DF kernel goes to zero for large separations,
the first self-consistent implementations of vdW-DF used a spatial 
cutoff to limit the computational effort of evaluating those integrals.
While such a spatial cut-off provides some form of ``linear-scaling,''
the prefactor still makes most simulations computationally very
expensive.

This bottle-neck was overcome by Rom\'{a}n-P\'{e}rez and Soler, who
rewrote $E_c^\text{nl}[n]$ as an integral convolution using splines
\cite{RoSo09}. In this way, the dependence of the kernel on $\rr$ and
$\rr'$ can be approximated as
\begin{eqnarray}
\lefteqn{\Phi_0(\rr,\rr')= \Phi_0\big(q_0(\rr), q_0(\rr'),|\rr-\rr'|\big)
   \approx\nonumber}\\
&& \qquad\sum_{\alpha\beta} \Phi_0\big(q_\alpha, q_\beta, |\rr-\rr'|\big)\;
   p_\alpha\big(q_0(\rr)\big)\;p_\beta\big(q_0(\rr')\big)\,,
\end{eqnarray}
where $q_\alpha$ are fixed values and $p_\alpha$ are cubic splines. It
follows that the original nonlocal functional can be written as
\begin{eqnarray}
\Enl &=& \frac{1}{2}\sum_{\alpha\beta}\int\dr d^3r'\,
   \theta_\alpha(\rr)\;\phi_{\alpha\beta}(|\rr-\rr'|)\;
   \theta_\beta(\rr')\nonumber\\
   &=& \frac{1}{2}\sum_{\alpha\beta}\int\dk\;
       \theta_\alpha^*(\rk)\;\phi_{\alpha\beta}(k)\;\theta_\beta(\rk)\;,
       \label{equ:finalfft}
\end{eqnarray}
where $\theta_\alpha(\rr)=n(\rr)\,p_\alpha\big(q_0(\rr)\big)$ and
$\theta_\alpha(\rk)$ is its Fourier transform. In the same sense,
$\phi_{\alpha\beta}(k)$ is the Fourier transform of
$\phi_{\alpha\beta}(r)\equiv\phi(q_\alpha,q_\beta, |\rr-\rr'|)$.

At first sight, the benefit of using equation (\ref{equ:finalfft}) might
seem minor---a six-dimensional integral has been traded for a
three-dimensional one that contains Fourier transformed quantities.
However, many of the implementations of vdW-DF have been in standard
planewave DFT codes. As such, fast Fourier transforms are readily
available and highly optimised. The overall computational speedup is
dramatic and today vdW-DF calculations on large systems take barely any
longer than standard GGA calculations.

%%%%%%%%%%%%%%%%%%%%%%%%%%%%%%%%%%%%%%%%%%%%%%%%%%%%%%%%%%%%%%%%%%%%%%%%
\subsection{Refinements of the first general-geometry vdW-DF version}
\label{sec:refinements}
%%%%%%%%%%%%%%%%%%%%%%%%%%%%%%%%%%%%%%%%%%%%%%%%%%%%%%%%%%%%%%%%%%%%%%%%

Numerous applications have demonstrated that the first general-geometry
vdW-DF version, vdW-DF1, is both robust and versatile. We we will return to in the next chapter. This success indicates just how
potent vdW-DF1's underlying constraint-based construction is. 
It is only natural then that when designing new functionals, the original vdW-DF1 construction is often the starting point. 
There are several different aspects of vdW-DF1 one might want to improve, such
as the typical overestimation of binding
separations~\cite{cooper10p161104}; the account of
binding energies, in particular for small molecules~\cite{Lee10p081101} and for hydrogen bonds
strengths like those of water~\cite{optX}; the description of
covalently-bounded solids~\cite{vdWsolids}; and consistency between exchange and correlation functionals~\cite{behy14}. Considering that 
vdW-DF is rooted in the electron gas tradition,
we believe that it should be refined  by
carefully assessing constraints and first-principle input parameters.
However, researchers outside the Rutgers-Chalmers program have 
attempted to improve the functionals by instead optimizing 
the functional based on selected reference data sets, that is, making semi-empirical variants of vdW-DF~\cite{optX,vdW-DF-09}.

The vdW-DF1 overestimation of binding separations was noted from
the onset~\cite{Dion} and attributed to the choice of revPBE as exchange
functional. Despite good reasons for starting out with
revPBE, as discussed in chapter~\ref{sec:layered}, there is no intrinsic
reason why this specific exchange functional should be the only appropriate one for vdW-DF1. Several attempts
to refine vdW-DF1 have therefore kept the correlation part fixed
and focused on the exchange. For transparency, we encourage the use of a standard nomenclature vdW-DF-$E_x$; where $E_x$ refers to the exchange functional. This has the added benefit of allowing one to specify different nonlocal correlation functions, e.g. vdW-DF vs. vdW-DF2; while clearly identifying the form of exchange being employed.

\begin{figure}
\includegraphics[width=\columnwidth]{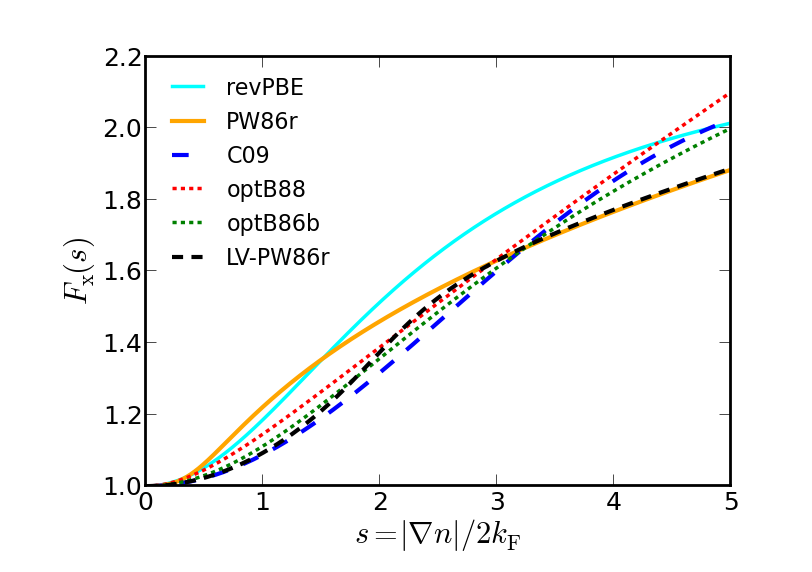}
\caption{\label{fig:Fs_x}Enhancement factors $F_x(s)$ of exchange
functionals suggested for vdW-DF1 and vdW-DF2 correlation.}
\end{figure}

Figure~\ref{fig:Fs_x} shows several different exchange enhancement
factors $F_x(s)$ suggested as partners for the vdW-DF1 and vdW-DF2
correlation. Here, $s=|\nabla n|/2k_F n$ is the reduced gradient and the
exchange energy is
\begin{equation} E_{\rm x} = \int \diff^3 r \;
n(\rr)\; \eps_{\rm x}^{\rm LDA}(n) \;F_x(s)\;.
\end{equation}

The exchange functional C09~\cite{cooper10p161104}, suggested as exchange partner for the vdW-DF1 correlation, is designed to counteract the
overestimation of binding separations while also avoiding
spurious exchange effects. This was achieved by using an enhancement
factor $F_x(s)$ that interpolated between the gradient expansion of the
slowly-varying electron gas, as used in PW86~\cite{PeWa86}, at small
reduced gradients~$s$ and the revPBE form at large $s$.
This large-$s$
form was also chosen to retain the good binding
energies of vdW-DF. This functional (vdW-DF-C09) demonstrated that the issue of overestimation of separation distances in vdW-DF could be solved by assessing the constraints of the exchange functional.

A different approach was taken by Klime\v{s} and
co-workers~\cite{optX} to design an exchange functional for vdW-DF1 correlation. They tuned the parameters of a set of familiar exchange
functionals to optimise the binding energies of the S22 set of dimers.
This is an example of what we call reference-system optimisation, as distinguished from constraint-based functionals. In practice, the refitting reduces the overestimation of
separations. Over time, the optB88 variant (vdW-DF-optB88) has revealed itself to provide an accurate account of many kinds of systems and thus has become widely used.  
In some
codes, care must be taken when implementing this functional because its
aggressive large-$s$ behaviour makes it noise sensitive~\cite{Londero11p1805}.

Murray and coworkers~\cite{MuLeLa09} also made an important exchange development by 
analysing what would be the most
suited exchange functional for sparse matter at the GGA level. They found that the PW86 had the best
agreement with Hartree-Fock
interaction curves between many molecules, in particular beyond binding
separations. This was attributed in part to the $s^{2/5}$ form of the
enhancement factor $F_x(s)$ at large $s$. Others have reached similar
conclusions~\cite{Kannemann}. This exponent form arises in the
numerical GGA construction~\cite{PeWa86,PeBuWa96}, which is built around
conservation of the exchange hole. Finally, PW86 was also refitted to
update the small-$s$ behaviour and to make the large-$s$ evaluation
of the numerical GGA construction exact (PW86r).

For some covalently-bound solids such as heavy transition metals, 
vdW-DF1's account can be inferior to that of standard GGAs.
Motivated by this shortcoming, Klime\v{s} and co-workers~\cite{vdWsolids} designed 
an exchange functional for vdW-DF1 called optB86b.  This functional is based on 
a \lq minimalistic\rq\ one-parameter
refitting of B86b~\cite{Becke1986p7184} using a small-$s$ form corresponding to the gradient
expansion of the slowly-varying electron gas. 
This makes the small-$s$
form similar to that of C09, yet the B86b form ensures that it crosses
over to the $s^{2/5}$ form at large $s$, though with a quite different prefactor than PW86r or B86b. 
This functional (vdW-DF-optB86) performs well for many other kinds of systems and together with vdW-DF-optB88 it has played an important role in showing that vdW-DF can handle systems characterized by weak chemisorption.

Most recently, a formally appropriate exchange functional to pair with vdW-DF1 was derived \cite{behy14}. 
This exchange functional is in part motivated by the fact that deviations between the outer and internal 
exchange form give rise to correction terms unaccounted for in the original derivation. It is
also motivated by the robustness of vdW-DF1 for van der Waals bonded systems. This robustness 
is linked to the success of the plasmon description underpinning the non-local part of the functional. 
The new exchange functional is therefore designed to resemble the internal functional of vdW-DF1 in 
as large of an $s$ regime as feasible, while smoothly crossing over to the PW86r form at larger $s$ (LV-PW86r). 
This exchange and correlation combination is referred to as vdW-DF-cx~\cite{behy14,BeArCoLeLuScThHy14}, 
where cx stands for \emph{consistent exchange};  In essence, the new exchange functional is quite 
similar to C09 and optB86b and thus partially validates their usage as well.  Tests so far indicate 
that vdW-DF-cx has  excellent performance, further supporting the quality of the plasmon model 
underpinning vdW-DF1.

%With a vdW-DF-cx, the poorer results for small molecules 
%For small molecules vdW-DF1 and related variants often tend to significantly overestimate binding energies. 

Separate from the development of exchange functionals for vdW-DF1, a 
second version of vdW-DF for general geometries called vdW-DF2 was also developed~\cite{Lee10p081101} which updates both the exchange and correlation
functional. For the nonlocal
correlation, it was recognised that the small-$s$ exchange parameter
$\beta$ of the B88~\cite{Be88} functional would provide a more
appropriate parametrisation of the plasmon response of molecules than
the slowly-varying electron-gas result underlying vdW-DF1. 
Indeed, this
parameter is used in many successful functionals and can be derived from
first principles using the large-N asymptote of neutral
atoms~\cite{Schwinger}.
For the exchange functional, the well-founded choice of PW86r~\cite{PeWa86} was made.  
Migrating
from vdW-DF1 to vdW-DF2 simply entails setting $Z_{\rm ab} = −1.887$ in
the internal functional (\ref{eq:Fex}). The exchange and correlation of
vdW-DF2 are reasonably consistent with each other because in the
significant $s$ regime beyond 0.1, the enhancement factor of PW86r
agrees well with that of the internal exchange of vdW-DF2.
The vdW-DF2 greatly improves both binding energies and separations  for systems of small molecules, but has also been criticized for poorly describing the asymptotic interaction between molecules. 
However, in a fairly wide region beyond binding separation, vdW-DF2 does in fact predict interaction curves in good agreement with coupled cluster results for the S22 set of dimers~\cite{Lee10p081101}. 

While vdW-DF2 greatly improves the description of small molecules compared to vdW-DF1, 
 issues with bulk matter and weakly chemisorbed systems remain. An alternative exchange functional for vdW-DF2 was very recently developed by Hamada~\cite{Hamada14}. 
Drawing on earlier experience with testing of the ad-hoc combination of vdW-DF2 correlation and C09 exchange, which works well for some systems~\cite{Hamada2010p153412}, 
he reparameterized the enhancement factor of the B86b functional to better describe the small-$s$ form of the slowly varying electron gas and approximately retain the large-$s$ form of B86b~\cite{Becke1986p7184}, a functional designed for highly inhomogeneous systems. 
Initial benchmarking demonstrates that the combination of B86R and vdW-DF2, named rev-vdW-DF2 or vdW-DF2-B86R, results in good performance for small molecules as well as an improved description of bulk and weakly chemisorbed systems. With such promise, this functional deserves further benchmarking. 

An almost opposite approach to how vdW-DF originally was developed is used in the ``BEEF--vdW model compromise''~\cite{vdwBEEF}. This is a reference-system optimised method taken to its extreme: The authors develop a methodology for semiempirical density functional optimization, using regularization and cross-validation methods from machine learning. The general idea is to minimize errors through a survey of a wide range of functionals through an understanding of the expected error bars. To date, there have been a few examples demonstrating moderate success. 
It still remains to be seem as to whether or not computer driven optimization strategies will surpass scientific insight when designing better functionals.

\subsection{The Vydrov and van Voorhis functionals}

\label{sec:VV}

Remaining within the vdW-DF framework, Vydrov and van Voorhis designed a functional called vdW-DF-09~\cite{vdW-DF-09} by introducing 
reference-system optimization 
even for the nonlocal part. Even though they abandon vdW-DFs close connection to GGA, this method retains all the essential constraints of vdW-DF. 

Closely following this work was the development of the 
two offspring of vdW-DF called VV09 \cite{VV09} and VV10 \cite{VV10}. 
These two functionals 
 inherit many of the features of vdW-DF, but also include additional physical mechanisms at the cost of adhering to fewer exact constraints~\cite{VV09comment}. 
Using physical arguments 
to improve the account of long-range interactions between small molecules, they rely on
reference-system optimization to parameterize one (VV09) or two (VV10) fixed parameters. 
Both functional are designed for simplicity, VV10 radically so.
The VV10 variant is also the most flexible and is more recently emphasized by the authors. The VV09 on the other hand is constructed in a manner most reminiscent of vdW-DF. 

Both rely on a spatially varying gap 
$\omega_{\text{g}}(\rr)$ in the plasmon dispersion model that is given by~\cite{Gutle99p885,VV09} 
\begin{equation}
  \omega_g^2(\vv{r})=\frac{C}{m^2} \, \left| \frac{\nabla n}{n(\vv{r})} \right|^4\, .
\label{eq:gabvariation}
\end{equation}
The parameter $C$ was fit to optimize the $C_6$ coefficients, 
describing the long-range interaction between molecules.

Similar to vdW-DF, VV09 builds upon the plasmon propagator $S \approx S_{\rm xc}=\ln \epsilon$  defined as  
\begin{equation}
\tilde{S}_{\vv{q},\vv{q'}}(iu) = \int d\rr\, e^{-i\vv{r}\cdot (\vv{q}-\vv{q}')}
\frac{\omega_\p^2(\vv{r})}{\omega_0(\vv{r})+u^2} F_{q,q'}(\vv{r})\; ,
\label{eq:vv09form}
\end{equation}
with plasmon poles given by
\begin{equation}
  \omega_0^2(\vv{r})=\omega_g^2(\vv{r})+\omega^2_{\rm p}(\vv{r})/3 \; .
  \label{eq:VVpole}
\end{equation}
Here $\omp=\sqrt{4\pi n e^2/m}$ is the classical plasma frequency.
Unlike vdW-DF, these poles do not shift as a function of wavevector; 
rather VV09 relies on damping factors $F_{q,q'} = \exp [- (q^2+q'^2)/(k_s^2 \phi^2)]$ to reduce the weight of dispersive states~\cite{VV09}. 
Here $k_s$ is the Thomas-Fermi wavevector and $\phi$ is the spin-scaling factor $\phi=[(1+\zeta)^{2/3}+(1-\zeta)^{2/3}]/2$.
 
This choice causes VV09 to break charge conservation, an essential constraint of vdW-DF~\cite{Dion,VV09comment}. 
%The VV09, as do VV10
%by a related mechanism discussed below, thus achieves
%a range-separation so that their accounts of the nonlocal
%correlation energy can be combined with general range of semilocal 
%functionals with only a limited degree of double counting.
%TheVV09 response response description,  equiation
%(\ref{eq:vv09form}), while there are also formal limitations.\cite{VV09comment}
The $1/3$ factor is motivated by the Clausius-Mossoti relation
and the appropriate screening relation for jellium spheres~\cite{VV09}.
%The method is also built around a model dielectric function $\epsilon(iu)=1+\omega_p^2/(\omega_g^2+u^2)$ that is well

%for larger molecules.
%there exist arguments that plasmon contributions, and their dispersion, becomes 
%important.\cite{fullerenewisdom,BeHy13} 
%The VV09 has a plasmon-pole structure
%that is motivated by the physical behavior, i.e., an electrodynamical coupling
%as it is described at $\lambda=1$ and it is so closer in nature to the
%ALL and vdW-DF0 descriptions.\cite{AnLaLu96,Rydberg03p126402} In contrast,
%the general-geometry methods seeks a plasmon-dispersion that represents the
%$\lambda$-everaged behavior that is relevant in the ACF; one should, for example,
%observe that the vdW-DF1 and vdW-DF2 model plasmons dispersion for which
%$\omega_{q\to 0} \neq \omega_p$.

Rather than attempting to approximate the exact ACF, 
the VV10 construction starts directly by assuming a universal kernel (\ref{eq:enlstd}) given by a  simple ansatz,
\begin{equation}
\Phi^{\rm VV10}(\rr,\rr',R=|\rr-\rr'|) = -\frac{3e^4}{2m^2\, gg'(g+g')}\,,
\label{eq:vv10form}
\end{equation}
where $g=\omega_0(\rr) R^2 + \kappa(\rr)$ (similar for $g'$) and $R=|\rr-\rr'| $.
This kernel is designed to give the same asymptotic form as that of VV09.
%\begin{align}
%g(\vv{p}; R) & =  \omega_0(\vv{p})R^2 + \kappa(\vv{p})} \, ,
%  \kappa{\vv{p}) & =  3 b \frac{\omega_p(\vv{p})}{k_{TF}^2(\vv{p})} \, .
%  \label{eq:vv10kappafactor}\\
%\end{align}
The $\kappa$ function serves to dampen 
nonlocal-correlation energy contributions at shorter separations.
It depends on a scaling parameter $b$ which is fixed for a given semilocal partner by optimizing binding energies to the 
S22 benchmark~\cite{S22}. 
In this respect,  like the dispersion-corrected DFT methods, VV10 is reference-system optimised, though the number of input parameters is drastically reduced.  
 In fact, in a separate development, the long-range account serves as input in such a method \cite{Vydrov10p62708} (section~\ref{sec:asympDamp}). 
The evaluation of the non-local correlation energy in VV10 can also be sped up in a similar manner to that of vdW-DF~\cite{RoSo09} by introducing a small modification to the kernel~\cite{Sabatini2013p041108}.

VV10 has been tested for a range  
of problems including noble-gas and molecular systems, traditional bulk and 
layered vdW systems~\cite{noblegasdimers,BjorkmannLayered1,BjorkmannLayered2,Bjorkman12p165109,Sabatini2013p041108,torbjorn14}.
It works particularly well for interactions between small molecules~\cite{VV10,noblegasdimers,Vydrov2012p1929}. 
%and it can outperform vdW-DF variants.
However, results for layered systems indicate that VV10's transferability across length scales may be inferior to that of vdW-DF \cite{BjorkmannLayered1,BjorkmannLayered2}. 
This shortcoming may be related to the crude mechanism used to account for the saturation of vdW forces at shorter separations, as it lacks the constraint-based mechanisms inherent to vdW-DF.
%For the closely related VV09 that attempt 
%a proper connection to constraints to account for the effect of plasmon dispersion. 
%The closely related VV09 fails to to cancel self-energy terms and breaks charge conservation \cite{VV09comment}, VV10 
On the other hand, the VV10 framework can readily be adapted to accurately describe special classes of systems. 
For instance,  Bj\"orkman crafted a special purpose functional for layered systems~\cite{Bjorkman12p165109} (VV10sol) that accurately describes the binding of layered systems. 
VV10 can also readily be paired up with hybrid functionals; thus producing the correct asymptotic exchange behaviour~\cite{VV10,Waldemar2011p3866,Vydrov2012p1929}.

The overall success of VV09 and VV10 further illustrates the potency of using a density functional framework for including van der Waals forces.
Being so closely related to vdW-DF, comparing their functionality to that of vdW-DF can be illuminating and may even trigger new developments.

%%%%%%%%%%%%%%%%%%%%%%%%%%%%%%%%%%%%%%%%%%%%%%%%%%%%%%%%%%%%%%%%%%%%%%%%
\section{Applications}
\label{sec:applications}
%%%%%%%%%%%%%%%%%%%%%%%%%%%%%%%%%%%%%%%%%%%%%%%%%%%%%%%%%%%%%%%%%%%%%%%%

After a decade of theory development and model calculations which
culminated in the development of vdW-DF, a surge of computations on
sparse matter has followed. An early review summarised the status in
2009~\cite{langrethjpcm2009}. Since then the number and variety of
applications have grown tremendously. A set of recent perspective papers
\cite{rev8,BurkePerspective,BeckePerspective} give a broader overview of
the current situation for method and applications also in the vdW
computation arena. Being such a widely used method, it is impossible to
cover every application in a single review. Here, we attempt to
illustrate the depth and extent of modern applications of vdW-DF.
This overview will necessarily have with some bias towards work related to our own research.
Several benchmark studies, such as that illustrated in figure~\ref{fig:S22set}, indicate that its accuracy has improved with more recent developments. 
Naturally, the race for higher accuracy functionals will
continue.  Additionally, since vdW forces are present in numerous
systems such as organic, inorganic, polymeric, and bio-organic systems,
the future will be full of interesting fundamental and applied studies.

\subsection{Early applications}
\label{sec:early}
%%%%%%%%%%%%%%%%%%%%%%%%%%%%%%%%%%%%%%%%%%%%%%%%%%%%%%%%%%%%%%%%%%%%%%%%

\begin{figure}
\includegraphics[width=\columnwidth]{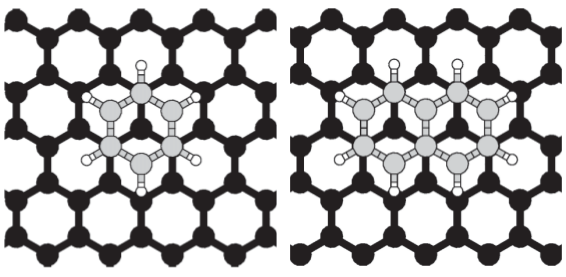}
\caption{\label{fig:BzNapht_geometry}Binding geometries for benzene and
naphthalene on graphene. Reprinted with permission from
\cite{Chakarova-Kack06p146107}, \copyright\ 2006 American Physical
Society.}\vspace{4ex}
\includegraphics[width=\columnwidth]{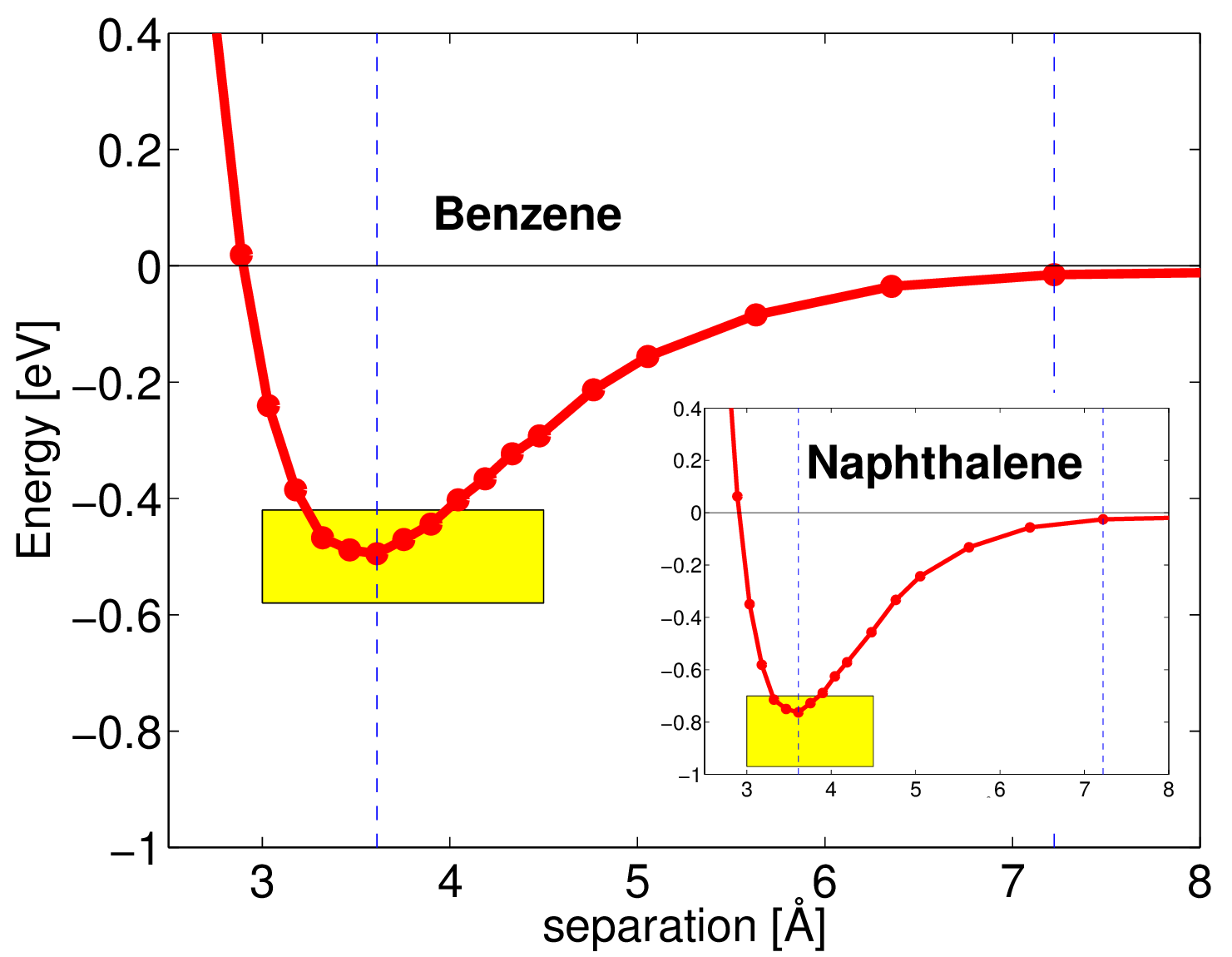}
\caption{\label{fig:BzGraphene_curve}Binding curve for benzene (and
naphthalene) on graphene, calculated with vdW-DF1 and compared to
experimental estimates. Figure adapted from
\cite{Chakarova-Kack06p146107}, \copyright\ 2006 American Physical
Society.}
\end{figure}

As seen in this review, early applications in the 90s focused on well-established
vdW results at that time, namely asymptotic vdW interactions at large
separations between fragments. To calculate, for instance, the $C_6$
coefficient of the vdW asymptote of two small fragments the traditional
method required quantum-mechanical calculations of numerous dipole
matrix elements for each fragment. In this light, it might be
understandable that figure \ref{fig:milkyway} created almost euphoric
feelings within the ALL collaboration; it did not only provide a simple formula for
$C_6$ in terms of the electron density $n$, but also a ``milky way'' (figure~\ref{fig:milkyway})
that got narrower as the approach was successively improved.

The first application, including non-asymptotics was made
for jellium \cite{RyLuLaDi00,RydbergThesis}. For this simple model
system accurate calculations were available, and good agreement between
these results and those with the early variant of vdW-DF0 showed promise.
For a modern practitioner of vdW methods, the optimistic tone described
in chapter~\ref{sec:layered} for graphite, using the year 2000
numbers~\cite{2001SurfScience}, might seem surprising. However, one
should keep in mind that in 2000 the leading DFT, namely the GGA, gave
almost no binding at all and if so only at unphysically large separations.
Later vdW-DF0 results were also found to be promising \cite{Rydberg03p606}.
Figure~\ref{fig:graphite} illustrates that this promise has been kept by
further developments along the vdW-DF track, that is vdW-DF1, vdW-DF2
and vdW-DF-cx, the latter overlapping with results of accurate quantum
Monte-Carlo simulations.

An early pivotal application was on the adsorption of benzene and
naphthalene on graphene \cite{Chakarova-Kack06p146107}.
Figure~\ref{fig:BzNapht_geometry} depicts the corresponding binding
geometries. This application provided a valuable comparison between
theory and experiment because a far-sighted experimental group
\cite{Zacharia} had measured thermal desorption-energy values. vdW-DF1
predicts binding energies and separations that agree well with experiment.
For the adsorption of benzene and naphthalene, figure~\ref{fig:BzGraphene_curve} shows
the binding curve and the experimental estimated ranges. 
The good agreement stimulated successive works with the vdW-DF as well as later theory development and numerous new applications of the functional.
For instance, this work triggered many
other early applications on adsorption of other molecules and on other
surfaces~\cite{Chakarova-Kack06p155402}. A few years later, the general
applicability of vdW-DF was further established with promising results
for the adsorption of molecules on metallic surfaces, such as for
benzene on coinage metals \cite{Berland09p155431,Toyoda09p2912}.

%\begin{figure}
%\includegraphics[width=\columnwidth]{BenzeneDimers.png}
%\caption{\label{fig:bzdimer_conf}Benzene dimer configurations. Reprinted
%with permission from \cite{Puzder06p164105}, \copyright\ 2009
%American Institute of Physics.}
%\end{figure}

Even on the molecular side, the benzene molecule provides the prototype.
For the interaction between two benzene molecules, orientation must also
be considered. An early vdW-DF study on a simple molecular dimer was on
the benzene dimer, in the sandwich, slip-parallel, and T-shape
configurations, detailed in 
\cite{Puzder06p164105}.  In all cases, the vdW-DF1 curves were typically between the coupled
cluster results at the CCSD(T) level and those of M\o ller-Plesset
perturbation theory (MP2). 
It is gratifying that vdW-DF seems to improve upon the MP2 results which can overestimate the dispersion binding
\cite{Cybulski07p141102,S22}.

%\begin{figure*}
%\includegraphics[width=0.8\textwidth]{BzDimers_curve.png}
%\caption{\label{fig:bzdimer_curve}Binding energy curves of benzene
%dimers in the sandwich and T-shape configuration (shown in
%figure~\ref{fig:bzdimer_conf}). The figure compares results of three
%different theoretical methods indicated in the figure legend. Similar
%curves exist for the slip-parallel
%configuration~\cite{Puzder06p164105,Thonhauser06p164106}. Reprinted
%with permission from \cite{Puzder06p164105}, \copyright\ 2006 American
%Institute of Physics.}
%\end{figure*}

The same is true when one hydrogen atom of a benzene ring is replaced
by such groups as OH, CH$_3$, F, and CN~\cite{Thonhauser06p164106}.
These ``monosubstituted'' benzene dimers are necessary precursors for
applications of vdW-DF to the stacking of nucleobases and DNA base pairs
reviewed in section~\ref{sec:biomolecules}. Just as with the benzene
dimer \cite{Puzder06p164105}, the monosubstituted benzene dimer
\cite{Thonhauser06p164106}, the benzene-water complex \cite{Li08p9031},
and the methane-benzene system and related dimers \cite{Hooper08p891}
have been shown to give results that lie between CCSD(T) and MP2 for the
binding energies as a function of separation.

\begin{figure}
\includegraphics[width=\columnwidth]{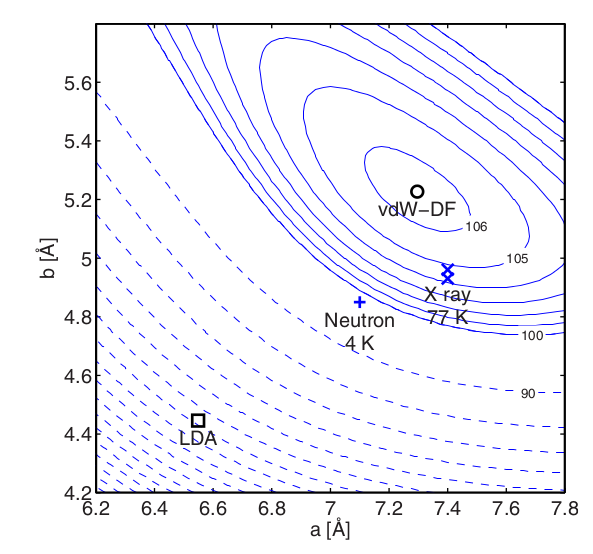}
\includegraphics[width=\columnwidth]{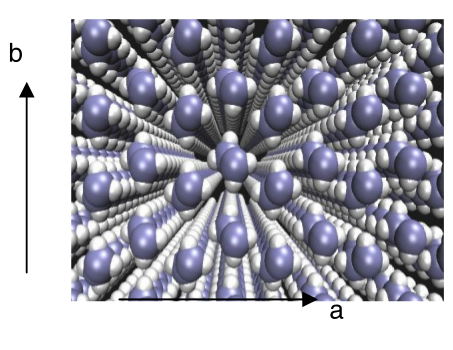}
\caption{\label{fig:poly}Binding energy contours for a polyethylene
crystal compared with LDA and results from diffraction experiments.
Reprinted with permission from \cite{Kleis07p100201}, \copyright\
2007 American Physical Society.}
\end{figure}

Originally, van der Waals proposed his interaction to describe real
gases, as distinguished from the ideal gas. In this spirit, replacing
empirical force-field methods by vdW-DF is an interesting challenge.
Such methods are widely used to determine the structure of molecular
crystals and for larger disordered systems such as polymers. The
prospect of using vdW-DF for such purposes was first investigated in a
study of a solid polymer, polyethylene \cite{Kleis07p100201}, in a low
temperature, crystalline phase, for which experimental structures are
available to compare with. The good agreement shown in
figure~\ref{fig:poly} gave hope for using vdW-DF1 also in simulations on
both ordered and disordered systems with molecules interacting with vdW
forces. This promise has continued to grow with many new successful
applications on related systems such as nanotubes \cite{Kleis08p205422}
and molecular crystals \cite{Berland10p134705,Berland11p1800}.

The early findings eventually lead to much further research, resulting
in the release of vdW-DF2 and other variants and numerous applications,
such as those highlighted in chapter~\ref{sec:applications}, which are
indicative of the present status.

%%%%%%%%%%%%%%%%%%%%%%%%%%%%%%%%%%%%%%%%%%%%%%%%%%%%%%%%%%%%%%%%%%%%%%%%
\subsection{Benchmark calculations}
\label{sec:benchmark}
%%%%%%%%%%%%%%%%%%%%%%%%%%%%%%%%%%%%%%%%%%%%%%%%%%%%%%%%%%%%%%%%%%%%%%%%

To assess the applicability of vdW-DF and other methods, we can rely on
results from accurate theoretical and experimental methods. Common
\textit{theoretical} methods for this purpose are many-body quantum
chemical methods such as coupled-cluster, e.g.\ CCSD(T), and perturbative
methods like MP2, which show high accuracy, in particular for relatively
small systems. Early benchmark calculations of this kind are mentioned
in chapter~\ref{sec:early}. From \textit{experiment} carefully
determined quantities describing structural and binding properties are
often used (refer to section~\ref{sec:H2Cu111case}). The literature is full of comparative assessments of the
performance of various sparse-matter methods.

\begin{figure*}
\includegraphics[width=6.5in]{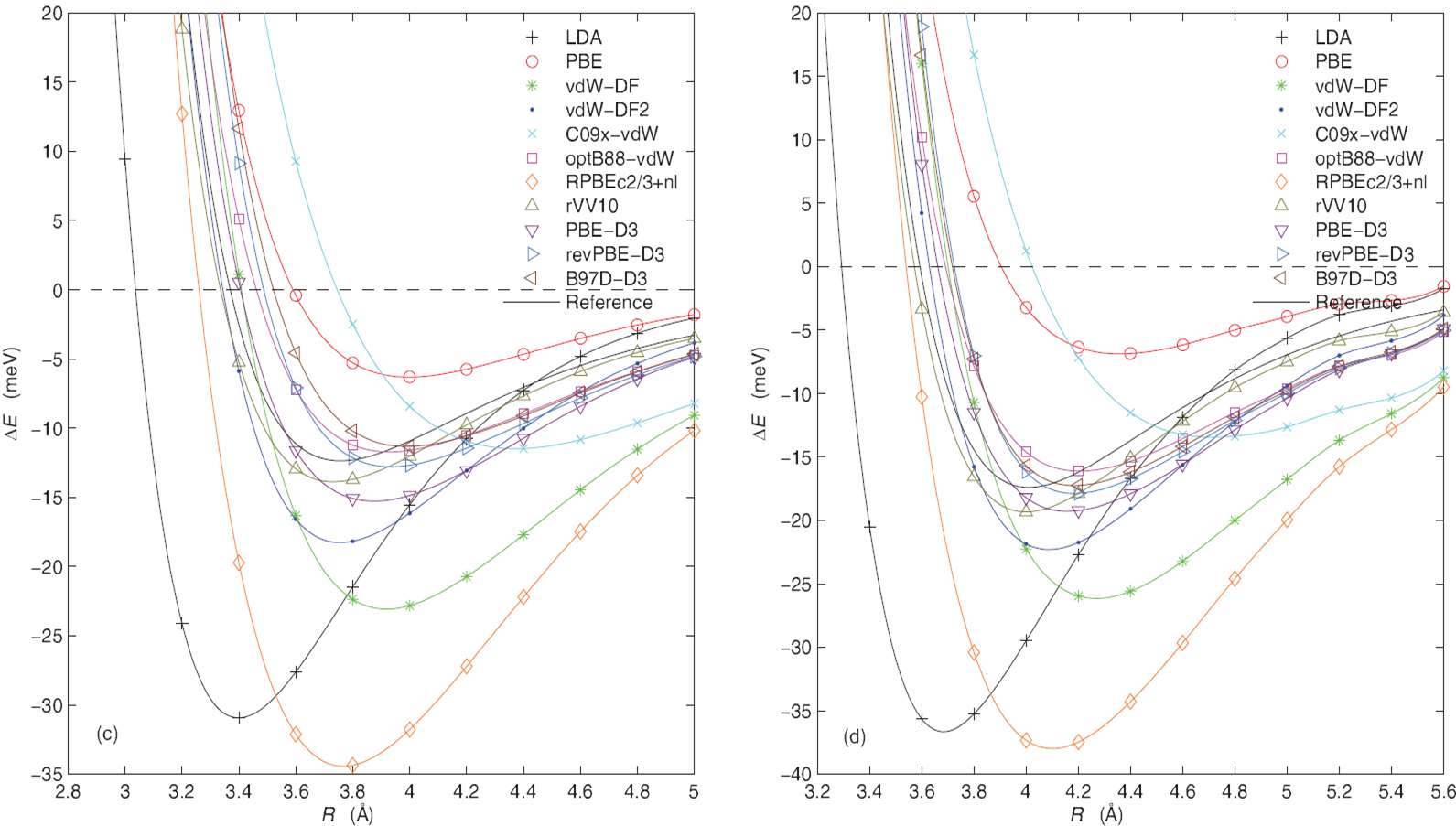}
\caption{\label{fig:nobel_gas_dimers} Interaction energy curves for (a)
(left) Ar$_2$ and (right) Kr$_2$ obtained from various
functionals and compared to reference results (black line without
symbols).   (N.B. the curves for vdW-DF-C09 in figure~\ref{fig:nobel_gas_dimers} and the data for this functional were updated in an erratum~\cite{Tran13p033903}; now showing much better agreement with other functionals like vdW-DF-optB88). Adapted with permission from~\cite{noblegasdimers},
\copyright\ 2013 American Institute of Physics.}
\end{figure*}

For the vdW-DF method there are both virtues and vices. Already from the
beginning \cite{Dion} the rare-gas dimers were perturbing examples of
the latter: vdW-DF1 was applied to the ``rare-gas dimers,
where it is shown to give a realistic description.'' The ``moderate
successes'' are quantified by results for $C_6$ coefficients of rare-gas
dimers and binding-energy curves of Ar$_2$ and Kr$_2$ \cite{Dion}. This
is certainly progress compared to GGA results, but its limitations are
obvious and partly analysed in further functional developments.  The
rare-gas systems are the prototypical van der Waals systems, where the
dispersion interactions are the only source of attraction between atoms
and for which highly accurate \emph{ab initio} or empirical results are
available. The rare-gas dimers, accounted for in this review, have been
used numerous times for the testing of functionals for weak
interactions.

Small atomic and molecular dimers traditionally serve as benchmarks or stepping stones for larger applications. The classic cases of noble-gas dimers were originally treated with the vdW-DF1 method and signaled problems with the lighter atoms, while for the Ar and Kr dimers the binding energies get reasonable values, though binding distances are overestimated by a few percent~\cite{Dion}. Recently \cite{noblegasdimers, Tran13p033903}, several variants of the vdW functionals have been tested on rare-gas dimers (from He$_2$ to Kr$_2$) and solids (Ne, Ar, and Kr) and their accuracy was compared to standard semilocal approximations, supplemented by an atom-pairwise dispersion correction~\cite{Grimme3}.  As depicted in figure~\ref{fig:nobel_gas_dimers}, in general modern variants such as vdW-DF-optB88 \cite{optX}, vdW-DF2~\cite{Lee10p081101}, vdW-DF-C09~\cite{cooper10p161104} (see erratum:~\cite{Tran13p033903}), and related functionals like rVV10~\cite{Sabatini2013p041108}, exhibit significant improvements in their treatment of noble-gas dimers with respect to binding separation distance, albeit not at the desired chemical accuracy for binding energies. Nevertheless, it is important to note that the noble-gas dimers are not systems one would expect the vdW-DF method to perform well as its emphasis on electron-gas many-body physics makes it ideally suited for systems with extended states, not for strongly localised ones. The later developed vdW-DF2 has a design that makes it better suited for smaller molecules. 

%We understand that there is an interest in comparing numbers like those in figure \ref{fig:nobel_gas_dimers} and also that empirical or semiempirical functionals can be useful. However, we wish to also assess a functional on how solid its physical foundations are. 
%Even with such a strategy in mind, the issues exposed for rare-gas dimers \cite{noblegasdimers} can be helpful in designing an even better vdW-DF.

\begin{figure*}
\includegraphics[width=0.95\textwidth]{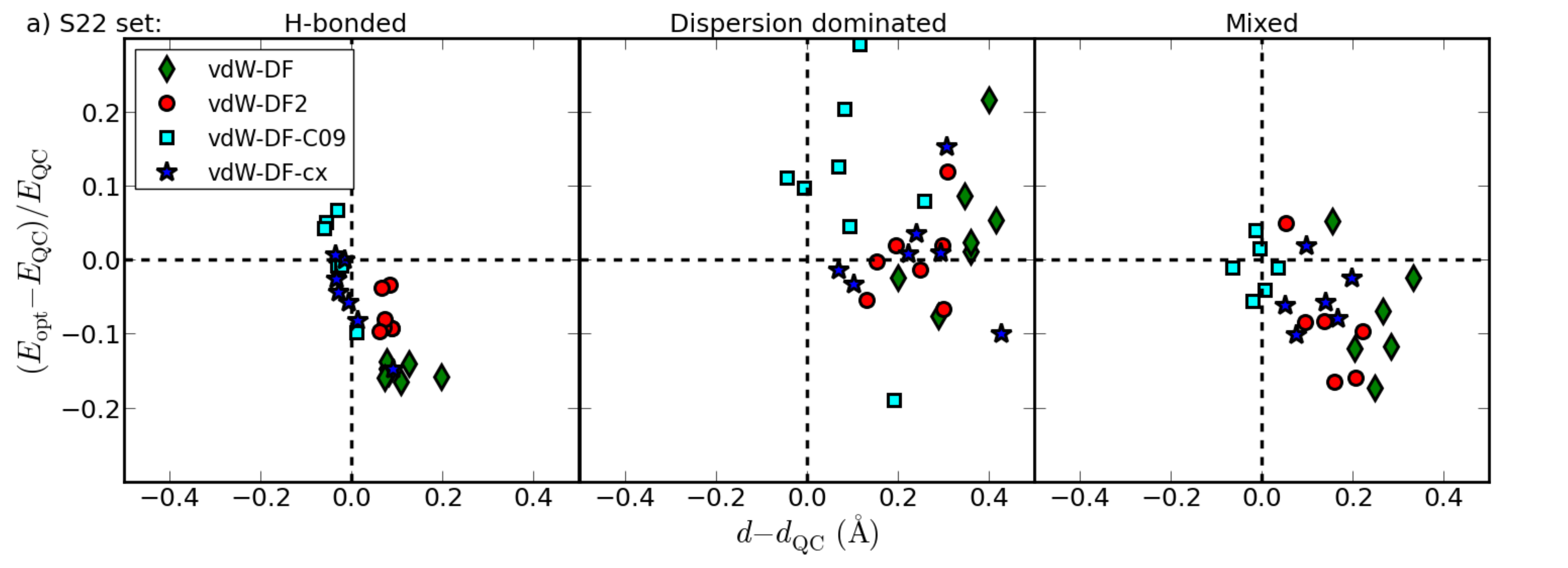}
\caption{\label{fig:S22set}Comparison of vdW-DF versions for the S22
data set. vdW-DF-C09, vdW-DF2, and vdW-DF-cx represent an improved
overall performance compared to vdW-DF1, in particular the
overestimation of separations is reduced. Reprinted with permission
from~\cite{BeArCoLeLuScThHy14}, \copyright\ 2014 American Institute of
Physics.}
\end{figure*}

Another popular theoretical benchmark, the so-called S22 data
set~\cite{S22}, comprised of CCSD(T) calculations of 22 molecular
duplexes including H-bonded, dispersion dominated, and mixed systems,
has been extensively studied to assess how techniques for including
dispersion interactions perform. vdW-DF calculations have been
performed on all duplexes in the S22 data set and results are presented
in~\cite{Lee10p081101, LinearScaling, cooper10p161104, vdWsolids, optX,
BeLoScHy13, behy14}. Figure~\ref{fig:S22set} presents an overview of the
performance of vdW-DF versions developed by us for the three kinds of
systems. This figure and the publications listed highlight how the
performance of vdW-DF has improved over time, either through
rederivation of the nonlocal correlation term or through the
development of exchange functionals, as discussed in section
\ref{sec:refinements}. In all cases, the mean absolute deviation (MAD)
has been reduced from 65 meV for the original functional to 10 -- 20~meV,
i.e.\ to within chemical accuracy. Similar improvements in accuracy have also been observed
for vdW-DF-optB86~\cite{vdWsolids} as well as the recent vdW-DF2-B86R functional~\cite{Hamada14} 
and with reference-optimized functionals like vdW-DF-optB88~\cite{optX}.  We note 
that there certainly exists limits for what can be gained from reference-system optimisation 
and from benchmarking against such datasets.  For example, reference systems like the 
S22 and rare-gas dimer sets tell very little about contributions arising from the $d$-electrons.

Comparisons with experiment also give new insights into how well vdW-DF
performs both qualitatively and quantitatively as the many examples of
this chapter will illustrate. Additionally, some experimental results
provide particularly accurate data and therefore constitute accurate
benchmarks.

%\begin{figure}
%\includegraphics[width=\columnwidth]{MolecularCrystals.png}
%\caption{\label{fig:molcrys1}Schematics of the structures of two
%high-symmetry molecular crystals. The left are for hexamine; the right
%for dodecahedrane. Reprinted with permission from
%\cite{Berland10p134705}, \copyright\ 2013 American Institute of
%Physics.}
%\end{figure}

Crystalline solids stand out as excellent case studies for how well a
method can predict structures, because x-ray and neutron scattering can
precisely determine the atomic positions in such materials. The good
results for polymer crystals mentioned in section \ref{sec:early} were
therefore encouraging during the early testing phase of vdW-DF.
Similarly, molecular crystals serve as particularly good benchmarks for
vdW-DF because they are held together by non-covalent forces. An early
such comparison for the high-symmetry molecular crystals hexamine,
dodecahedrane, cubane, and C$_{60}$~\cite{Berland10p134705, Berland11p1800}
showed that vdW-DF describes the structure, cohesive energies, and bulk
moduli of these molecular crystals well, though vdW-DF1 consistently
overestimates the crystal cell volume. Similar
conclusions have been drawn in other studies on molecular
crystals~\cite{rev8, Benchmark:molcrys, Sabatini12p424209, pi_conjmolcrys,
Lee12p104102}.

Surface adsorption studies also provide a unique avenue for
benchmarking. For example, the adsorption of molecules on metal surfaces,
which are closer to jellium surfaces, can be insightful in testing and
designing functionals with dispersion interactions. Section
\ref{sec:H2Cu111case} exemplifies this with a case study on the
adsorption of H$_2$ on copper. It is shown that vdW-DF2, 
 in particular, gives
potential-energy curves for different Cu facets that agree well with
experiment~\cite{lee11p193408, lee12p424213}. This is a feat, which techniques that employ pairwise
corrections are unable to reproduce, presumably due to the inability of
such potentials in distinguishing between bulk and surface density
regions.

The performance of vdW-DF has also been benchmarked against
QMC results for molecular-hydrogen phases \cite{Clay14p184106} and for 
bulk water \cite{Morales14p2355}. The studies test the vdW-DF method as a 
description of a full potential-energy variation for molecular dynamics,
sharing a benchmarking philosophy with our H$_2$ physiosorption 
studies \cite{lee11p193408,lee12p424213}. The molecular-hydrogen and
water studies compare QMC and vdW-DF calculations for a rich set of 
configurations  both within and among different phases and allow for 
flexible molecular configurations. The vdW-DF1 and vdW-DF2 versions are 
found to have very good transferability across length scales, tending to 
overestimate the binding separation but being reliable on the total-energy 
variation.

The positive feedback from benchmarks as well as from quantitative and
qualitative comparison between theory and experiment signals a new era
for nonlocal functionals in which they can be used to understand the
role that dispersion interactions play in real materials. Many such
applications are mentioned next.

%%%%%%%%%%%%%%%%%%%%%%%%%%%%%%%%%%%%%%%%%%%%%%%%%%%%%%%%%%%%%%%%%%%%%%%%
\subsection{Adsorption}
\label{sec:adsorption}
%%%%%%%%%%%%%%%%%%%%%%%%%%%%%%%%%%%%%%%%%%%%%%%%%%%%%%%%%%%%%%%%%%%%%%%%

The adsorption of molecules to surfaces and within porous media is a
defining feature of numerous industrial, chemical, and energy relevant
processes. For example, the self-assembly of organic molecules and
catalytic reactivity are mediated by molecular chemisorption to
surfaces. Similarly, carbon sequestration and H$_2$ storage in
carbon-based structures are driven by the dispersion-dominated
physisorption of guest molecules. In addition, the charging and
discharging of lithium ion batteries depend on the movement of metal
ions between weakly bound planes of a graphitic anode. These examples
highlight the need for a method that can cross the traditional boundary
of chemisorption to weaker physisorption. In this regard, vdW-DF has
been proven to be both sufficiently accurate and computationally
efficient for simulating the intricate details of dispersion
interactions at surfaces and within porous materials.

%%%%%%%%%%%%%%%%%%%%%%%%%%%%%%%%%%%%%%%%%%%%%%%%%%%%%%%%%%%%%%%%%%%%%%%%
\subsubsection{On surfaces}
\label{sec:surfaces}
%%%%%%%%%%%%%%%%%%%%%%%%%%%%%%%%%%%%%%%%%%%%%%%%%%%%%%%%%%%%%%%%%%%%%%%%

The physisorption of molecules to graphite and two-dimensional surfaces,
including graphene, metal dichalcogenides like MoS$_2$, and PAHs,
has been a long-standing focus of vdW-DF
calculations~\cite{Chakarova-Kack06p146107,
Moses09p104709,berland11p135001, BeHy13, Chakarova-Kack10p013017,
Londero12p424212, Akesson12p174702, Schroder13p871706, Jiang09p4019,
Cooper12p34, Chakarova-Kack06p155402,
Bergvall11p155451,anti-aromatic,CoinageOnGraphene}. These surfaces
often allow for meaningful comparisons with quantum-chemical
calculations. For instance, a comparison of the adsorption energy of
H$_2$ interacting with PAHs shows excellent agreement with previous MP2
calculations---with only small deviations at large separation
distances~\cite{Cooper12p34}. This is in dramatic contrast to the
behaviour of GGA functionals that show little to no binding, or even LDA
functionals which overbind molecules to graphene.

\begin{figure}
\includegraphics[width=\columnwidth]{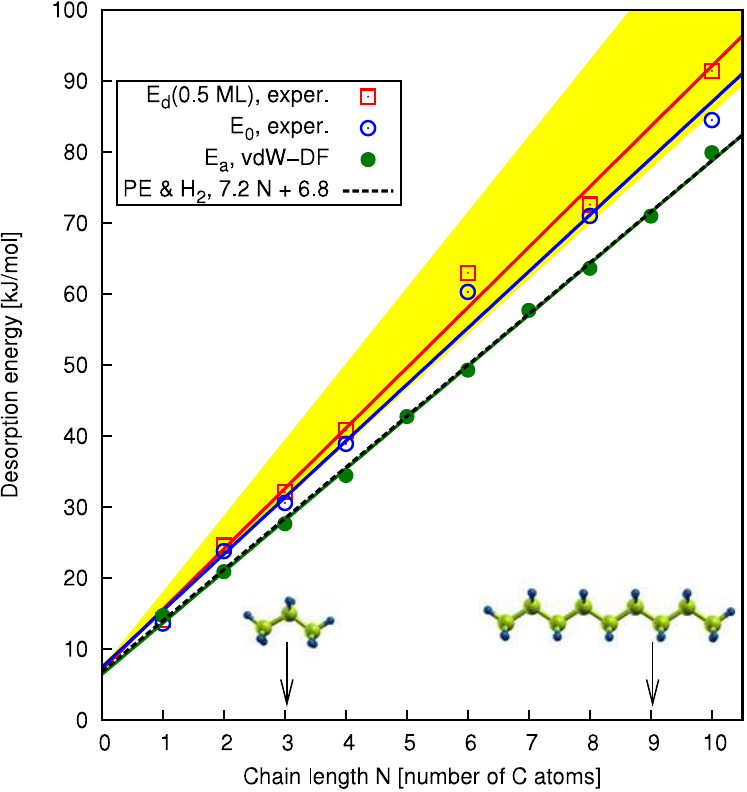}
\caption{\label{fig:alkanes}Desorption energy as a function of the
length of an $n$-alkane chain. Solid points are vdW-DF1 results, open
points are from the TPD measurements by Tait et al.~\cite{Tait}. Linear
fits for the three sets of data points are also shown. Study detailed
in~\cite{Londero12p424212}. }
\end{figure}

\begin{figure}
\includegraphics[width=\columnwidth]{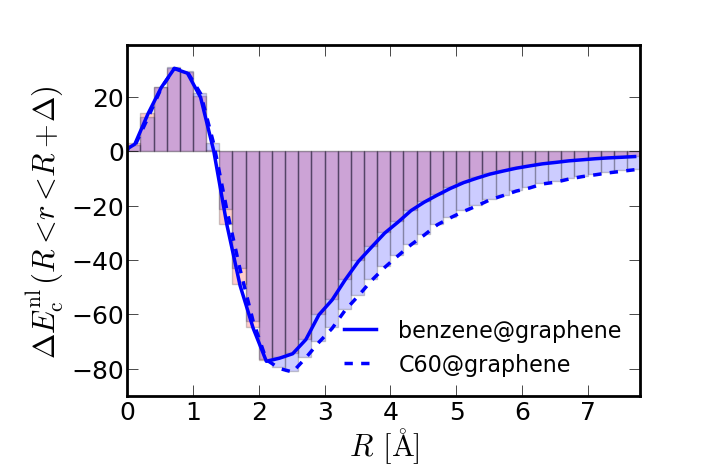}
\caption{\label{fig:C60grapheneContrib}The nonlocal correlation energy
contributions [equation~(\ref{eq:enlstd})] to the adsorption energy for
different separations between density regions $R=|r-r'|$. The two curves
compare this analysis for benzene and C$_{60}$ on graphene at the same
molecule-to-graphene separation. Study detailed in \cite{BeHy13}.
Reprinted with permission from \cite{BeHy13}, \copyright\
2013 American Physical Society.}
\end{figure}

\begin{figure}
\includegraphics[width=0.8\columnwidth]{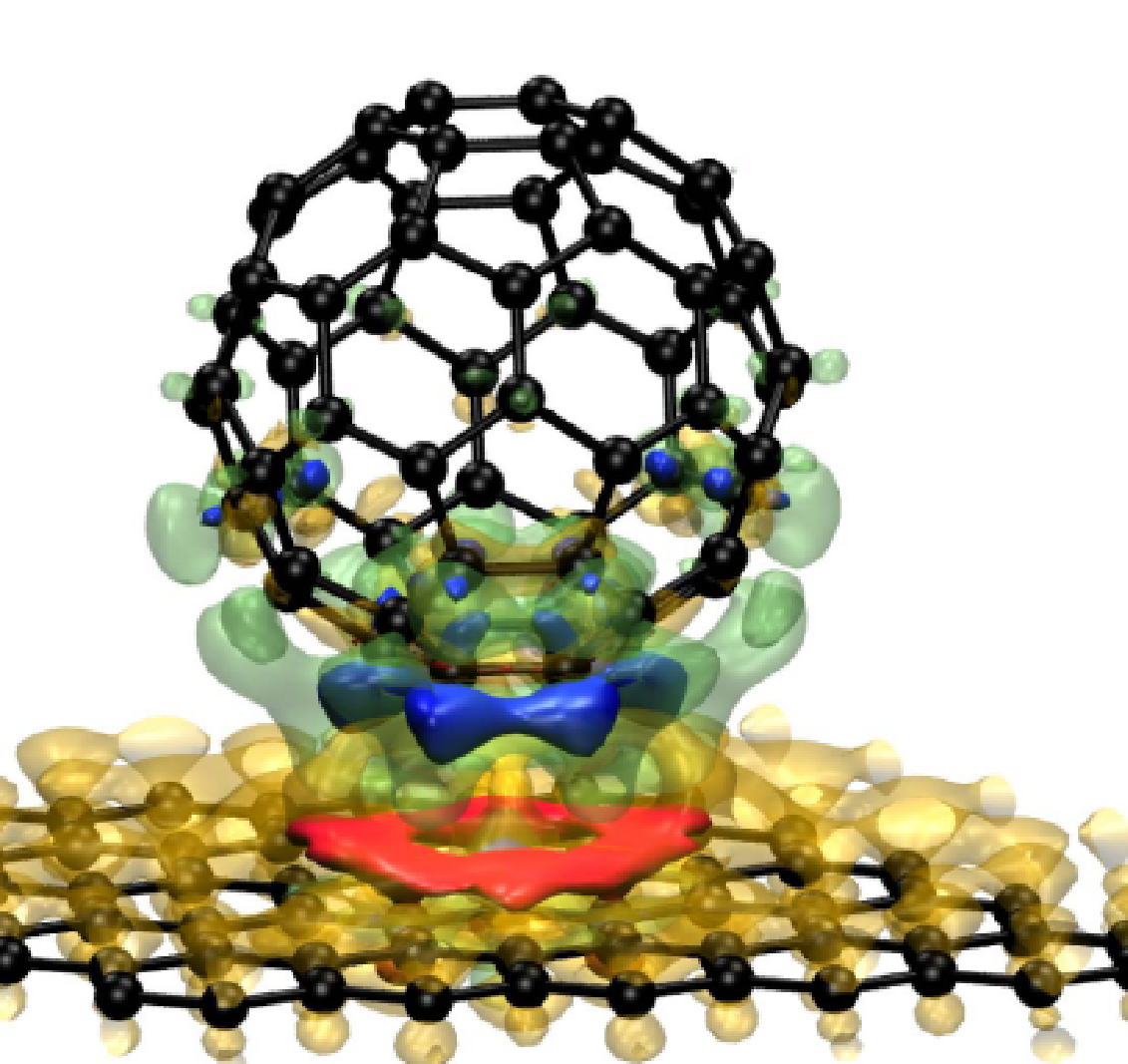}
\caption{\label{fig:C60graphene} Charge transfer isosurfaces for
C$_{60}$ on graphene. Blue isosurfaces indicate loss of charge. Red isosurfaces indicate
gain in charge. Reprinted with permission from \cite{BeHy13},
\copyright\ 2013 American Physical Society.}
\end{figure}

Figure~\ref{fig:alkanes} compares experimental data on  $n$-alkane
adsorption on C(0001) deposited on Pt(111) with vdW-DF1
results~\cite{Londero12p424212}. Experimentally it is found that the
adsorption energy varies linearly with $N$ the number of units in the alkane
 for several different surfaces, but with a small offset at $N=0$.
This result might be puzzling in a naive picture of vdW contributions to
binding. However, the calculations of Londero et al.~\cite{Londero12p424212} show that this trend and offset is
well reproduced in vdW-DF. This shift can be attributed to the role of
the end groups of the alkane chains. In fact, they find that the offset
of 6.44~kJ/mol in vdW-DF agrees well with the calculated adsorption
energy of an H$_2$ molecule of 6.48~kJ/mol. Similar comparisons between
vdW-DF and experimental data \cite{Lee10p155461} have also been
performed for $n$-butane on Cu, Au, and Pt surfaces. Only a
small part of the polymer chain can contribute significantly to the
binding. This is related to the fact that low-density regions dominate 
 the nonlocal correlation energy~\cite{MuLeLa09}.

%\begin{figure*}
%\includegraphics[width=\textwidth]{thiophene_on_Cu}
%\caption{\label{fig:enldensanalysis}
%(a) Charge-density difference upon adsorption of a single thiophene
%molecule on the Cu(111) surface. The charge transferred from the
%molecule accumulates between the molecule and the surface. In (b) the
%LDA correlation binding energy density, in (c) the semilocal correction,
%and in (d) the nonlocal correction for the relaxed geometry are
%depicted; study details can be found in Ref.~\cite{callsen12p085439}.
%Panels (c) and (d) contrast the nonlocal-energy-density variation as
%described by the semilocal component in GGA and by the vdW-DF1
%nonlocal-correlation energy [equation~(\ref{eq:enlstd})]. The comparison
%shows that vdW-DF1 collects weak-chemisorption energy contributions from
%a much wider region than GGA. This observation is further corroborated
%by the analysis in \cite{BeHy13,behy14,hybesc14}. Reprinted with
%permission from \cite{callsen12p085439}, \copyright\ 2012 American
%Physical Society.}
%\end{figure*}

Adsorption systems, such as the above examples, are well
suited for analysing the properties of vdW-DF. A study in this vein
\cite{BeHy13} compares the adsorption of benzene and C$_{60}$ on
graphene and hexagonal-BN ($h$-BN). C$_{60}$ is bigger but its interface with graphene is similar to benzene. This comparison casts light on
how different separation distances contribute to the nonlocal
correlation energy. At a fixed molecule-graphene distance, vdW-DF1 shows (figures \ref{fig:C60grapheneContrib} and
\ref{fig:C60graphene}) that the binding contributions of benzene and C$_{60}$ are almost identical at short separations but different at larger separations. The repulsive contributions for short
separations reflect the oscillatory shape of the vdW-DF kernel in
figure~\ref{fig:kernel}.  At asymptotic separations the vdW-DF
description of C$_{60}$ has short-comings (\ref{sec:unified}). In spite of this, vdW-DF provides a good description of 
the binding energy of C$_{60}$, relative to benzene. The vdW-DF1 values, 0.49~eV to 0.85~eV, agree with experiment. The higher binding energy of C$_{60}$ comes primarily from the larger
vdW attraction of the bigger C$_{60}$ molecule; yet, a small
contribution also comes from a charge transfer induced dipole (figure~\ref{fig:C60graphene}). The presence of such a dipole
suggests that C$_{60}$ can act as a contact to graphene in molecular
electronics~\cite{Bergvall11p155451}.

Understanding metal-organic interfaces is important for the development
of organic-light emitting diodes and molecular electronics. The nature
of such interfaces is determined by the combination of different
physical effects. For flat aromatic molecules on (111) metal surfaces
the type of adsorption can range from weak (dominated by van der Waals
forces but softened by a small amount of charge transfer, typical for coinage
metals) to strong chemical binding (typical for transition metals),
with van der Waals forces significantly contributing to the
binding~\cite{Kelkkanen11p113401,Liu12p245405}.  Capturing the fine
balance between attractive and repulsive contributions to the binding is
challenging for theory. Early calculations for benzene on coinage metals
gave promising binding energies with
vdW-DF1~\cite{Berland09p155431,Toyoda09p2912}. However, comparison with
experimental vacuum-level shifts for benzene~\cite{Toyoda09p2912} and
bigger molecules~\cite{Romaner09p053010,Toyoda09p78,Toyoda10p134703}
indicate that vdW-DF1 overestimates the separation by as much as 0.4 --
0.8~\AA, more than twice of what is typical for vdW-DF1. Normal
incidence x-ray standing wave experiments for the PTCDA molecule
adsorbed on Ag(111) provide further evidence for this overestimation,
with vdW-DF1 predicting 3.6~\AA~\cite{TS2,BeArCoLeLuScThHy14} compared
to an experimental value of 2.86~\AA~\cite{ExpPTCDAg111}. Aromatic
molecules on coinage metals were one kind of system where improvements
were sorely needed.  Work by Hamada and Tsukada~\cite{Hamada11p245437}
and later by us~\cite{lee12p424213} showed that this issue can be
resolved by updating the exchange choice. However, the excellent
capabilities of vdW-DF were only recently established with a string of
extensive studies combining vdW-DF1 correlation with optB88 and optB86b
exchange, providing accurate results for binding separations and
energies both for the weaker adsorption on coinage metals and the
stronger adsorption on transition
metals~\cite{Liu12p245405,Yildrim13p20572,Yildrim13p2893,Carrasco14p084704,Bjork2014}.
New studies also indicate that vdW-DF-cx~\cite{BeArCoLeLuScThHy14} and
vdW-DF2-B86R~\cite{Bjork2014} are well suited for describing these kind of
systems.

%Figure~\ref{fig:enldensanalysis} partly summarizes a combined GGA,
Callsen and coworkers compared the results of a combined GGA and DFT-D calculation with those of vdW-DF1 for the
weak-chemisorption of thiophene on Cu(111) \cite{callsen12p085439}. The nonlocal-correlation
energy density, derived as the spatially-resolved contribution to the
vdW-DF1 equation~(\ref{eq:enlstd}) is determined. Using this approach, the binding is shown to arise from a wide region 
that exists between the adsorbate and the substrate.

The
finding is fully corroborated by an analysis of the nonlocal-correlation
binding presented (by other means) in \cite{BeHy13,hybesc14}.  The saddle-point or trough-like regions between interacting fragments are important to this and related effects. This insight and the realisation that these regions correspond
to low-to-moderate values of the scaled-density gradient were instrumental in the design of the recent consistent-exchange functional in vdW-DF-cx~\cite{behy14,BeArCoLeLuScThHy14, hybesc14}.

%%%%%%%%%%%%%%%%%%%%%%%%%%%%%%%%%%%%%%%%%%%%%%%%%%%%%%%%%%%%%%%%%%%%%%%%
\subsubsection{In porous materials}
\label{sec:porous}
%%%%%%%%%%%%%%%%%%%%%%%%%%%%%%%%%%%%%%%%%%%%%%%%%%%%%%%%%%%%%%%%%%%%%%%%

Adsorption on surfaces builds the foundation for exploring the
adsorption within porous materials. Nanoporous materials have seen a
surge in interest over the past few decades. A driving force is their potential for wide
applicability in practical devices ranging from sensing to gas
separation and storage. Of particular interest are applications for hydrogen
storage and carbon capture, where relevant materials challenge 
first-principles materials modelling: the adsorbate typically binds to
the host via physisorption, making vdW interactions important. On the
other hand, the host material itself is typically an extended system of
considerable size. As such, methods are needed that can treat extended
systems and vdW interactions on the same footing---a perfect
application for vdW-DF.

The first studies of vdW-DF on porous materials investigated
 the binding energy of H$_2$ in metal organic framework (MOF) materials~\cite{Kong09p081407,Lan09p7165}. MOFs consist
of metal clusters connected by organic linkers, typically creating
networks of cavities or channels inside. Due to the vast number of
choices for both the metals and the linkers, the number of MOFs that can
be synthesised is beyond counting. In the recent past, due to size
limitations, quantum-chemical calculations have focused on understanding
how molecules such as hydrogen interact with the linker fragments as a
gauge of how a particular MOF would perform. Similar considerations can
be applied to other porous materials, such as clathrates and nanoporous
carbons. vdW-DF calculations have shown that the binding in the true
porous structure is significantly enhanced beyond that predicted by
simple arguments based on linker fragments. 

Furthermore, vdW-DF
calculations show remarkable agreement with experimental signatures such
as infrared (IR) frequency and nuclear-magnetic resonance signal shifts
and heats of adsorption \cite{Nijem10p14834, Lopez13p154704,
Nijem12p424203, Yao12p064302, Tan12p3153, poloni_co2_2012,
poloni_ligand-assisted_2012, poloni_understanding_2014}; for a short
review on the integration of experiment and vdW-DF calculations see
\cite{Nijem12p15201, Zuluaga2014}. The ability to monitor changes in IR spectra in
situ, combined with vdW-DF calculations can be crucial for developing a
complete atomistic understanding of the dynamics of molecules such as
H$_2$, CO$_2$, and H$_2$O within porous media~\cite{Canepa13p026102}.
Understanding these mechanisms can be powerful tools for designing MOFs
with new functionalities or preventing degradation.  For example, based
on knowledge of how water molecules interact with the framework, a new
class of MOF, called F-MOF, was designed to overcome this problem by
fluorinating the inside of its cavity~\cite{Nijem13p12615}. Fluoride
ligands repel water molecules, which then start to form small water
clusters in the MOF cavity rather than causing the MOF to degrade.
Another study related to the hydrogen storage capacity of
MOFs~\cite{Li12p424204} investigated the filling of such MOFs with
(H$_2$)$_4$CH$_4$, a vdW crystal itself. (H$_2$)$_4$CH$_4$ has the
highest hydrogen mass-storage density of all materials, except pure
hydrogen itself, but is unfortunately not stable under practical
conditions. The results of the study showed that these MOFs can be used
to significantly improve the stability window of (H$_2$)$_4$CH$_4$ by
providing external pressure to such clusters in its cavity.

\begin{figure}
\includegraphics[width=\columnwidth]{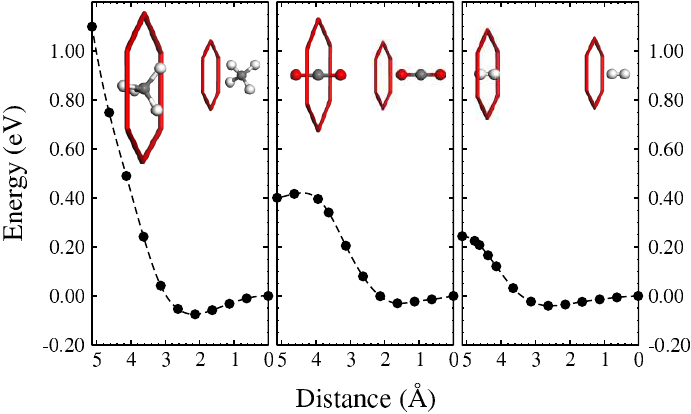}
\caption{\label{fig:clathrate_diff}Diffusion barriers for (left) CH$_4$,
(middle) CO$_2$, and (right) H$_2$ molecules through a clathrate cage. All energies are reported with respect to adsorption energy of the molecule at the centre of the clathrate cage. The relaxation of
the framework (shown in the inset, not to scale) is essential for obtaining
accurate barriers for diffusion. Reprinted with permission from
\cite{Perez10p145901}, \copyright\ 2010 American Physical Society.}
\end{figure}

Clathrates, another class of porous materials, are similar to MOFs with
large cages which can be used for gas storage. They are formed under low
temperatures or high pressures and are found at the bottom of the ocean
in huge deposits around the globe. Naturally occurring clathrates
typically have CH$_4$ trapped inside, which is believed to be
responsible for their stability. A first study of the binding and
diffusion of CH$_4$, CO$_2$, and H$_2$ in the clathrate structures SI
and SH is reported in \cite{Perez10p145901}. The authors show that the
adsorption energy is dominated by van der Waals interactions and that,
even more interesting, without them, gas hydrates would not be stable.
The authors further find that the calculated maximum adsorption capacities as well as
the maximum hydrocarbon size that can be adsorbed are in good agreement
with experiment. One particularly interesting discovery is
that the relaxation of the host lattice is crucial for an accurate
description of molecule diffusion, as the framework deforms
significantly when molecules diffuse from one cage to another (see
figure~\ref{fig:clathrate_diff}). A follow-up study \cite{Li11p153103}
expands the same ideas to rotational barriers of molecules in various
types of cages and also includes the structure SII---which is
particularly challenging to model, as its unit cell contains 408
atoms---demonstrating the efficient scaling of vdW-DF.

Adsorption in nanoporous carbons brings a new challenge to theory and
computation. Unlike clathrates and MOFs, these do not contain regularly
sized pores. In fact, depending on starting materials and synthesis
conditions they can have varied distributions of pores. Modern
approaches, particularly neutron scattering, have been developed to
probe the microstructure of pores~\cite{Morris13p9341}. Using vdW-DF
combined with an efficient continuum model it was shown that adsorption
capacities and heats of adsorption could be predicted for a wide range
of carbons and for both H$_2$ and methane, with knowledge of only the pore
distribution~\cite{Ihm14p1, Ihm12p424205}. This method can also be used
to define the optimal pore size range that enhances
adsorption~\cite{Ihm12p424205}. This study further provides insight into
why carbons that seem similar adsorb very different amounts of a
particular molecule. The approach mentioned here could be the foundation
for rapid screening for the design of highly adsorbing nanoporous
materials.
%%%%%%%%%%%%%%%%%%%%%%%%%%%%%%%%%%%%%%%%%%%%%%%%%%%%%%%%%%%%%%%%%%%%%%%%
\subsubsection{Strong vs. weak adsorption}
\label{sec:strongweak}
%%%%%%%%%%%%%%%%%%%%%%%%%%%%%%%%%%%%%%%%%%%%%%%%%%%%%%%%%%%%%%%%%%%%%%%%

\begin{figure}
\includegraphics[width=\columnwidth]{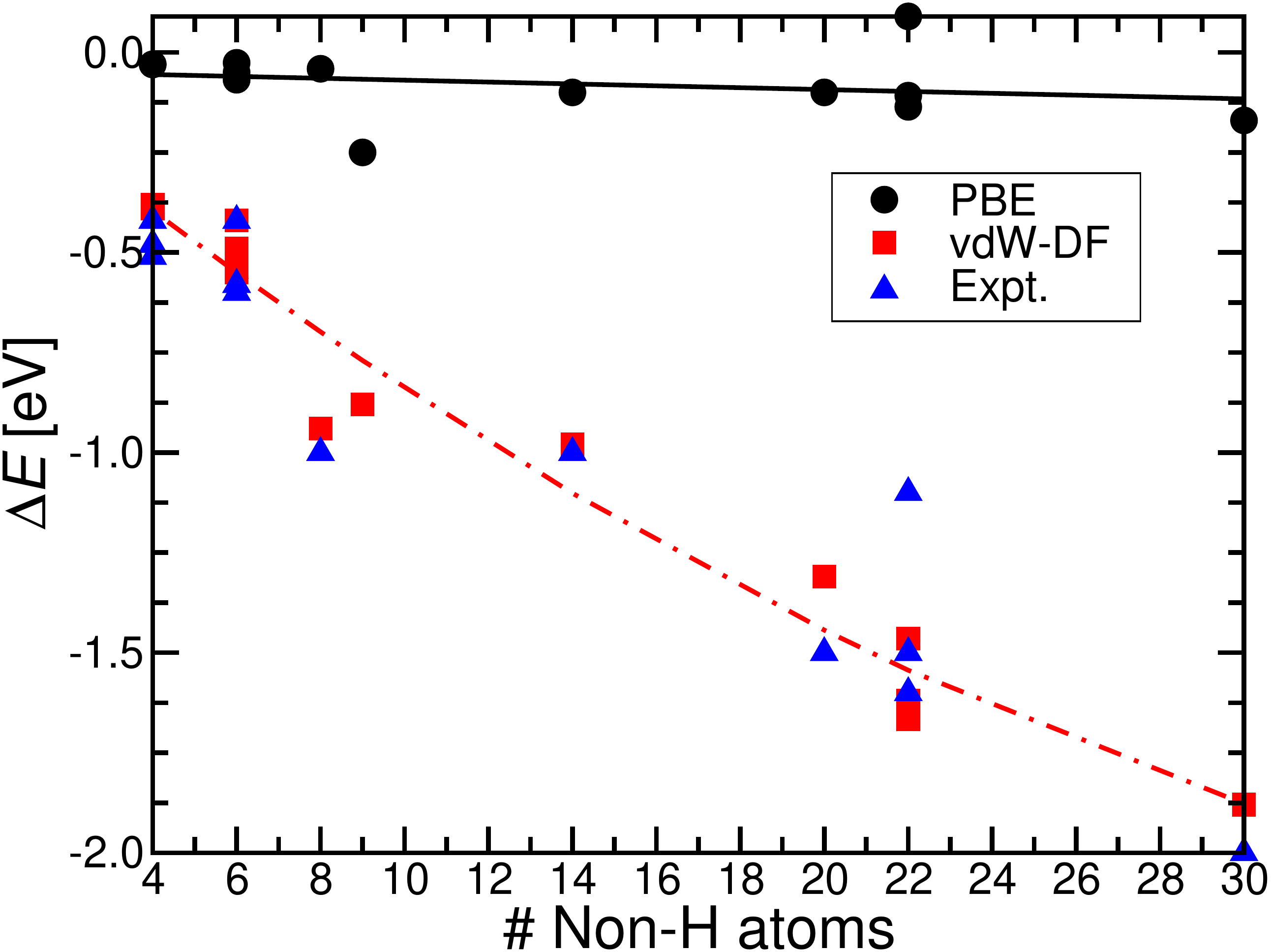}
\caption{\label{OrgBind}Interaction energy as a function of non-hydrogen
atoms for a range of organic molecules on metal surfaces. Black circles,
red squares, and blue triangles represent PBE, vdW-DF, and experimental
data, respectively. The solid black line and red dashed lines are fits
to the data.}
\end{figure}

The adsorption of molecules to metal and semiconducting surfaces is of
tremendous practical importance for catalysis, self-assembly, and the
formation of molecule-metal junctions or templates for the growth of
porous materials. In many cases, these involve interactions that span
the range of strong chemisorption bonds to what has been thought of as
weakly physisorbed states. Traditionally, it was thought that the strongly
chemisorbed states could be adequately represented by standard GGA-type
functionals and that dispersion interactions would only play a small
role in physisorbed states. Recent work however using vdW-DF has
illustrated that dispersion interactions play important roles in both
regimes. For example, many studies have explored the physisorption of
organic molecules to different noble metal surfaces (i.e.\ Ag, Cu, Au and
Pt)~\cite{Berland09p155431, Toyoda09p2912, Toyoda09p78, Mura10p4759,
Toyoda10p134703, Sun10p201410, Lee10p155461, Wellendorff10p378,
Kelkkanen11p113401, Li12p121409, Morikawa12pS2,
Wyrick11p2944,Romaner09p053010,Yildrim13p20572}. These papers find that
vdW-DF brings the adsorption energies of these molecules into the
correct order of magnitude as compared with experiment, a vast
improvement over GGA. A survey of adsorption energies demonstrates that
they depend strongly on the size of the molecule. Figure~\ref{OrgBind}
illustrates this trend, relating binding energy to molecular size
($\sim N-N^2$). This simple trend arises even if the molecules all
have different structures---some chains, some with connected
six-membered rings, some without and some with additional molecules such
as N and O---and it is irrespective of the surface being considered.
This contrasts PBE results that show very little binding and no
dependence on molecular size.

Furthermore, some adsorption studies highlight the importance of atomic
relaxations. Results for melanine, PTCDA, and NTCDA adsorption show that
the difference in binding energy between fully relaxed and non
self-consistent calculations, or calculations with static molecule and
surface geometries, may be as large as 20\%~\cite{Mura10p4759}.
Adsorption studies also provide an understanding of the organisation of
overlayers of organic
molecules~\cite{Berland09p155431,Wyrick11p2944,Sun10p201410,
Lee10p155461, Yanagisawa11p235412, Li11p241406} and
water~\cite{Hamada10p115452,Carrasco11p026101} as well as their
spectroscopic signatures~\cite{Toyoda09p2912,
Toyoda09p78,Toyoda10p134703}.

Some of the earlier studies on adsorption employed vdW-DF1 and therefore
overestimate molecule-surface separation distances. This overestimation
can be larger for adsorption on some metals than what would be typically
expected for other systems \cite{Toyoda09p2912,rev8,BeArCoLeLuScThHy14}.

\begin{figure}
\includegraphics[width=0.8\columnwidth]{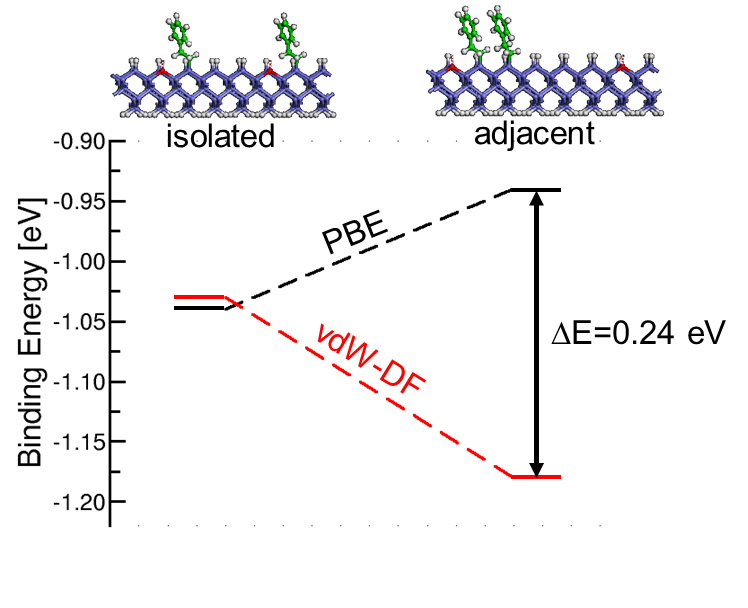}
\caption{\label{fig:styrene-H-Si}Interaction energies for isolated monomers and adjacent dimers adsorbed on the H-terminated
Si surface. The fact that the difference in energy between PBE and
vdW-DF for the dimers is roughly equal to that of gas phases dimers demonstrates the crucial role that  dispersion interactions play in the self-assembly of styrene wires on these surfaces.}
\end{figure}

The chemisorption of molecules on surfaces is also affected by
dispersion interactions. In some cases, the adsorption of a molecule can
be dramatically stabilised, allowing the molecule to stick to the
surface. For example, the adsorption of Alq$_3$ on the Mg(001)
surface~\cite{Yanagisawa11p235412} requires a delicate balance between
structural deformations and weaker surface adsorption. PBE
predicts that these molecules would be weakly bound, whereas vdW-DF
calculations stabilise the binding by up to 900 meV over the PBE
energies. In other cases, this interaction can stabilise a particular
adsorption configuration. For instance, traditional DFT calculations
predict that for benzene adsorbed to the silicon surface the
tight-bridge configuration is the most stable. In contrast, vdW-DF, in
spite of predicting a binding energy that differs merely by $\sim$0.2~eV
from the PBE result, shows a preference for the butterfly configuration, in agreement
with MP2 calculations and experiment~\cite{Johnston08p121404}. This
change in interaction energy is primarily due to molecule-surface
interactions mediated by dispersion forces~\cite{Chakarova-Kack06p155402}. In addition to the
interaction between the molecule and surface, interactions between
molecules on a surface can significantly affect how molecules arrange
themselves~\cite{berland11p135001}. For instance, vdW
interactions between styrene molecules drive their self-assembly on
H-terminated Si~\cite{Li11p241406}. Figure~\ref{fig:styrene-H-Si}
depicts the interaction energies for isolated
styrene molecules and dimers chemisorbed on the Si surface obtained with PBE and vdW-DF. As can be
seen from the plot, vdW interactions stabilise
the dimer configuration over the isolated monomers. Here, the energy
difference of 0.24~eV between PBE and the vdW-DF calculations for the
dimer configuration is roughly equal to the difference in binding energy
for gas-phase dimers ($\approx$ 0.21 eV) between the two functionals. They also play a
key role in the self-assembly of aromatic molecules on surfaces with low
corrugation, such as coinage metal surfaces, where the competition with
surface-mediated interactions~\cite{Berland09p155431, Wyrick11p2944,
Sun10p201410}---which has an even longer range than vdW
forces---ultimately controls adsorption geometries and energies.

\begin{figure*}
\includegraphics[width=\textwidth]{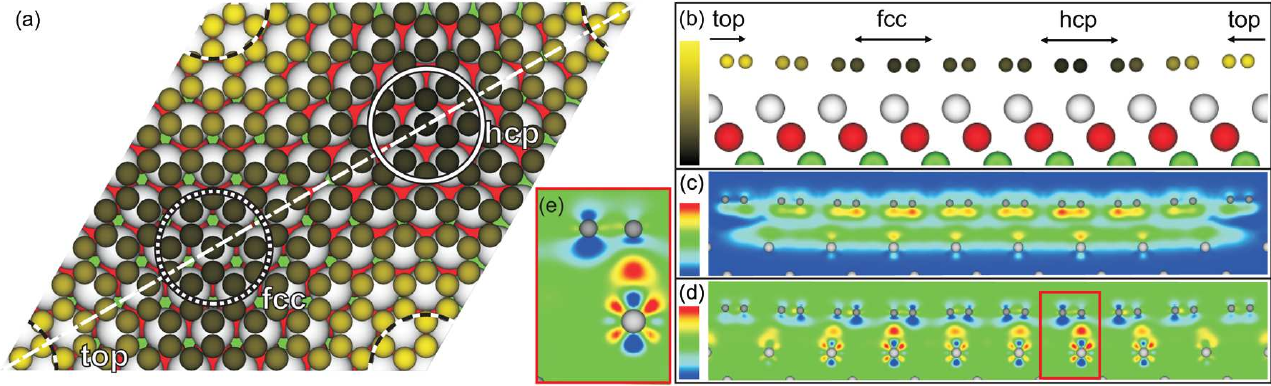}
\caption{\label{fig:Ir111}(a) Top view and (b) side view [cut along the
dashed line in (a)] of the relaxed structure of graphene/Ir(111)
obtained by vdW-DF. Regions of high-symmetry stacking (fcc, hcp, top)
are marked by circles (a) or arrows (b)--(d). (c) Visualization of the
nonlocal-correlation binding-energy density caused by adsorption. (d)
Charge transfer upon adsorption. A negative value indicates loss of
electron density. (e) Magnified view of red box in (d). Reprinted with
permission from \cite{Busse11}, \copyright\ 2011 American Physical
Society.}
\end{figure*}

Such interplay between strong and weak interactions are even more evident when considering the crossover from molecules to
graphene adsorbed on metal surfaces. In a concerted effort of experiment
(standing-wave x-ray diffraction) and theory (vdW-DF), epitaxy has been
studied in Ir(111) \cite{Busse11}. In fact, epitaxial growth on metals
is a key method for producing high-quality graphene on large scales. In the
interface region the strength of the C--C bonds varies. Large
incommensurate or weakly commensurate  superstructures are found for
lattice-mismatched systems. In the case of graphene on Ir(111), 
precise agreement between calculated (3.41~\AA\ of the C atom; DFT-GGA
calculations give 3.9~\AA) and measured (3.38~\AA) values for the mean
height has been obtained. This allows for the interpretation that the bonding of graphene to Ir(111) is due to vdW interactions with an additional antibonding average contribution arising from chemical interactions \cite{Busse11}. Despite its globally repulsive character, in certain areas of the large graphene moir{\'e} unit cell, charge accumulation between the Ir substrate and graphene C atoms is observed, indicating the formation of a weak covalent bond. In other words,
graphene on Ir(111) can be described as physisorption
with chemical modulation. Thanks to the vdW-DF analysis this can be
clearly illustrated, as in figure~\ref{fig:Ir111}. Here, the variation
over the unit cell of the nonlocal-correlation energy density and charge
transfer caused by adsorption is depicted, indicating a weak covalent
bond. This example emphasises the fact that dispersion interactions are
essential in weakly bound systems, while having significant
contributions to the binding in systems that are chemically bound to a
surface. These interactions thus play an important role in many
processes at surfaces that are fundamental to many modern applications
such as catalysis and molecular self-assembly. Therefore, this result provides a useful benchmark for the applicability of the nonlocal functional. \cite{Busse11} shows also another important aspect, namely that the vdW-DF is a handy tool in the everyday collaboration with experimentalists.

A final poignant example is that of the so-called CO-adsorption puzzle~\cite{Feibelman01p4018}. For CO chemisorption on metals, GGAs almost always favours a hollow site, whereas experiments reveal that top-site adsorption is typical, for example, on Pt, Rh, Cu.  This failure of GGAs was particularly surprising given the fact that CO bonds to a metal surface should be covalent in character and should thus be adequately described by GGAs. In this regard, Lazi{\'c} and co-workers reported an important early demonstration of vdW-DF, showing that the inclusion of van der Waals interactions offered a promising solution to this long standing issue~\cite{lazic10p045401}. Appearing in parallel were demonstrations that RPA also corrects GGA errors in 
finding a top-site CO-adsorption preference on Cu(111)
\cite{JarlKresse,Ren09p045402}, further indicating the importance
of nonlocal correlations. The behaviour of RPA was anticipated by Ref.\ \cite{lazic10p045401}. The CO-adsorption progress was important to development efforts as it gave early indications that vdW-DF could serve as a general-purpose functional
\cite{Paper6-01,rationalevdwdfBlugel12,BeArCoLeLuScThHy14,hybesc14}.
Thus emphasising the
the aim of the vdW-DF method to work both
for sparse matter (where GGA fails) as well as for dense matter (where
GGA often but not always succeeds).  

We note that the vdW-DF method and its variants are gaining acceptance
as a first-principle method that can be trusted to predict properties of
a wide array of materials. For example, Kokott and coworkers used vdW-DF-optB86b to provide first-principle predictions of the impact of nonmetallic substrates on the bandstructure of silicene overlayers \cite{kokott14p185002}. They demonstrate that a cleaved CaF$_2$
surface will leave the electron dynamics in the silicene overlayer
unchanged from that of a buckled free standing film and suggest
intensive experimental studies based on their first-principles vdW-DF
exploration. Also, Sun and co-workers have combined both vdW-DF1 and vdW-DF-optPBE with the DFT+U formalism to map out the hydrogenation of Pu and Pu-oxides \cite{sun14p164709}.
Here, the vdW-DF+U description is used in an \emph{ab initio} molecular
dynamics exploration of how H$_2$ molecules penetrate various Pu-oxide
surfaces. Generally, trust in the overall results seems to now be built
(when experiments are scarce) on the fact that vdW-DF represents a
parameter-free approach.

In general, the extensive use and benchmarking of vdW-DF suggests that it is indeed capable of accounting for dispersion interactions involved in the adhesion of molecules (and materials) to surfaces and within pores. These results give promise for extending the method to examining the properties of densely packed materials and large molecular systems.

%%%%%%%%%%%%%%%%%%%%%%%%%%%%%%%%%%%%%%%%%%%%%%%%%%%%%%%%%%%%%%%%%%%%%%%%
\subsection{Solids and liquids}
\label{sec:solids_liquids}
%%%%%%%%%%%%%%%%%%%%%%%%%%%%%%%%%%%%%%%%%%%%%%%%%%%%%%%%%%%%%%%%%%%%%%%%

In organic crystals, physical and chemical properties are strongly
influenced by the structure of the material. The structure depends on mutual forces between the participating atoms and often on growth conditions, which lead to differences in crystal packing. In such a thermal process weak forces also play a role,  for instance in the arrangement of hydrogen bonds and $\pi$-$\pi$ stackings within the crystal. This often makes it hard to consistently produce high quality
samples. Knowledge of which polymorphs can form and their respective
properties is therefore of great importance for both the synthesis and
application of organic crystals.

Given the number of different crystal structures that can form and how
close their corresponding cohesive energies are to the ground-state
structure, theoretical crystal structure prediction is a daunting task.
This challenge is even more dramatic for liquids, where there are no
periodic structures and innumerable different configurations can form.
As such, the study of vdW bonded solids and liquids is yet another good
application for vdW-DF.

Perhaps the prototypical class of van der Waals bonded solid are layered
materials such as graphite, hexagonal boron nitride ($h$-BN), and metal
dichalcogenides like MoS$_2$. Here, covalently bonded 2D layers are
attracted to each other via interplanar dispersion interactions. In
fact, initial attempts at developing vdW-DF were mostly focused on the
interactions between layered materials~\cite{Rydberg03p126402,
Rydberg03p606}, as discussed in chapter~\ref{sec:layered}. Successive
improvements of vdW-DF have led to an even better description of the
binding of graphite, as seen in figure \ref{fig:graphite}, thus setting
vdW-DF up for extensive studies of other van der Waals bonded layered
materials~\cite{BjorkmannLayered1,BjorkmannLayered2,SoftLayered,behy14},
such as V$_2$O$_5$~\cite{Londero10p054116, Londero11p1805} and
graphane~\cite{Rohrer11p165423}.

Luo and co-workers used vdW-DF-C09 to examine the atomic, electronic, and thermoelectric properties of Bi$_2$Se$_3$ and Bi$_2$Te$_3$ \cite{Luo12p184111}. They showed that by including van der Waals interactions they were able to obtain much better agreement with the experiment for the structural properties of the two materials. Furthermore, using this structural model their band structures gave equally good qualitative agreement with photoemission spectra. Even more interestingly, they demonstrated that strain could be used to tune the thermoelectric properties, with the $n$-type Seebeck coefficient of Bi$_2$Te$_3$ increasing under compressive inplane strain and Bi$_2$Se$_3$ increasing under tensile strain.

The ability to accurately model the interactions between planes opens up
the potential for studying real world applications such as the
intercalation of ions between graphene
sheets~\cite{Ziambaras07p155425,BeArCoLeLuScThHy14}, which is relevant
to battery technology, and understanding the pore size dependent
adsorption of neutral non-polar molecules~\cite{Ihm12p424205,Ihm14p1},
such as H$_2$ and CH$_4$, which is critical for alternative energy
technologies. The spintronics of a ferromagnet/graphene junction,
Co(0001) with graphene, was studied in Ref.~\cite{sipahi14p104204}. This
system is another example in which there is strong charge rearrangement
at short distances but the binding comes from nonlocal correlation.

Complementary to the 2D layered materials, crystal structures comprised
of carbon nanotubes and long-chain hydrocarbons have also been studied
using vdW-DF~\cite{Schroder03p721, Kleis05p192, Kleis05p164902,
Kleis07p100201, Kleis08p205422}. These studies were discussed in more
detail in a previous review~\cite{langrethjpcm2009}. It is interesting
to note, however, that the intertube interaction energy was determined
to be on the order of that found in graphite.

Due to the many possible molecules, a huge number of potential molecular
crystals are believed to exist, out of which only a comparatively few
have been synthesised. The inability to predict molecular crystal
structures is one of the most notorious failures of traditional DFT
methods. With this in mind, vdW-DF has been applied to the study of
organic crystal structures, exhibiting significant predictive
capabilities~\cite{pi_conjmolcrys,rev8,draxl09p125010}.

In addition, vdW-DF has been used to explore the physical properties of
functional materials. For instance, a ferroelectric organic crystal comprised of
phenazine and chloranilic acid~\cite{Lee12p104102} and boron based
hydrogen-storage materials, such as ammonia borane
(NH$_3$BH$_3$)~\cite{Lin12p2172} and magnesium borohydride
(Mg(BH$_4$)$_2$)~\cite{Bil11p224103}, have been studied. In the case of
Mg(BH$_4$)$_2$ it is not obvious that dispersion interactions would play
a role in defining the crystal structure. It has long been assumed that
covalent interactions dominate and are responsible for its structure.
However, the results of \cite{Bil11p224103} show that the inclusion of
vdW interactions between the BH$_4$ units is crucial for getting the
correct ground-state structure. The application of vdW-DF led---for
the first time---to good agreement with experiment, favouring the
$\alpha$-Mg(BH$_4$)$_2$ phase (P6122) and a closely related
Mn(BH$_4$)$_2$-prototype phase (P3112) over a large set of polymorphs at
low temperatures. This study thus demonstrates the need to go beyond
semilocal density functional approximations for a reliable description
of crystalline high valent metal borohydrides.

A particularly interesting non-standard dimer that forms a molecular
crystal is presented in~\cite{Kolb13p3642}, i.e.\ the phenalenyl dimer
and closed-shell analogues. Phenalenyl---an open-shell neutral radical
that can form both $\pi$-stacked dimers and conducting molecular
crystals---has gained attention for its interesting and potentially
useful electrical and magnetic properties. The results indicate that
vdW-DF is capable of qualitatively describing the interaction between
two neutral radicals in the $\pi$-stacked configuration, giving binding
distances that are significantly below the sum of the van der Waals
radii, in agreement with experiment.

A recent derivation of the appropriate formalism for calculating stress
in vdW-DF can be a useful tool in many studies of systems at finite pressure.
  For instance,
  the pressure-dependent phase transitions of amino acid crystals have been
explored~\cite{Sabatini12p424209}.

Bulk water itself has long been a particularly difficult material to
model. Results of vastly overestimated LDA and GGA freezing temperatures
of water triggered the first applications of vdW-DF to small water
clusters. The smallest water system, i.e., the water dimer, is part of
the S22 data set and its vdW-DF results have been reported as part of
the benchmark calculations \cite{optX,Lee10p081101, cooper10p161104, behy14}.
Further exploratory studies have focused on the energetic, structural,
and vibrational properties of small water clusters (H$_2$O)$_n$ with $n$
$\le$ 6 and standard ice I$_h$~\cite{Kelkkanen09p46102,Hamada10p214503,
Kolb11p045116}. In addition, while vdW-DF shows significant improvements with
regards to the structure and binding energies over LDA and GGA, a
remarkable improvement is also found for the vibrational frequencies.
In-depth analyses shed light on why LDA and GGA fail in describing
water.

The improvements of vdW-DF for small water clusters gives hope that the
freezing temperature of bulk water predicted with DFT calculations might
also be improved. Indeed, recent vdW-DF simulations show remarkable
differences in the predicted structure of water~\cite{Mogelhoj11p14149,
Wang11p024516}, again better aligned with experiment.

Finally, it should be noted that vdW-DF is also capable of describing
traditional densely packed solids. Whereas the vdW-DF1 and vdW-DF2
descriptions predict somewhat too large volumes for some inorganic
crystals, other modern vdW-DF variants actually perform better than a
standard semilocal functional like PBE~\cite{vdWsolids, BeHy13, behy14, hybesc14}.

%%%%%%%%%%%%%%%%%%%%%%%%%%%%%%%%%%%%%%%%%%%%%%%%%%%%%%%%%%%%%%%%%%%%%%%%
\subsection{Biological molecules}
\label{sec:biomolecules}
%%%%%%%%%%%%%%%%%%%%%%%%%%%%%%%%%%%%%%%%%%%%%%%%%%%%%%%%%%%%%%%%%%%%%%%%

Soft matter is fundamental for all life. Not only do dispersion forces
play a critical role in defining form and function of genetic material,
but they are crucial in the interactions of organic molecules with the
genetic material and thus critical for our understanding of how to treat
many diseases.

For a review on molecular simulations of biomaterials, see
\cite{Kolb12p1230006}. Indeed, initial vdW-DF studies focused on
examining the interactions between and within DNA base
pairs~\cite{Cooper08p204102, Cooper08p1304} and between base pairs and
anti-cancer drugs~\cite{LiCoThLuLa09}. Even with the systematic
overestimation of separation distances typical of vdW-DF1, the
importance of including dispersion interactions was evident. For
instance, the twist and rise of DNA clearly emerge as a result of
nonlocal interactions, with unprecedented agreement with x-ray
crystallography data and computationally costly quantum-chemical
calculations. A more detailed summary of our early studies can be found
in \cite{langrethjpcm2009}. Recently, a Harris-type
scheme~\cite{BeLoScHy13} has been explored for speeding up vdW-DF
calculations of large sparse matter system, such as the interactions
between strands of DNA~\cite{LoHySc13}. The results demonstrate the
ability to scale such calculations to large numbers of atoms with
limited effects on accuracy.

There has been recent interest in using DNA molecules as molecular
wires. This involves the adsorption and interaction of nucleobases with
metal surfaces and other substrates such as graphene. In theoretical
studies, graphene is also used as model for carbon nanotubes, which can
be used in applications in medicine and as sensors. The adsorption of
nucleobases and other biological molecules on graphene provides a venue
for studying the complicated interactions between nucleobases in
biological matter and thus casts light on processes such as molecular
recognition and self assembly. As a first step, the adsorption of
adenine on graphene was studied and compared with the two-dimensional
crystal forming in the denser phase~\cite{berland11p135001}. This work
was later followed by studies of all five nucleobases~\cite{Le12p424210}

\begin{figure}
\includegraphics[width=\columnwidth]{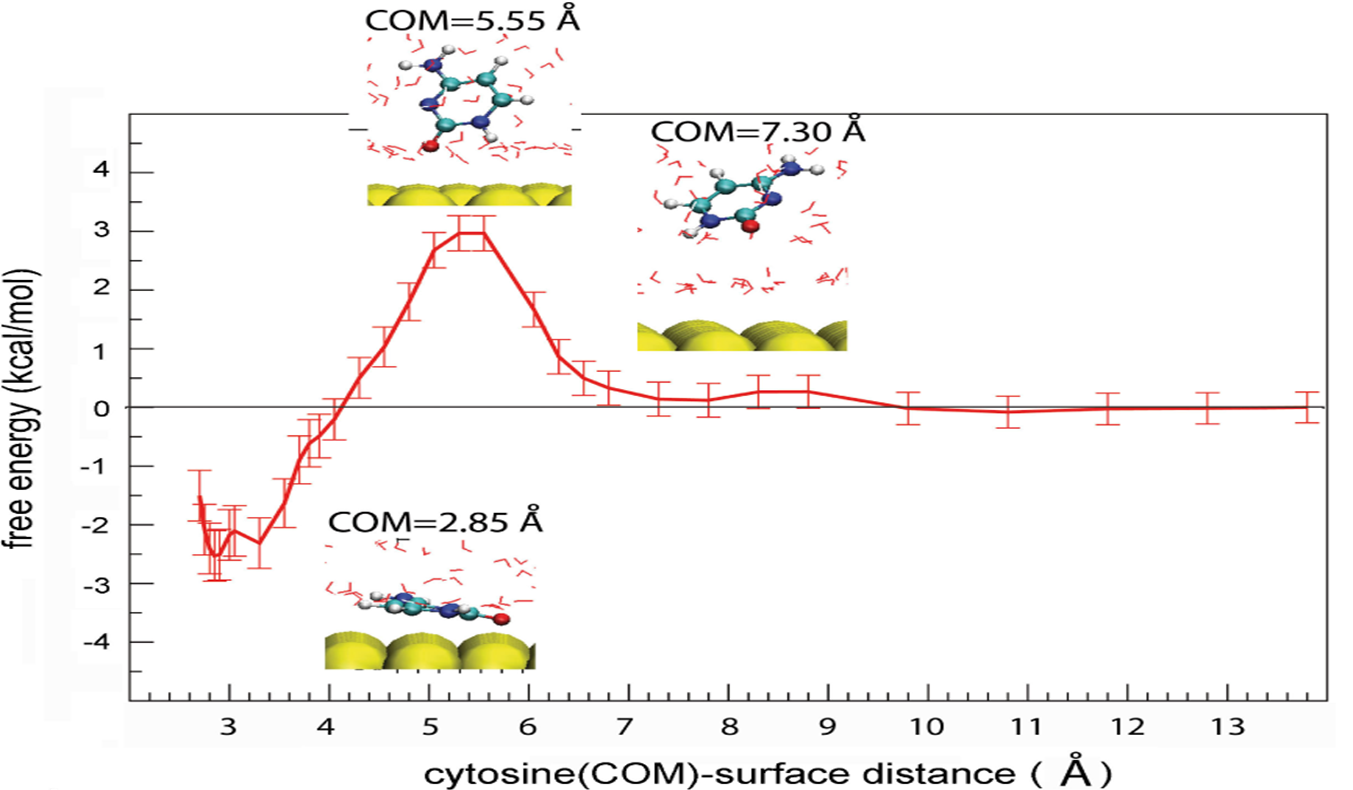}
\caption{\label{fig:cytosine} Potential energy surface for the adsorption of cytosine to the Au(111) surface obtained with GolDNA-AMBER force field derived from vdW-DF calculations. Adapted with permission from
\cite{rosa14p1707}, \copyright\ 2013 American Chemical Society.}
\end{figure}

It is often necessary to study biomolecules in solution, possibly
including the effect of entropy. To this end, classical molecular
dynamics atomistic simulations are used, utilizing force fields to
describe the atomic interactions. An example where vdW-DF calculations
were used to tune parts of such a force field is the study of DNA bases
in aqueous solution adsorbed on Au(111), defining the new
GolDNA-AMBER force field \cite{rosa14p1707}. Using this vdW-DF derived potential they were able to map out the potential energy surface for the adsorption/desorption of DNA bases from the Au (111) surface (see figure~\ref{fig:cytosine}).

Naturally, given the societal benefits of understanding the interactions
of biomolecules, either from a biological or technological perspective,
it is expected that applications in this arena will surely increase in
years to come. This is especially true, given advances in computers and
algorithms that will allow for the study of materials on biologically
relevant scales.

Dispersion interactions are ubiquitous and our understanding of them is
still developing. As demonstrated in many recent applications, these
forces are necessary not only for prototypical weakly bound systems,
but also play a vital role in materials where they were traditionally
thought of as negligible. As new methods, algorithms, and computational
resources continue to develop and evolve, we expect to see methods that
include these interactions becoming even more widely used.

%%%%%%%%%%%%%%%%%%%%%%%%%%%%%%%%%%%%%%%%%%%%%%%%%%%%%%%%%%%%%%%%%%%%%%%%
\section{Conclusions and outlook}
%%%%%%%%%%%%%%%%%%%%%%%%%%%%%%%%%%%%%%%%%%%%%%%%%%%%%%%%%%%%%%%%%%%%%%%%

Today's emphasis in computational materials science has shifted from
semiconductors and metals to nanomaterials and biological materials.
With this new emphasis, the standard tools of the past---LDA- and
GGA-based DFT---often fall short because of the many low-density
regions, i.e.\ van der Waals regions. Several pictures of van der
Waals forces have been touched upon in this review, starting with the
London picture and later continuing with the vdW-DF method which
unifies these different pictures of vdW forces within a DFT framework
derived from the coupling-constant integration of the adiabatic connection formula. The efforts to develop vdW-DF have been
driven by the need for general purpose theoretical tools that can
describe molecules, bulk materials, and surfaces on equal footing. To do so, we need to describe both dense and sparse electron systems. 

For ground-state properties DFT has been used for half a century, yet only within the most recent decade has its applicability to sparse, and hence general matter, been significantly improved. There are many  examples of systems where such methods are needed, far beyond those indicated in this review. With the aim of designing general-purpose methods,  non-empirical criteria are preferable. vdW-DF is such a method, delivering simple yet accurate and robust functionals.

Initial vdW-DF developments focused on the nonlocal correlation. Contact had to be made with available vdW results at that time; typically the asymptotic behaviours of simple model systems. The ALL and related descriptions gave the correct $R^{-6}$, $z^{-3}$, and $d^{-2}$ forms for model systems and simple formulas for the vdW parameters. These led to promising quantitative results.
The later-developed vdW-DF0 and vdW-DF1 functionals further advanced the
field with significant improvements in the description of vdW-bonded
regions. The vdW-DF1, usually called vdW-DF, is in common use and has
given many valuable results, especially for adsorbed molecules and bulk
materials. Physically
motivated by issues of overestimation of separations and
underestimation of H-bond energies of small molecule and
molecule-molecule interactions, vdW-DF2, with its update to the exchange energy and the
nonlocal correlation energy, brought further enhancements.

More recently, the focus has switched to exchange with the purpose of
improving performance and for internal consistency reasons. Issues
related to the overestimation of separation distances were solved with
new exchange variants paired with vdW-DF1. 
Modern variants to the exchange also provide a good account of
systems beyond the vdW regime, such as covalent solids and systems with
a mix of different binding characteristics, giving interaction energies
that now approach chemical accuracy. For internal consistency, one could also use an exchange functional derived from the same plasmon-based model from which the nonlocal correlation energy was derived. The recent functional vdW-DF-cx stands
out in its attempt to do so.

A major advance on the computational side was the move
from slow, non self-consistent, in-house codes to stable, efficient,
FFT-based codes implemented in mainstream DFT packages.
Now, the many diverse and successful applications of the vdW-DF method have helped to paint a modern picture of van der Waals forces, where they play an essential role, even in systems traditionally thought to be dominated by ionic or covalent bonding.

The excellent performance for bulk systems for modern variants makes
vdW-DF a contender for a spot in the standard repertoire of the contemporary materials scientist. There are phenomena and processes where the flexibility of DFT is called for: For instance, when the extended nature of the polarised electrons really matters, like for molecules on metal surfaces where the characteristics of the bulk and surface are drastically different; or for charge transfer and screening at, for example, grain boundaries. Catalytic processes on surfaces and in sparse matter, such as MOFs is yet another example. 

As for the future of vdW-DF, a good starting point is to consider its weaknesses, as we see them. To begin with, current variants of vdW-DF lack spin and the explicit exchange mechanisms typical of hybrid functionals. Other issues include low accuracy for noble-gas dimers, incorrect asymptotic power laws of low-dimensional structures and metals, and $d$- and $f$-electron effects. 

Spin properties can be introduced by extending the treatment through first-principle arguments. Like in the local spin density approximation, this means that up and down spins will be described by different potentials. Such a development will make it possible to describe, for instance, ionization potentials of atoms and molecules, and magnetic materials, which in fact are often bound by van der Waals forces.

The lack of hybrid features might be solved by attempting to apply arguments used to extend GGA. Hybrid functionals often only change the exchange. The fact that PW86r was chosen for vdW-DF2 because it mimics Hartree-Fock could be helpful in making such an extension.

For the future, higher accuracy will likely be obtainable for many physical effects.
A possible solution for the problem of dealing with noble gases could be to carefully design a gap mechanism, like that in dielectric semiconductor models, and then possibly learn from the VV functionals. For the asymptotic behaviour of low-dimensional structures and metals possible solutions require updating the plasmon model to capture sophisticated many-particle effects.

The ultimate solution would be to generalize the theory underlying the unified treatment in~\cite{hurylula99}, involving an explicit solution of the electrodynamics to smaller separations and to solve it efficiently. In fact, the vdW-DF0 attempts to do such a generalization, but for the very restrictred layered geometry and with an inaccurate plasmon model. Ultimately, strong connections with these earlier models would be extremely beneficial for the future development of vdW-DFs.

As more and more systems are revealed to be affected by van der Waals forces,  we believe that vdW-DF could replace GGA.  On the other hand, methods such  as RPA and developments in that direction might replace vdW-DF for smaller-to-medium sized van der Waals bonded systems. However, bigger systems and  time-dependent  calculations will still require DFT-based methods.  Nevertheless, new functionals within the vdW-DF family are expected to be derived or present variants may be extended (as mentioned, spin is currently being developed). It is also likely that vdW-DF will continue to inspire the development of other nonlocal correlation functionals that might capture additional physical effects or be specialized for particular kinds of systems.

If we dare to speculate, we believe that nonlocal correlation functionals will replace dispersion-corrected DFT in practical use. Nonlocal correlation functionals rest on firmer physical foundations, are more flexible, can be 
as fast as local and semi-local functionals and are available in all major codes. For systems outside the reach of DFT, force field methods will 
still have its place.

In conclusion, dispersion interactions are ubiquitous and our understanding of them is still developing. As demonstrated in many recent applications, these forces are necessary not only for prototypical weakly bound systems, but also play a vital role in materials where they were traditionally thought of as negligible.  New systems for applying the vdW-DF method are probably bigger, more
extensive, and sparser than today's materials and more often involve time-dependent phenomena. Disordered systems,
liquids, systems with several length scales, systems with several kinds
of competing interactions, and biological systems are cases of such systems. The
vdW-DF method adheres to important conservation rules---a feature that
has good potential for transferability and lays the foundation
for broader conclusions about the capacity of the method. With positive
results in test cases that can be seen as difficult and that cover a
range of problems where interactions compete \cite{BeArCoLeLuScThHy14},
there is also potential for good performance for general problems. The
vdW-DF method so far provides competitive density functionals for sparse
matter and its robust and flexible formulation offers a promise for further
improvements \cite{BeArCoLeLuScThHy14,
langrethjpcm2009}. As new methods, algorithms, and computational resources continue to develop and evolve, we expect to see methods that include these interactions becoming even more widely used. That said, there is truly an open playground
ready to be explored---where dispersion interactions are no longer an
afterthought, but a key interaction that must be understood.

%%%%%%%%%%%%%%%%%%%%%%%%%%%%%%%%%%%%%%%%%%%%%%%%%%%%%%%%%%%%%%%%%%%%%%%%
\begin{acknowledgements}
Our late colleague David Langreth was first invited by Reports on Progress in Physics (RoPP) to review our vdW-DF method. Sadly, David barely got time to start it. We owe him a lot and  therefore dedicate this review to him. We thank K.\,Z.\ Soliman for the careful checking of the
references. We also thank M.\ Kuisma and T.\,L.\ Einstein for comments and discussions. Work by KB, ES, and PH was supported by the Swedish research council (VR) under grants VR-2011-4052 and VR-2010-4149 and by
the Chalmers Area of Advance, Materials. TT acknowledges support
from US NSF Grant No.\ DMR-1145968. VRC was supported by the Materials
Sciences and Engineering Division, Office of Basic Energy Sciences, US
Department of Energy. KL was supported by the US Department of Energy,
Office of Basic Energy Sciences, Division of Chemical Sciences,
Geosciences, and Biosciences under award DE-FG02-12ER16362.
\end{acknowledgements}
%%%%%%%%%%%%%%%%%%%%%%%%%%%%%%%%%%%%%%%%%%%%%%%%%%%%%%%%%%%%%%%%%%%%%%%%

%%%%%%%%%%%%%%%%%%%%%%%%%%%%%%%%%%%%%%%%%%%%%%%%%%%%%%%%%%%%%%%%%%%%%%%%
\bibliography{references}
\bibliographystyle{ROPP}
%%%%%%%%%%%%%%%%%%%%%%%%%%%%%%%%%%%%%%%%%%%%%%%%%%%%%%%%%%%%%%%%%%%%%%%%

%%%%%%%%%%%%%%%%%%%%%%%%%%%%%%%%%%%%%%%%%%%%%%%%%%%%%%%%%%%%%%%%%%%%%%%%
\end{document}